\documentclass[10pt]{article} 

\PassOptionsToPackage{numbers, compress}{natbib}

 \usepackage[preprint]{neurips_2025}
\usepackage[table]{xcolor}
\usepackage{todonotes}
\usepackage{hyperref}
\usepackage{wrapfig}
\usepackage{soul} 
\usepackage{amsmath}
\usepackage{amssymb}
\usepackage{graphicx}
\usepackage{algorithm}
\usepackage{algpseudocode}
\usepackage{placeins}
\usepackage{xurl}
\usepackage{pgfplotstable} 
\pgfplotsset{compat=1.17} 
\usepackage{caption} 
\usepackage{subcaption} 
\usepackage{multirow} 
\usepackage{rotating} 
\usepackage{tabularx} 
\usepackage{booktabs} 
\usepackage{comment}
\excludecomment{draft} 
\usepackage{multibib}
\newcites{appI}{References for Extended Technical Appendix} 

\newcommand{\Xtrue}{X_{\text{true}}}
\newcommand{\Xpred}{X_{\text{pred}}}
\newcommand{\Xref}{X_{\text{ref}}}
\newcommand{\Zlatent}{Z} 
\newcommand{\Zref}{Z_{\text{ref}}}
\newcommand{\R}{\mathbb{R}}
\newcommand{\Ibb}{\mathcal{I}_{\text{bb}}}
\newcommand{\Isc}{\mathcal{I}_{\text{sc}}}
\newcommand{\Nbb}{N_{\text{bb}}}
\newcommand{\Nsc}{N_{\text{sc}}}
\newcommand{\Lhno}{\mathcal{L}_{\text{HNO}}}
\newcommand{\Lbb}{\mathcal{L}_{\text{BB}}}
\newcommand{\Lsc}{\mathcal{L}_{\text{SC}}}
\newcommand{\Ldec}{\mathcal{L}_{\text{Dec}}} 
\newcommand{\Lcoord}{\mathcal{L}_{\text{coord}}}
\newcommand{\Lmseh}{\mathcal{L}_{\text{mse\_dih}}} 
\newcommand{\Ldivh}{\mathcal{L}_{\text{div\_dih}}} 

\begin{document}

\title{Generative Modeling of Full-Atom Protein Conformations using Latent Diffusion on Graph Embeddings}

\author{
  Aditya Sengar\thanks{Signal Processing Laboratory (LTS2), EPFL, Lausanne, Switzerland}
                 \thanks{Institute of Bioengineering, EPFL, Lausanne, Switzerland}
                 \thanks{Corresponding author} \\
  \texttt{aditya.sengar@epfl.ch}
  \And
  Ali Hariri\footnotemark[1] \\
  \texttt{ali.hariri@epfl.ch}
  \And
  Daniel Probst\thanks{Bioinformatics Group, University \& Research Wageningen, The Netherlands} \\
  \texttt{daniel.probst@wur.nl}
  \And
  Patrick Barth\footnotemark[2] \thanks{Ludwig Institute for Cancer Research, Lausanne, Switzerland} \footnotemark[3]\\
  \texttt{patrick.barth@epfl.ch}
  \And
  Pierre Vandergheynst\footnotemark[1] \footnotemark[3]\\
  \texttt{pierre.vandergheynst@epfl.ch}
}

\maketitle

\begin{abstract}
Generating diverse, all-atom conformational ensembles of dynamic proteins such as G-protein-coupled receptors (GPCRs) is critical for understanding their function, yet most generative models simplify atomic detail or ignore conformational diversity altogether. We present latent diffusion for full protein generation (LD-FPG), a framework that constructs complete all-atom protein structures, including every side-chain heavy atom, directly from molecular dynamics (MD) trajectories. LD-FPG employs a Chebyshev graph neural network (ChebNet) to obtain low-dimensional latent embeddings of protein conformations, which are processed using three pooling strategies: blind, sequential and residue-based. A diffusion model trained on these latent representations generates new samples that a decoder, optionally regularized by dihedral-angle losses, maps back to Cartesian coordinates. Using D2R-MD, a $2\,\mu\text{s}$ MD trajectory (12 000 frames) of the human dopamine D$2$ receptor in a membrane environment, the sequential and residue-based pooling strategy reproduces the reference ensemble with high structural fidelity (all-atom lDDT $\sim 0.7$; $C\alpha$-lDDT $\sim 0.8$) and recovers backbone and side-chain dihedral-angle distributions with a Jensen--Shannon divergence $<0.03$ compared to the MD data. LD-FPG thereby offers a practical route to system-specific, all-atom ensemble generation for large proteins, providing a promising tool for structure-based therapeutic design on complex, dynamic targets. The D2R-MD dataset and our implementation are freely available to facilitate further research.

\end{abstract}

\section{Introduction}
\label{sec:introduction}

Proteins function as dynamic molecular machines whose biological activities critically depend on transitioning between distinct conformational states~\citep{karplus2005molecular, henzler2007dynamic}. Landmark artificial intelligence methods like AlphaFold2~\citep{Jumper2021AlphaFold2} and others~\citep{Baek2021RoseTTAFold, lin2023evolutionary, Abramson2024, wohlwend2024boltz} have advanced structure prediction but predominantly predict single static conformations, limiting their utility for systems with conformational heterogeneity. Accurate modeling of an ensemble of accessible conformations is essential to elucidate protein function and guide therapeutic design~\citep{boehr2009role, teague2003implications, Chen2020, chen2025computational}. Crucially, these ensembles must explicitly represent all atomic details, particularly side chains, whose subtle conformational rearrangements often govern molecular recognition and catalytic mechanisms \cite{najmanovich2000side}.

Despite rapid advancements, existing generative models frequently fall short in capturing the detailed dynamics of side-chain movements specific to particular proteins~\citep{rudden2022deep}. Many powerful methodologies have focused either on \textit{de novo} backbone designs~\citep{watson2023novo, wu2024protein, Geffner2025Proteina} or on generating static all-atom structures~\citep{chu2024all, Ingraham2023}, but neither produce comprehensive conformational ensembles. Ensemble-generating approaches—via perturbations of static predictors~\citep{vani2023alphafold2, lu2023str2str}, flow-matched variants~\citep{jing2024alphafold}, or general MD-trained generators~\citep{Lewis2024BioEmu, zheng2024predicting}—typically operate at the backbone or coarse-grained level. Consequently, they fail to capture the intricate all-atom rearrangements, particularly involving side chains, that are critical for function. Although promising latent space~\citep{Fu2023LatentDiff, Lu2024PLAID, Janson2024-idpSAM} and physics-informed models~\citep{Wang2024ConfDiff, Liu2024EGDiff} have emerged, their capability to generate high-resolution all-atom ensembles from MD data reflecting functional transitions remains unproven, highlighting a significant unmet need for specialized generative frameworks.

G protein-coupled receptors (GPCRs) provide a compelling example of dynamic systems wherein precise all-atom modeling is indispensable~\citep{Latorraca2017}. This large family of transmembrane receptors, comprising over 800 human members, is responsible for mediating most known transmembrane signal transduction~\citep{Zhang2024, Casad2023} and is targeted by approximately 50\% of all marketed drugs~\citep{Yang2021, Knox2023}. GPCR signaling involves conformational transitions among multiple states, frequently induced by ligand binding~\citep{hilger2018structure}, and occurs through intricate allosteric mechanisms where specific side-chain interactions are crucial~\citep{Monod1965, latorraca2017gpcr}. Capturing such dynamic, atomically detailed landscapes is vital to understanding receptor signaling specificity~\citep{jefferson2023computational}, biased agonism~\citep{kenakin2019biased}, and the design of drugs targeting unique allosteric sites~\citep{Goupil2012, JeffreyConn2009, Do2023GPCRallostery}, yet studying these events computationally remains formidable~\citep{Kapla2021, He2022, rodriguez2020gpcrmd}. Current predictive methods for GPCR dynamics~\citep{LopezCorrea2024} do not generate the comprehensive all-atom conformational landscapes essential for mechanistic understanding.

To address this critical requirement, we introduce \textit{Latent Diffusion for Full Protein Generation} (\textbf{LD-FPG}), a generative framework designed to \textbf{learn and generate diverse, all-atom conformational ensembles from existing MD simulation data of a target protein}, explicitly including side-chain details. Rather than simulating new trajectories, LD-FPG leverages extensive MD datasets. Our approach employs a Chebyshev Spectral Graph Convolutional Network (ChebNet)~\citep{Defferrard2016} to encode all-atom MD snapshots into a compact latent representation. A Denoising Diffusion Probabilistic Model (DDPM)~\citep{ho2020denoising} is then trained to explore this learned latent manifold, and latent representations are decoded back into full all-atom Cartesian coordinates. We demonstrate our framework on extensive MD simulations of the human dopamine D2 receptor (D2R), systematically evaluating distinct decoder pooling strategies.
Our primary contributions in this paper are:
\begin{enumerate}
    \item To the best of our knowledge, we present the first latent diffusion modeling framework specifically tailored to generate \textit{complete, all-atom protein conformational ensembles}, capturing both backbone and \textbf{side-chain} dynamics directly from MD simulations.
    \item We introduce and critically evaluate a novel graph-based autoencoder architecture utilizing ChebNet combined with distinct decoder pooling strategies, offering insights into dynamic protein ensemble generation.
    \item Using the D2R system, we demonstrate our method’s ability to generate high-fidelity ensembles, highlight the advantages of residue-based pooling, and assess the impact of auxiliary dihedral loss terms on generative accuracy.
\end{enumerate}

Our approach provides a computationally efficient tool for exploring complex dynamics in switchable proteins, supporting both fundamental mechanistic studies and drug discovery applications. The remainder of this paper is structured as follows: Section~\ref{sec:related_work} reviews related work; Section~\ref{sec:methodology} details our proposed methodology; Section~\ref{sec:experiments} presents the experimental setup and results; and Section~\ref{sec:conclusion_future} summarizes our findings and outlines future directions.

\section{Related Work}
\label{sec:related_work}

\textbf{Generative Models for Protein Design and Static Structure Prediction.}
Deep generative models have made significant strides in protein science, employing techniques such as diffusion \citep{ho2020denoising}, flow-matching \citep{Lipman2022, Huguet2024FoldFlowPP}, and learned latent spaces \cite{Fu2023LatentDiff, Eguchi2022IgVAE} to tackle complex structural tasks. For \textit{de novo} backbone design, various methods, often inspired by the success of predictors like AlphaFold2 \citep{Jumper2021AlphaFold2} and diffusion-based approaches like RFdiffusion \citep{watson2023novo} have emerged. Notable examples include FoldingDiff \citep{wu2024protein}, alongside flow-matching models such as FoldFlow2 \citep{Huguet2024FoldFlowPP}. Latent space strategies have also been pivotal in this domain; for instance, LatentDiff \citep{Fu2023LatentDiff} generates novel protein backbones using an equivariant diffusion model within a condensed latent space, while Ig-VAE \citep{Eguchi2022IgVAE} utilizes a variational autoencoder for class-specific backbone generation (e.g., for immunoglobulins). While powerful for creating new folds (e.g., Proteina \citep{Geffner2025Proteina}) or specific components, these methods typically target soluble proteins, often generate only backbone coordinates, and are not primarily designed for sampling multiple conformations of a specific, existing protein. The generation of complete static all-atom structures has also seen considerable progress. Models like Protpardelle \citep{chu2024all} and Chroma \citep{Ingraham2023} can produce full static structures from sequence information. Diffusion-based generative models such as AlphaFold3 \citep{Abramson2024} and Boltz-1 \citep{wohlwend2024boltz} also provide detailed single-state predictions of all-atom structures and complexes. Other approaches, like PLAID \citep{Lu2024PLAID}, integrate predictors with diffusion samplers. To complement backbone or static generation, methods including FlowPacker \citep{lee2025flowpacker} and SidechainDiff \citep{liu2023predicting} focus on side-chain packing or prediction. However, these tools predominantly yield single static structures. Furthermore, decoupling backbone and side-chain generation risks overlooking their critical interplay during the complex dynamic transitions \citep{Latorraca2017} relevant to the conformational landscapes our work aims to capture.

\textbf{Modeling Protein Conformational Diversity: From General Strategies to MD-Informed Approaches.}
Beyond single static structures, capturing a protein's conformational diversity is crucial for understanding its function. 
Initial strategies to explore this diversity include learning from structural variations in experimental databases (e.g., Str2Str \citep{lu2023str2str}) or perturbing static predictions to sample conformational space, as seen with AF2-RAVE \citep{vani2023alphafold2} for GPCRs and AlphaFlow/ESMFlow \citep{jing2024alphafold} more broadly. While these methods effectively broaden sampling, capturing system-specific, native-like dynamics often benefits from models trained directly on simulation data, which can provide a richer representation of a given protein's accessible states.
Among such simulation-informed approaches, latent space models have shown promise. For example, EnsembleVAE \citep{Mansoor2024EnsembleVAE}, trained on MD snapshots and crystal structures of the soluble protein K-Ras, generates C$_\alpha$ ensembles from sampled latent features with full-atom picture subsequently produced by RoseTTAFold \citep{Baek2021RoseTTAFold}. Similarly, idpSAM \citep{Janson2024-idpSAM}, trained on extensive simulations of intrinsically disordered regions (IDRs) with an implicit solvent model, produces C$_\alpha$ trace ensembles that can then be converted to all-atom representations. Such approaches demonstrate the power of leveraging simulation data within generative frameworks, though their application has often focused on particular protein classes (e.g., soluble proteins, IDRs) or involved multi-stage processes for generating final all-atom structures.
Further advancements in learning ensembles directly from MD simulations encompass a range of techniques. These include methods like ConfDiff \citep{Wang2024ConfDiff} (force-guided diffusion), P2DFlow \citep{Jin2025P2DFlow} (SE(3) flow matching), and MD-Gen \citep{Jing2024MDGen} (continuous trajectories). Larger-scale models such as BioEmu \citep{Lewis2024BioEmu} and Distributional Graphormer (DiG) \citep{zheng2024predicting} aim to learn from vast MD datasets or equilibrium distributions, while experimentally-guided approaches like EGDiff \citep{Liu2024EGDiff} integrate diverse data types. Collectively, these MD-informed methods signify substantial progress in modeling protein dynamics. However, a persistent challenge, particularly when these methods are applied generally or to very large datasets, is the consistent generation of high-resolution, all-atom ensembles that fully capture system-specific details. This is especially true for intricate side-chain rearrangements within native environments, such as lipid membranes for G protein-coupled receptors (GPCRs), which are essential for elucidating their functional transitions \citep{Latorraca2017}. The development of generative models that can directly learn and sample such specific, all-atom conformational landscapes from relevant MD data, particularly for complex targets like GPCRs \citep{rodriguez2020gpcrmd}, thus remains an important frontier. 

\section{Methodology}
\label{sec:methodology}

\begin{figure*}[t]
    \centering
    \includegraphics[
        height=0.27\textheight
    ]{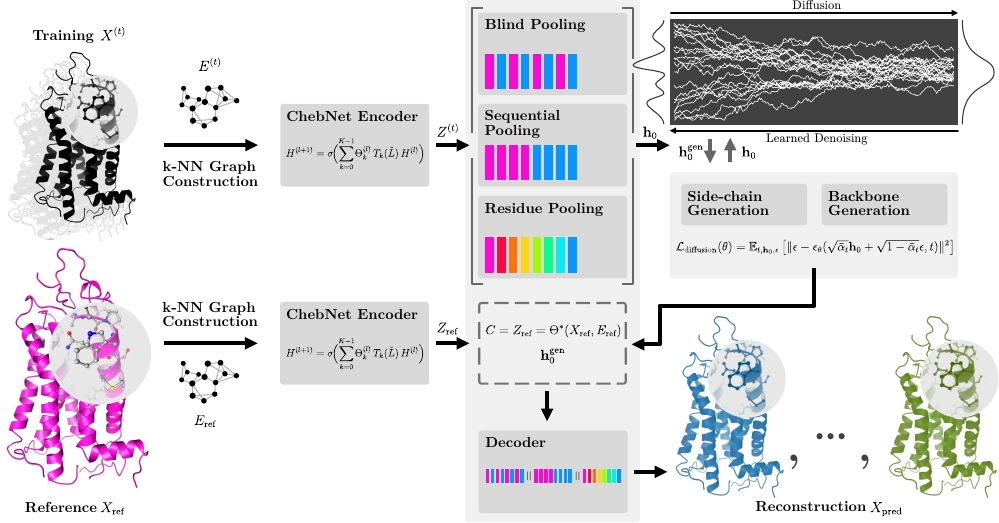}
    \caption{\textbf{Schematic of the LD-FPG framework.} The workflow depicts: (1) ChebNet encoding of MD frames to atom-wise latent embeddings ($\Zlatent^{(t)}$). (2) Pooling of $\Zlatent^{(t)}$ into a compact latent context ($\mathbf{h}_0$). (3) DDPM-based sampling of this pooled latent space to generate $\mathbf{h}_0^{\text{gen}}$. (4) Decoding of $\mathbf{h}_0^{\text{gen}}$, conditioned on reference latents ($\Zref$), to all-atom coordinates ($\Xpred$). The conceptual details of the framework are described in Section~\ref{sec:methodology}.}
    \label{fig:scheme} 
\end{figure*}

The LD-FPG framework (Fig.~\ref{fig:scheme}) generates diverse, all-atom conformational ensembles of a specific protein by learning to sample deformations relative to a reference structure, $\Xref$. Our approach employs an encoder (Section~\ref{sec:encoder_method_main}) to map protein conformations into atom-wise latent embeddings representing these deformations. These high-dimensional embeddings are then processed by a pooling strategy (detailed in Section~\ref{sec:decoders}) to create a compact latent representation, $\mathbf{h}_0$. A diffusion model (Section~\ref{sec:diffusion_method_main}) is trained to sample this lower-dimensional pooled latent space. Finally, a decoder (Section~\ref{sec:decoders}), conditioned on reference information derived from $\Xref$ (see Conditioning Mechanism in Section~\ref{sec:encoder_method_main}), maps these sampled pooled latents back to full all-atom Cartesian coordinates. This strategy simplifies generation by focusing on learning dynamic changes around a known fold. Appendix~\ref{app:method_overview} details the overall algorithm and notation used throughout this work. Further methodological details for each component are provided across Appendices~\ref{app:input_processing_details} through \ref{app:loss_function_formulations}, covering aspects from input representation to loss definitions. All specific architectural parameters, model configurations, and hyperparameter optimization scans are consolidated in the Extended Technical Appendix (Appendix~\ref{app:extended_technical}).

\subsection{Input Representation and Preprocessing}
\label{sec:input_preprocessing_main}

\noindent Each MD snapshot \(t\) is represented by a graph \(G^{(t)}=(V,E^{(t)})\), where the node set \(V\) comprises the \(N\) heavy atoms and the node features are their 3D coordinates \(X^{(t)}\in\mathbb{R}^{N\times3}\). For each frame, the edge index \(E^{(t)}\) is built on the fly by applying a \(k\)-Nearest Neighbors search to the aligned coordinates \(X^{(t)}\) with \(k=4\), see Extended Technical Appendix \ref{app:extended_technical}.  The node positions serve both as input features and regression targets. Prior to graph construction, the raw MD coordinates \(\tilde X^{(t)}\) are rigid-body aligned to the first frame using the Kabsch algorithm~\citep{kabsch1976} to remove global rotation and translation. 

\subsection{Latent learning of conformations}
\label{sec:encoder_method_main}

\textbf{Multi-hop encoding:} For each MD frame $t$, the encoder $\Theta$ maps the Kabsch‑aligned heavy‑atom coordinates $X^{(t)}$ and their $k$‑NN graph $E^{(t)}$ to latent embeddings
$Z^{(t)}\!\in\!\mathbb{R}^{N\times d_z}$, one $d_z$‑dimensional vector per heavy atom.
We implement $\Theta$ as a four‑layer ChebNet~\citep{Defferrard2016} whose layers perform spectral graph convolutions with Chebyshev polynomials of order $K=4$:
\begin{equation}
H^{(l+1)}
  = \sigma\!\Bigl(\sum_{k=0}^{K-1}\!
      \Theta^{(l)}_{k}\,T_k(\tilde{L})\,H^{(l)}\Bigr),
\label{eq:chebconv}
\end{equation}
where $H^{(l)}\!\in\!\mathbb{R}^{N\times F_l}$ are the node features at layer $l$ ($H^{(0)}\!=\!X^{(t)}$),
$\tilde{L}$ is the scaled graph Laplacian,
$T_k(\cdot)$ is the $k^{\text{th}}$ Chebyshev polynomial,
$\Theta^{(l)}_{k}\!\in\!\mathbb{R}^{F_l\times F_{l+1}}$ are learnable weights,
and $\sigma$ denotes a Leaky/ReLU non‑linearity.
Each layer is followed by BatchNorm, and the final output is $L_{2}$‑normalised per atom to yield $Z^{(t)}$.
The embedding dimension $d_z$ was tuned; the best trade‑off was obtained with $d_z{=}16$ for blind pooling, $8$ for sequential pooling, and $4$ for residue pooling. 

\textbf{Conditioning Mechanism: }
Each generated conformation is expressed as a deformation of a \emph{reference} structure: the first Kabsch‑aligned MD frame, $(X_{\text{ref}},E_{\text{ref}})$.
Rather than conditioning on raw Cartesian coordinates, we feed the decoder the reference’s latent representation $C=Z_{\text{ref}}=\Theta^{*}(X_{\text{ref}},E_{\text{ref}})$ where $\Theta^{*}$ denotes the frozen, pre‑trained encoder parameters. This embedding compactly summarizes both 3‑D geometry and graph topology, and in ablation studies outperformed using $X_{\text{ref}}$ directly.
At generation time we copy $C$ to every sample in the batch, $C_{ex}$; the diffusion model then predicts only the atomic \emph{displacements} from this common reference.
This simplifies learning and guarantees that all sampled conformations stay anchored in the same chemical frame.

\subsection{Decoder Architectures and Pooling Strategies} \label{sec:decoders}
The decoder maps atom-wise latent embeddings $\Zlatent^{(t)} \in \R^{N \times d_z}$ (representing conformational deformations from $\Xref$, output by the encoder detailed in Section~\ref{sec:encoder_method_main}) and a conditioner $C \in \R^{N \times d_c}$ to all-atom coordinates $\Xpred \in \R^{N \times 3}$. While these $\Zlatent^{(t)}$ embeddings are information-rich (as demonstrated in Section~\ref{sec:results_analysis}), their high dimensionality (e.g., up to $35k$ for D2R with $d_{z}=16$) makes them computationally challenging for direct use as input to a diffusion model. Therefore, $\Zlatent^{(t)}$ is processed via a pooling strategy to yield a much more compact latent representation, $\mathbf{h}_0$ (typically $d_p \approx 60-100$ for Blind and Sequential strategies), which serves as the substrate for the diffusion model (Section~\ref{sec:diffusion_method_main}). This compression is crucial, as preliminary experiments showed that $d_p > 200-300$ hampered diffusion training (for blind and sequential), while $d_p < 50$ degraded reconstruction quality. The efficacy of LD-FPG thus relies on the pooling strategy's ability to generate an informative yet compact $\mathbf{h}_0$. We investigate three strategies: Blind pooling, sequential pooling, and residue-based pooling.

\textbf{Blind pooling:} Atom-wise embeddings are globally pooled across all $N$ atoms using 2D adaptive average pooling $\mathcal{P}_{\text{global}}$ (reshaping $Z^{(t)}$ as an image-like tensor of size $N \times d_z$), yielding one context vector $z_{\mathrm{global}}\!\in\!\mathbb{R}^{d_p}$ per sample in the batch (where $d_p=H \times W$ from the pooling dimensions). This global vector is tiled for each atom and concatenated with the corresponding broadcast conditioner vector $C^{(i)}$ to form the input $M_{\mathrm{in}}^{(i)}$ for a shared MLP, $\mathrm{MLP}_{\text{blind}}$, which predicts all atom coordinates $\Xpred$ simultaneously. 

\textbf{Sequential pooling:} Decoding is split into two stages. A \emph{BackboneDecoder} first processes $Z^{(t)}$ and $C$ to output backbone coordinates $X_{\mathrm{bb}}$. It typically pools backbone-specific embeddings to form a backbone context. Subsequently, a \emph{SidechainDecoder} predicts side‑chain coordinates $X_{\mathrm{sc}}$ using $Z^{(t)}$, $C$, and the predicted $X_{\mathrm{bb}}$. This stage often involves pooling sidechain-specific embeddings and combining this with backbone information and parts of the conditioner to form features for an MLP. The final structure is $\Xpred=[X_{\mathrm{bb}}\;\|\;X_{\mathrm{sc}}]$. Three SidechainDecoder variants (arch‑types 0–2) explore different feature constructions for the sidechain prediction MLP. 

\textbf{Residue‑based pooling:} This strategy models conformational changes as residue-level deformations relative to $\Xref$. For each residue $R_j$, its constituent atom embeddings $\Zlatent_{R_j}^{(t)}$ (a subset of the overall atom-wise deformations $\Zlatent^{(t)}$) represent its specific deformation from the reference state implicitly provided by $\Zref$. These $\Zlatent_{R_j}^{(t)}$ are pooled via $\mathcal{P}_{\text{res}}$ into a local context vector $z_{\mathrm{res},j} \in \R^{d_p}$, which summarizes residue $R_j$'s deformation. Each atom $i$ (in residue $R_{f(i)}$) then receives $z_{\mathrm{res},f(i)}$ concatenated with its reference latent $C^{(i)}$ (from $\Zref$) as input to $\mathrm{MLP}_{\text{res}}$ for coordinate prediction. Thus, the decoder reconstructs atom positions from these summaries of residue-specific deformations relative to the reference. 

\subsection{Latent Diffusion Model for Generation}
\label{sec:diffusion_method_main}
A Denoising Diffusion Probabilistic Model (DDPM)~\citep{ho2020denoising} operates on the pooled latent embeddings $\mathbf{h}_0$ derived from the chosen decoder pooling strategy. The model is trained to predict the noise $\epsilon$ that was added during a forward diffusion process. This training minimizes the standard DDPM loss function, $\mathcal{L}_{\text{diffusion}}$ (Eq.~\ref{eq:diffusion_loss_main}).
New latent representations, $\mathbf{h}_0^{\text{gen}}$, are then sampled by iteratively applying the learned denoising network in a reverse diffusion process.

\begin{equation}
\mathcal{L}_{\text{diffusion}}(\theta) = \mathbb{E}_{t, \mathbf{h}_0, \epsilon} \left[ \| \epsilon - \epsilon_\theta(\sqrt{\bar{\alpha}_t}\mathbf{h}_0 + \sqrt{1-\bar{\alpha}_t}\epsilon, t) \|^2 \right] \label{eq:diffusion_loss_main}
\end{equation}

\subsection{Loss Functions}
\label{sec:loss_funcs_main}

The LD-FPG framework uses a series of MSE-based losses to train its encoder and decoders. A pre-trained encoder $\Theta$ minimizes a coordinate reconstruction MSE ($\Lhno$), while all decoders focus on coordinate accuracy: Blind pooling begins with $\Lcoord$ and,if fine-tuned,applies a weighted composite $\Ldec = w_{\text{base}}\Lcoord + \lambda_{\text{mse}}\Lmseh + \lambda_{\text{div}}\Ldivh$ (the latter two dihedral terms used stochastically only for this strategy), Residue-based Pooling uses $\Ldec = \Lcoord$, and Sequential Pooling optimizes backbone and sidechain predictions in two stages via separate MSE losses $\Lbb$ and $\Lsc$.

\section{Experimental Setup}
\label{sec:experiments}

\subsection{Experimental Setup}
\label{sec:exp_setup}

\textbf{D2 Receptor Dynamics dataset: }
We perform extensive all-atom Molecular Dynamics (MD) simulations of the human Dopamine D2 receptor (D2R) to generate the input dataset. The system, comprising the D2R (2191 heavy atoms) embedded in a POPC membrane with water and ions, was simulated using GROMACS. The final dataset consists of 12,241 frames sampled every 100~ps after discarding initial equilibration. All frames were aligned to the first frame using the Kabsch algorithm. The data was then split into training (90\%) and test (10\%) sets. Static topology information, including atom indexing and dihedral angle definitions, was pre-processed. Further details on the simulation protocol, system preparation, and data pre-processing are provided in Appendix~\ref{app:experimental_dataset_details} and Appendix~\ref{app:input_processing_details}.

\textbf{Evaluation Metrics.}
Model performance is assessed using metrics evaluating coordinate accuracy (e.g., MSE, lDDT, TM-score), dihedral angle distributional accuracy ($\sum$JSD), physical plausibility (steric clashes), and conformational landscape sampling (e.g., A100 activation index~\citep{ibrahim2019universal}, PCA of latent embeddings). Auxiliary dihedral training losses are also reported where applicable for specific decoder configurations. Detailed definitions and calculation methods for all evaluation metrics are provided in Appendix~\ref{app:evaluation_metric_details}.

\textbf{Implementation Details.}
All models were implemented in PyTorch~\citep{paszke2019pytorch} and trained using the Adam optimizer~\citep{kingma2014adam}. The overall three-phase training and generation workflow (encoder pre-training, decoder training, and diffusion model training) is detailed in Appendix~\ref{app:method_overview} (Algorithm~\ref{alg:main_workflow_appendix}).

\subsection{Results and Analysis}
\label{sec:results_analysis}

We evaluated LD-FPG's ability to generate high-fidelity, all-atom protein conformational ensembles via a multi-stage analysis: (1) assessing ChebNet encoder quality to establish an upper fidelity bound; (2) analyzing decoder reconstruction from ground-truth latents to isolate decoder errors; and (3) evaluating conformational ensembles from the full latent diffusion pipeline. Model lDDT scores (vs. $\Xref$) are interpreted against the ``Ground Truth (MD) Ref'' lDDT (Tables~\ref{tab:decoder_reconstruction}, Appendix~\ref{app:encoder_performance_summary}), which reflects the MD ensemble's average internal diversity relative to $\Xref$. Scores near this MD benchmark suggest a good balance of structural fidelity and diversity; significantly lower scores imply poorer fidelity, while substantially higher scores (approaching 1.0) might indicate insufficient diversity and over-similarity to $\Xref$.

\begin{table}[b]
\centering
\caption{Decoder Reconstruction Performance}
\label{tab:decoder_reconstruction}
\footnotesize
\renewcommand{\arraystretch}{1.1}
\setlength{\tabcolsep}{3pt}
\begin{tabular}{@{}lcccccccc@{}}
\toprule
Decoder & lDDT\textsubscript{All} & lDDT\textsubscript{BB} & TM\textsubscript{All} & $\sum$JSD\textsubscript{bb} & $\sum$JSD\textsubscript{sc} & MSE\textsubscript{bb} & MSE\textsubscript{sc} & $\sum\mathcal{L}_{\text{dih}}$ \\
Configuration & $\uparrow$ & $\uparrow$ & $\uparrow$ & $\downarrow$ & $\downarrow$ & $\downarrow$ & $\downarrow$ & MSE $\downarrow$ \\
\midrule
Blind pooling ($d_z=16$) & 0.714 & 0.792 & 0.961 & 0.0032 & 0.0290 & 0.1102 & 0.3934 & 0.3802 \\
\rowcolor{gray!10}
\quad + Dih. Fine-tuning & 0.698 & 0.776 & 0.960 & 0.0029 & 0.0279 & 0.0971 & 0.3564 & 0.2849 \\
\addlinespace[0.5em]
Seq. pooling ($d_z=8$) & 0.718 & 0.800 & 0.961 & 0.0026 & 0.0192 & 0.1291 & 0.5130 & 0.5164 \\
\addlinespace[0.5em]
Residue pooling ($d_z=4$) & 0.704 & 0.777 & 0.962 & 0.0078 & 0.0125 & 0.083 & 0.2257 & 0.2163 \\
\midrule
Ground Truth (MD) Ref & 0.698 & 0.779 & 0.959 & - & - & - & - & - \\
\bottomrule
\end{tabular}
\end{table}

\textbf{Multi-hop encoding fidelity of ChebNet}
The ChebNet encoder generates high-fidelity atom-wise latent representations ($\Zlatent^{(t)}$) directly from input conformations. Reconstruction from these unpooled embeddings is excellent (details in Appendix~\ref{app:encoder_performance_summary}): for the D2R system ($N=2191$ atoms), an encoder configuration with a latent dimension $d_z=16$ achieves a backbone MSE (MSE\textsubscript{bb}) of $0.0008$ and dihedral JSDs around $0.00016$. This high-dimensional $\Zlatent^{(t)}$ (up to $2191 \times 16 = 35,056$ dimensions in this case) establishes a strong upper benchmark for the atomic detail initially captured. For efficient downstream diffusion modeling, these rich $\Zlatent^{(t)}$ embeddings are significantly compressed via a pooling step (Section~\ref{sec:decoders}) into a much more compact latent vector, $\mathbf{h}_0$. This pooling focuses the learning process on conformational \textit{deformations} relative to a reference structure ($\Xref$, provided to the decoder via $\Zref$), rather than encoding the entire static fold. While this necessary compression means the final generative pipeline's coordinate accuracy may not fully match the encoder's standalone reconstruction capabilities, the high intrinsic fidelity of the initial $\Zlatent^{(t)}$ ensures that $\mathbf{h}_0$ is distilled from a robust, deformation-rich representation.

\begin{figure}[t]
\centering
\includegraphics[width=\linewidth]{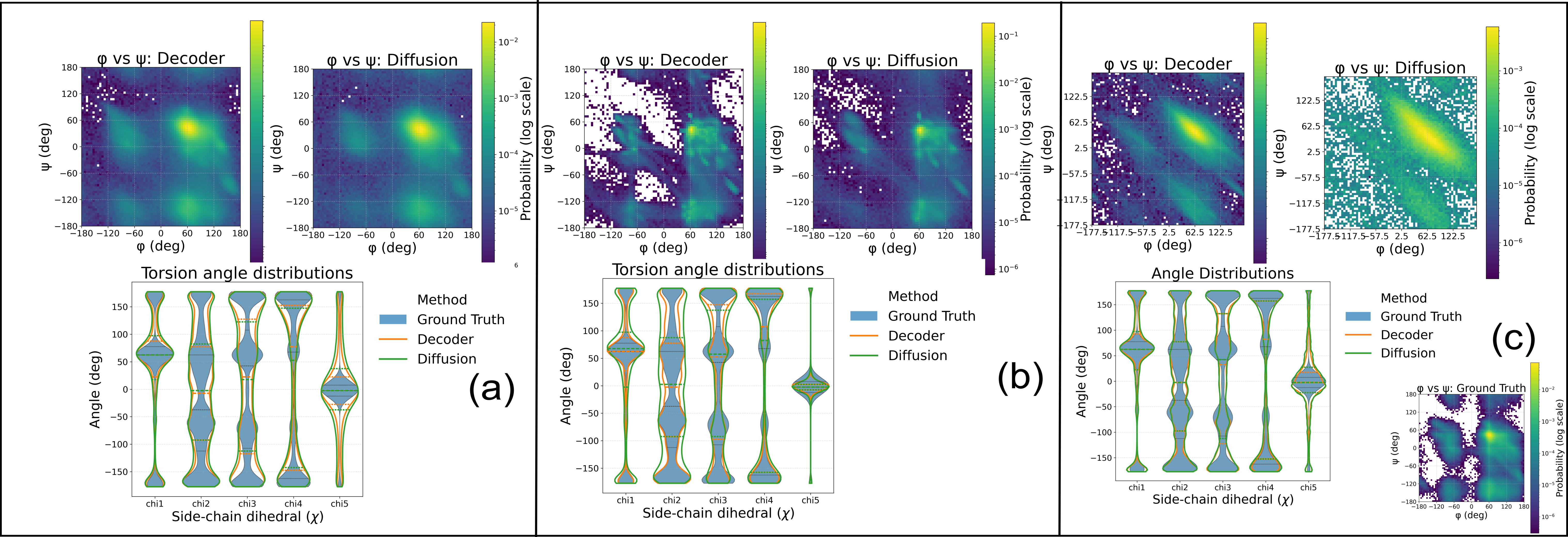}
\caption[Comparison of Dihedral Angle Distributions]{Comparison of dihedral angle distributions for different pooling strategies. (Top row) Ramachandran plots ($\phi$ vs $\psi$, log probability) comparing Decoder Reconstruction (left) and Diffusion Generation (right) outputs. (Bottom row) Violin plots comparing 1D sidechain dihedral angle ($\chi_1$–$\chi_5$) distributions for Ground Truth (MD, blue), Decoder Reconstruction (orange), and Diffusion Generation (green). (a) Blind pooling. (b) Sequential pooling. (c) Residue pooling.}
\label{fig:combined_dihedral_comparison}
\end{figure}

\textbf{Decoder Reconstruction Fidelity}
We next assessed decoder reconstruction of all-atom coordinates from these ground-truth ChebNet latents ($Z^{(t)}$), isolating decoder-specific errors (Table~\ref{tab:decoder_reconstruction}).
The \textbf{blind pooling} decoder (using $d_z=16$ atom features) achieved good coordinate accuracy (lDDT\textsubscript{All} 0.714) but had limited dihedral precision, evidenced by a blurred Ramachandran plot (visualizing backbone $\phi,\psi$ angles) and smoothed $\chi$-angle distributions ($\sum$JSD\textsubscript{sc} 0.0290; Figure~\ref{fig:combined_dihedral_comparison}a, orange traces/distributions). Optional dihedral fine-tuning yielded minimal JSD improvement while slightly reducing lDDTs.
In contrast, \textbf{sequential pooling} (from $d_z=8$ atom features) yielded excellent coordinate accuracy (lDDT\textsubscript{All} 0.718) and superior backbone geometry, marked by sharp Ramachandran plots (Figure~\ref{fig:combined_dihedral_comparison}b, orange trace) and the lowest backbone dihedral divergence ($\sum$JSD\textsubscript{bb} 0.0026). Its sidechain $\chi$-angle distributions were also well-reproduced ($\sum$JSD\textsubscript{sc} 0.0192; Figure~\ref{fig:combined_dihedral_comparison}b, orange distributions).
Intriguingly, \textbf{residue pooling} (from $d_z=4$ atom features) excelled locally, achieving the lowest backbone/sidechain MSEs (0.083/0.2257) and the best sidechain distributional fidelity ($\sum$JSD\textsubscript{sc} = 0.0125) with outstanding $\chi$-angle reproduction (Figure~\ref{fig:combined_dihedral_comparison}c, orange distributions). This local strength, despite a broader global backbone dihedral distribution ($\sum$JSD\textsubscript{bb} 0.0078) and a "hazier" Ramachandran plot (Figure~\ref{fig:combined_dihedral_comparison}c, orange trace) partly due to its smaller per-atom latent dimension, stems from its architecture. Pooling features within each of D2R's $N_{\text{res}}=273$ residues into $d_p=4$ local contexts ($z_{\mathrm{res},j}$) provides the MLP with access to a rich information space (effectively $N_{\text{res}} \times d_p \approx 1.1k$ dimensions describing overall residue deformations), boosting local performance. The quality of this decoder stage is key, as the diffusion model samples the distribution defined by these pooled latent embeddings $\mathbf{h}_0$.

\begin{table}[b]
\centering
\caption{Diffusion Generation Performance.}
\label{tab:diffusion_generation}
\footnotesize
\renewcommand{\arraystretch}{1.1}
\begin{tabular}{@{}lcccccc@{}}
\toprule
Model Configuration & lDDT\textsubscript{All} $\uparrow$ & lDDT\textsubscript{BB} $\uparrow$ & TM\textsubscript{All} $\uparrow$ & $\sum$JSD\textsubscript{bb} $\downarrow$ & $\sum$JSD\textsubscript{sc} $\downarrow$ & Avg. Clashes $\downarrow$ \\
\midrule
Blind pooling & 0.719 & 0.792 & 0.964 & 0.006582 & 0.04185 & 1350.5 \\
\rowcolor{gray!10}
\quad + Dih. Fine-tuning & 0.683 & 0.748 & 0.9321 & 0.00648 & 0.0409 & 1340.9 \\
\addlinespace[0.5em]
Sequential pooling & 0.712 & 0.801 & 0.942 & 0.0029 & 0.02895 & 1220.5 \\
\addlinespace[0.5em]
Residue pooling & 0.6880 & 0.7575 & 0.9570 & 0.0117 & 0.0224 & 1145.6 \\
\bottomrule
\end{tabular}
\end{table}

\begin{figure}[t]
\centering
\includegraphics[width=0.9\linewidth]{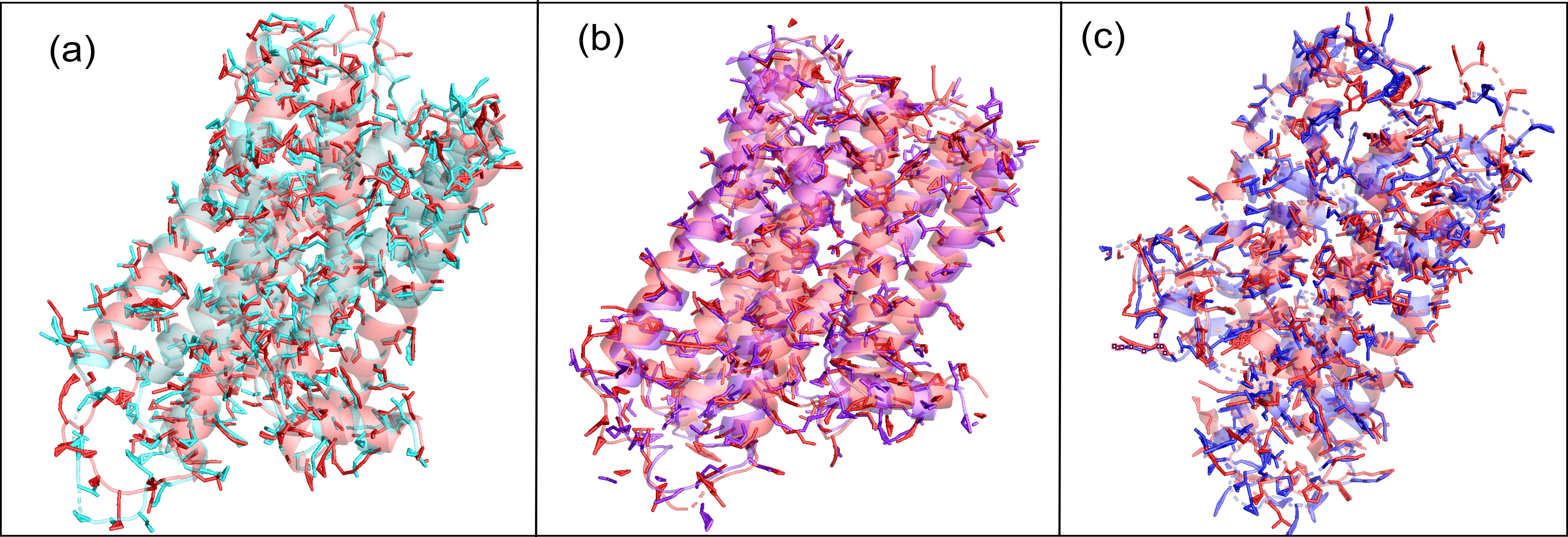}
\caption{Examples of generated D2R conformations using different pooling strategies. Panels likely correspond to: (a) Blind pooling, (b) sequential pooling, and (c) residue pooling. Structures are visualized to show overall fold and sidechain placement.}
\label{fig:generated_structures_comparison}
\end{figure}

\begin{figure}[b]
\centering
\includegraphics[width=\linewidth]{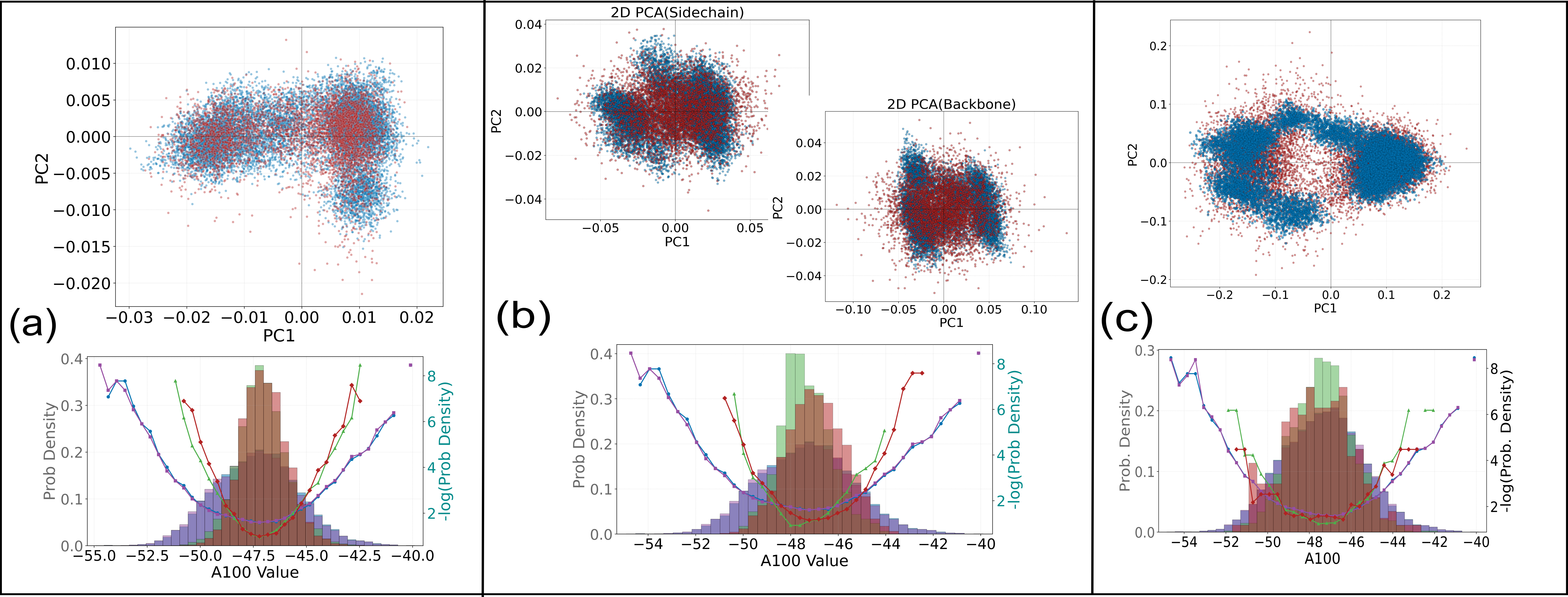}
\caption[Latent Space PCA and Collective Variable Analysis]{Latent space and collective variable (A100) analysis for Blind (a), Sequential (b), and Residue (c) pooling. Top rows: 2D PCA of dataset (blue) vs. diffusion-generated (red) pooled embeddings (Sequential shows separate backbone/sidechain PCAs). Bottom rows: A100 Value distributions (histograms) and PMFs (-log(Probability Density)) for dataset: ground truth (blue), autoencoder (purple), decoder (green), and generated ensembles (red).}
\label{fig:pca_pmf_analysis}
\end{figure}

\textbf{Diffusion Generation Quality}
The ultimate test is the quality of conformational ensembles from the full LD-FPG pipeline (Table~\ref{tab:diffusion_generation}), where pooling strategies yield distinct ensemble characteristics. \textbf{Blind pooling} produces structures with the highest global coordinate accuracy scores (lDDT\textsubscript{All} 0.719, TM\textsubscript{All} 0.964). However, this global fidelity, likely impacted by the aggressive compression to its final latent $\mathbf{h}_0$, sacrifices finer details: its Ramachandran trace (Figure~\ref{fig:combined_dihedral_comparison}a, green trace) is somewhat blurred, side-chain distributions are over-smoothed ($\sum$JSD\textsubscript{sc} 0.04185), and it averages 1350.5 clashes, notably higher than the $\approx 1023$ clashes averaged by the ground truth MD ensemble. Rendered structures (Figure~\ref{fig:generated_structures_comparison}a) show well-placed helices but wandering $\chi$-angles, leading to protruding or clashing side chains.
\textbf{Sequential pooling} offers an excellent all-round compromise. It maintains sharp backbone geometry (lowest $\sum$JSD\textsubscript{bb} 0.0029; clean Ramachandran in Figure~\ref{fig:combined_dihedral_comparison}b, green trace), respectable global scores (lDDT\textsubscript{All} 0.712), and good side-chain realism ($\sum$JSD\textsubscript{sc} 0.02895), with an average of 1220.5 clashes. Visually (Figure~\ref{fig:generated_structures_comparison}b), models are largely tidy. However, some surface side-chains may still flare out or collide. This aligns with its $\sum$JSD\textsubscript{sc} being better than blind but not residue pooling, suggesting some $\chi$-angles outside dense rotamer clouds, possibly due to the side-chain decoder's pooled backbone context leading to ambiguity that diffusion "averages". Thus, while delivering a sharp backbone and strong side-chain stats, a few rotamers (preferred side-chain conformations) may remain in incorrect basins. Its global backbone statistics are lower after diffusion (lDDT\textsubscript{All} 0.6880; $\sum$JSD\textsubscript{bb} 0.0117; Figure~\ref{fig:combined_dihedral_comparison}c, green trace), with occasional backbone kinks (Figure~\ref{fig:generated_structures_comparison}c), partly due to its use of a smaller ($d_z=4$) per-atom latent dimension. However, its per-residue focus excels at $\chi$-angle distributions (best $\sum$JSD\textsubscript{sc} 0.0224) and yields the lowest average clash count (1145.6, closest to the MD reference), producing the tightest side-chain packing (Figure~\ref{fig:generated_structures_comparison}c).
In essence, Figure~\ref{fig:combined_dihedral_comparison} (dihedrals) and Figure~\ref{fig:generated_structures_comparison} (structures) reveal these strategies' distinct behaviors: blind pooling prioritizes global fold over side-chain details; sequential pooling balances these, with minor outliers; residue pooling excels at rotamers, sometimes with less regular global backbones.

Figure~\ref{fig:pca_pmf_analysis} further explores conformational landscape coverage using PCA of latent embeddings and the A100 collective variable, an index of D2R activation states that proxies the MD-sampled landscape. It provides a dual analysis: PCA plots (top row) compare diffusion-generated latent distributions ($\mathbf{h}_0^{\text{gen}}$, red) against MD-derived ones (blue), while A100 value distributions (bottom row) assess the replication of the MD conformational landscape (and its autoencoder/decoder capture) by the full diffusion pipeline. The diffusion models' ability to cover major PCA regions of the MD latents positively indicates effective learning of the underlying data distribution.
When examining the A100 distributions, Blind and Sequential Pooling, operating on more compact, globally-pooled latent spaces ($\mathbf{h}_0$ with $d_p \approx 100$), show reasonable landscape coverage, with Sequential closely tracing the MD distribution. However, \textbf{residue pooling} (Figure~\ref{fig:pca_pmf_analysis}c, red curve) achieves the most comprehensive A100 landscape capture, populating both primary and subtler MD-observed states often missed by other methods. This superior recovery, despite moderate global backbone metrics, stems from its distinct latent space configuration. Unlike the compact, global latents of Blind/Sequential strategies, residue pooling's diffusion model leverages a significantly larger effective latent space formed by all per-residue contexts ($N_{\text{res}} \times d_p \approx 1.1k$ for D2R with $d_p=4$). This richer, structured space, even if derived from a smaller $d_z=4$ per-atom encoding, enables more nuanced representation of conformational substates. Its proficiency in local rotamer modeling is likely a key contributor, as accurate side-chain placement is crucial for the subtle cooperative shifts defining A100. The decoder's A100 distribution for residue pooling (Figure~\ref{fig:pca_pmf_analysis}c, green curve) already indicates robust MD landscape coverage, which the diffusion model effectively samples. Visualizations for residue pooling in Figure~\ref{fig:pca_pmf_analysis} employed multi-epoch sampling—aggregating samples from checkpoints across different DDPM training stages, Appendix~\ref{app:residue_pooling_inference_details}.

Furthermore, BioEmu~\citep{Lewis2024BioEmu}, a general MD model, struggled to capture the D2R-specific conformational landscape: its generated A100 distributions (mean $-17.19 \pm 7.45$) markedly diverged from our D2R-MD reference (mean $\approx -47.5 \pm 2.26$), highlighting limitations of such generalist approaches for specific membrane protein ensembles.

\section{Conclusion and Future Work}
\label{sec:conclusion_future}

We introduced LD-FPG, a latent diffusion framework generating all-atom protein conformational ensembles from MD data, demonstrated on D2R GPCR. It captures system-specific dynamics, including side-chain details, via learned deformations from a reference. Blind pooling offered global fidelity but compromised side-chain detail and clashes. Sequential pooling provided a strong balance, especially for backbone geometry. Residue pooling excelled in local side-chain accuracy and landscape coverage, despite some global backbone trade-offs and needing multi-epoch sampling for full diversity. A common challenge was increased steric clashes in generated structures versus MD.

Future work includes targeted enhancements. For \textbf{Blind and Sequential pooling}, exploring larger pooled latent dimensions ($d_p$) is immediate, potentially improving detail capture but needing more extensive, diverse training data (e.g., from multiple related protein systems). For \textbf{Residue pooling}, with its promising high-dimensional effective latent space ($N_{\text{res}} \times d_p$), better denoisers are key to overcome multi-epoch sampling limits. This might use complex MLP/convolutional architectures, attention for its structured latent vector, or alternative generative models (score-based, flow-matching). For \textbf{Sequential pooling}, improving denoiser coverage of the MD-derived latent space (Fig.\ref{fig:pca_pmf_analysis}b, backbone/sidechain overlap) could resolve side-chain misplacements and boost performance. Broader initiatives will enhance physical realism and architectural sophistication by incorporating lightweight energy surrogates, physics-guided diffusion schemes~\citep{wang2025diffpie,zhang2025physdock,Liu2024EGDiff}, and advanced architectures like attention-based fusion mechanisms~\citep{lombard2025petimot,fang2025atomica,Joshi2023ADiT}, SE(3)-equivariant GNNs/transformers, or flow-matching generators~\citep{Lipman2022,Zhan2024-P2DFlow}. A long-term goal is generalizing LD-FPG by training on data from multiple related proteins (e.g., Class A GPCRs in various states) for foundational models akin to specialized ``LLMs for protein dynamics.'' Efforts will also cover tailoring pooling for applications, scaling to larger systems~\citep{Ingraham2023}, and rigorous benchmarking.

\textbf{Societal Impact}: 
LD-FPG can offer  benefits by accelerating drug discovery, improving biological understanding, and enabling protein engineering for medicine/biotechnology. Ethical issues include dual-use, equitable access, data privacy for sensitive data, and rigorous validation to prevent misdirected efforts. Responsible development and ethical dialogue are crucial to maximize benefits and mitigate risks.

\section*{Acknowledgments}
A.S. was supported by a grant from the Center of Intelligence Systems (EPFL) to P.B. and P.V. This work was also supported by the Institute of Bioengineering (EPFL), the Signal Processing Laboratory (LTS2, EPFL), the Ludwig Institute for Cancer Research, and the Swiss National Science Foundation (grants 31003A\_182263 and 310030\_208179).

\newpage
\bibliographystyle{unsrtnat}
\bibliography{references} 

\newpage

\appendix

\section{Methodology Overview and Notation}
\label{app:method_overview}

This section provides a consolidated overview of the LD-FPG framework's workflow, key mathematical notation used throughout the paper, and supplementary details regarding data processing, model architectures, and training procedures.

\subsection{Notation}
\label{app:notation_details}
We define the following notation used throughout the main text and appendix:
\begin{itemize}
    \item $k$: Number of nearest neighbors used for graph construction.
    \item $\tilde X^{(t)} \in \R^{N \times 3}$: Raw (unaligned) Molecular Dynamics coordinates for frame $t$.
    \item $\Theta$: Encoder neural network function; $\Theta^*$ denotes frozen, pre-trained encoder parameters.
    \item $K$: Order of Chebyshev polynomials in the ChebNet encoder.
    \item $d_c$: Dimensionality of the conditioner tensor $C$.
    \item $H, W$: Output dimensions (height and width respectively) for pooling layers, where $d_p = H \cdot W$.
    \item $N_{\text{layers}}$: Number of layers in a Multi-Layer Perceptron (MLP).
    \item $\mathcal{L}_{\text{diffusion}}$: Loss function for training the Denoising Diffusion Probabilistic Model (DDPM).
    \item $N$: Total heavy atoms per structure.
    \item $V = \{1, ..., N\}$: Atom nodes.
    \item $G=(V, E)$: Graph representation.
    \item $\Ibb \subset V$: Backbone atom indices; $\Nbb = |\Ibb|$.
    \item $\Isc \subset V$: Sidechain atom indices; $\Nsc = |\Isc|$. ($V = \Ibb \cup \Isc$).
    \item $z_{\mathrm{res},j} \in \R^{d_p}$: Local pooled context vector summarizing the deformation of residue $R_j$ (this corresponds to $\mathbf{h}_{R_j}^{(b)}$ as used in Appendix~\ref{app:residue_pooling_details}).
    \item $\Xtrue \in \R^{N \times 3}$: Ground truth aligned coordinates. $\Xtrue^{(b, i)}$ for atom $i$, sample $b$.
    \item $\Xpred \in \R^{N \times 3}$: Predicted coordinates.
    \item $\Xref \in \R^{N \times 3}$: Reference structure coordinates (e.g., first frame).
    \item $d_z$: Encoder latent embedding dimensionality.
    \item $\Zlatent \in \R^{N \times d_z}$: Latent atom embeddings from encoder: $Z = \text{Encoder}(\Xtrue, E)$.
    \item $\Zref \in \R^{N \times d_z}$: Reference structure latent embeddings: $Z_{\text{ref}} = \text{Encoder}(\Xref, E_{\text{ref}})$.
    \item $C \in \R^{N \times d_c}$: General conditioning tensor (typically $C=\Zref$). $C_{ex}$ is batch-expanded.
    \item $d_p$: Pooled context vector dimension.
    \item $\mathcal{P}$: Generic pooling operator.
    \item $\mathcal{R} = \{R_1, ..., R_{N_{\text{res}}}\}$: Partition of atoms into residues.
    \item $f: V \to \{1, ..., N_{\text{res}}\}$: Atom-to-residue mapping.
    \item $\phi, \psi, \chi_k$: Dihedral angles; $\alpha \in \mathcal{A}$ is a generic type.
    \item $M_{\alpha}(j)$: Mask for valid angle $\alpha$ in residue $j$.
    \item $P(\alpha_{\text{pred}}), P(\alpha_{\text{true}})$: Empirical angle distributions.
    \item $D(\cdot \| \cdot)$: Divergence function (KL, JS).
    \item $w_{\text{base}}, \lambda_{\text{mse}}, \lambda_{\text{div}}$: Loss weights.
    \item $f_{\text{dih}}$: Probability of applying dihedral-based loss terms during stochastic fine-tuning.
    \item $\| \mathbf{v} \|^2$: Squared Euclidean norm.
    \item $T$: Diffusion timesteps.
    \item $\beta_t, \alpha_t, \bar{\alpha}_t$: Diffusion schedule parameters.
    \item $\mathbf{h}_0$: Initial pooled latent embedding for diffusion.
    \item $\mathbf{h}_t$: Noisy latent at step $t$.
    \item $\epsilon \sim \mathcal{N}(0, I)$: Standard Gaussian noise.
    \item $\epsilon_\theta(\mathbf{h}_t, t)$: Denoising network predicting noise $\epsilon$.
\end{itemize}

\subsection{Overall Training and Generation Workflow}
\label{app:workflow_details}
The LD-FPG framework follows a multi-phase training and generation procedure, outlined in Algorithm~\ref{alg:main_workflow_appendix}.

\begin{algorithm}[htbp] 
\caption{Overall Training and Generation Workflow (LD-FPG)}
\label{alg:main_workflow_appendix}
\begin{algorithmic}[1] 
\State \textbf{Input:} Aligned coordinates file (\texttt{my\_protein.json}), Topology file (\texttt{condensed\_residues.json}), Reference PDB (\texttt{heavy\_chain.pdb}).
\State \textbf{Parameters:} Encoder config ($d_z, K, \dots$), Decoder config (Pooling Type, $H, W, N_{layers}, \dots$), Optional fine-tuning weights ($\lambda_{mse}, \lambda_{div}$), Diffusion config ($T, \beta_{start}, \beta_{end}, \epsilon_\theta$ type).

\Statex \textbf{// --- Preprocessing (Details in Appendix~\ref{app:input_processing_details}) ---}
\State Load aligned coordinates $\{\Xtrue^{(f)}\}_{f=1}^{N_{\text{frames}}}$ from \texttt{my\_protein.json}.
\State Load topology (atom indices $\Ibb, \Isc$, residue map $f$, dihedral defs.) from \texttt{condensed\_residues.json}.
\State Build graph dataset $GraphDataset = \{\text{BuildGraph}(\Xtrue^{(f)}, k=4) \}_{f=1}^{N_{\text{frames}}}$.

\Statex \textbf{// --- Phase 1: Encoder Pre-training (Optional) (Details in Appendix~\ref{app:encoder_config_details_ext} ---}
\State Initialize $\text{Encoder}_{\theta}$ (ChebNet model).
\State Train $\text{Encoder}_{\theta}$ on $GraphDataset$ using coordinate reconstruction loss $\Lhno$ (Eq. in Appendix~\ref{app:loss_function_formulations}).
\State Save best encoder parameters $\theta^*$.

\Statex \textbf{// --- Phase 2: Decoder Training (Details in Appendix~\ref{app:decoder_architectures_summary}) ---}
\State Load pre-trained $\text{Encoder}_{\theta^*}$ and freeze weights.
\State Generate latent embeddings $Z^{(f)} = \text{Encoder}_{\theta^*}(\Xtrue^{(f)})$ for all frames $f$.
\State Create Decoder Input Dataset $D_{dec} = \{(Z^{(f)}, \Xtrue^{(f)}) \mid f=1..N_{\text{frames}}\}$.
\State Define conditioner $C = \Zref = \text{Encoder}_{\theta^*}(\Xtrue^{(1)})$ (Details in Appendix~\ref{app:conditioning_tuning_details_ext}). 

\State \textbf{Select Decoder Variation:}
\If{PoolingType is Blind}
    \State Initialize $\text{Decoder}_{\phi} \gets \text{BlindPoolingDecoder}(\dots)$.
    \State Train $\text{Decoder}_{\phi}$ on $D_{dec}$ using coordinate loss $\Ldec = \Lcoord$.
    \State \textit{Optional Fine-tuning:} Continue training with $\Ldec = w_{\text{base}}\Lcoord + \lambda_{mse}\Lmseh + \lambda_{div}\Ldivh$ (stochastically, see Appendix~\ref{app:loss_function_formulations}).
    \State Save best decoder parameters $\phi^*$.
\ElsIf{PoolingType is Sequential}
    \State Initialize $\text{BBDecoder}_{\phi_{bb}}, \text{SCDecoder}_{\phi_{sc}}(\dots)$.
    \State Train $\text{BBDecoder}_{\phi_{bb}}$ on $D_{dec}$ using $\Lbb = \Lcoord^{\text{bb}}$. Freeze $\phi_{bb}^*$.
    \State Train $\text{SCDecoder}_{\phi_{sc}}$ on $D_{dec}$ using $\Lsc = \Lcoord^{\text{full}}$ (requires frozen $\text{BBDecoder}_{\phi_{bb}^*}$).
    \State Save best parameters $\phi_{bb}^*$, $\phi_{sc}^*$.
\ElsIf{PoolingType is Residue-Based}
    \State Initialize $\text{Decoder}_{\phi} \gets \text{ResidueBasedDecoder}(\dots)$.
    \State Train $\text{Decoder}_{\phi}$ on $D_{dec}$ using coordinate loss $\Ldec = \Lcoord$.
    \State Save best decoder parameters $\phi^*$.
\EndIf
\Statex \textbf{// --- Phase 3: Latent Diffusion Training (Details in Appendix~\ref{app:diffusion_model_methodology}) ---}
\State Load best $\text{Encoder}_{\theta^*}$ and $\text{Decoder}_{\phi^*}$ (or relevant pooling part).
\State Generate pooled latent embeddings $\mathbf{h}_0^{(f)} = \text{Pool}(Z^{(f)})$ for all $f$, using the specific pooling mechanism of the chosen decoder.
\State Create $DiffusionInputDataset = \{\mathbf{h}_0^{(f)}\}$.
\State Initialize $\epsilon_\theta$ (Denoising model).
\State Train $\epsilon_\theta$ on $DiffusionInputDataset$ using $\mathcal{L}_{\text{diffusion}}$ (Eq. \ref{eq:diffusion_loss_main}).
\State Save best diffusion model parameters $\theta_{diff}^*$.

\Statex \textbf{// --- Output Generation (Sampling) ---}
\State Load best models: $\text{Encoder}_{\theta^*}$, $\text{Decoder}_{\phi^*}$ (or $\text{BBDecoder}_{\phi_{bb}^*}, \text{SCDecoder}_{\phi_{sc}^*}$), $\epsilon_{\theta_{diff}}^*$.
\State Sample novel pooled latent(s) $\mathbf{h}_0^{\text{gen}}$ using $\epsilon_{\theta_{diff}}^*$ (Algorithm in Appendix~\ref{app:diffusion_model_methodology}).
\State \textbf{Decode generated latent(s):} (Feed $\mathbf{h}_0^{\text{gen}}$ into appropriate decoder stage)
\State $\Xpred^{\text{gen}} \gets \text{Decoder}_{\phi^*}(\dots, \text{context}=\mathbf{h}_0^{\text{gen}}, C)$ 
\State \textbf{Output:} Generated coordinates $\{\Xpred^{\text{gen}}\}$
\end{algorithmic}
\end{algorithm}
\FloatBarrier 


\section{Input Data Processing and Representation}
\label{app:input_processing_details}

The raw Molecular Dynamics (MD) simulation data (trajectory processing detailed in Appendix~\ref{app:experimental_dataset_details}) is transformed into a structured format suitable for the machine learning pipeline. This involves generating two key JSON files using custom Python scripts (see Supplementary Code for \texttt{extract\_residues.py} and \texttt{condense\_json.py}): one defining the static topology of the protein with a consistent indexing scheme (\texttt{condensed\_residues.json}), and another containing per-frame coordinate and dihedral angle data.

\subsection{Static Topology and Consistent Indexing File (\texttt{condensed\_residues.json})}
\label{app:static_topology_file}
To provide a consistent structural map for the machine learning models, a static JSON file, typically named \texttt{condensed\_residues.json}, is generated. This file is crucial as it establishes a definitive and model-centric representation of the protein's topology:
\begin{itemize}
    \item \textbf{Zero-Based Contiguous Atom Indexing:} A new, zero-based, and contiguous indexing scheme ($0, 1, \dots, N-1$) is created for all $N$ heavy atoms in the protein system. This re-indexing maps original PDB atom identifiers to a consistent integer range, essential for constructing graph inputs and feature matrices for the neural network.
    \item \textbf{Residue Definitions:} Residues are also re-indexed contiguously (e.g., $0, \dots, N_{\text{res}}-1$). For each re-indexed residue, the file stores:
    \begin{itemize}
        \item The residue type (e.g., 'ALA', 'LYS').
        \item Lists of atom indices (using the \textit{new zero-based scheme}) that constitute the backbone atoms of that residue.
        \item Lists of atom indices (using the \textit{new zero-based scheme}) that constitute the sidechain heavy atoms of that residue.
    \end{itemize}
    \item \textbf{Dihedral Angle Definitions (Atom Quadruplets):} This is a critical component for calculating dihedral-based losses and analyses. For each residue, the file stores definitions for all applicable standard backbone angles ($\phi, \psi$) and sidechain angles ($\chi_1, \chi_2, \chi_3, \chi_4, \chi_5$). Each dihedral angle is defined by an ordered quadruplet of four atom indices. Crucially, these atom indices adhere to the \textit{new, zero-based, contiguous indexing scheme}. For example, a $\phi$ angle for a specific residue would be defined by four specific integer indices from the $0 \dots N-1$ range. This allows for unambiguous calculation of any dihedral angle directly from a set of $N$ atomic coordinates, whether they are ground truth or model-predicted. The definitions also account for residue types where certain $\chi$ angles are not present (e.g., Alanine has no $\chi$ angles, Glycine has no sidechain).
\end{itemize}

\subsection{Primary Per-Frame Data File}
This second JSON file stores the dynamic information extracted from each frame of the MD trajectory. For every snapshot:
\begin{itemize}
    \item \textbf{Heavy Atom Coordinates:} The 3D Cartesian coordinates of all heavy atoms are recorded after rigid-body alignment to a common reference frame (the first frame of the trajectory, as described in Section~\ref{sec:input_preprocessing_main}). These coordinates are stored in an order that corresponds to the new zero-based indexing defined in the static topology file.
    \item \textbf{Dihedral Angle Values:} The values for standard backbone dihedral angles ($\phi, \psi$) and sidechain dihedral angles ($\chi_1$ through $\chi_5$, where applicable for each residue type) are pre-calculated. These calculations initially use atom identifications based on the original PDB residue and atom naming conventions but are stored in a way that can be mapped to the new indexing if needed for direct comparison or analysis.
\end{itemize}
This primary data file is typically organized on a per-residue basis (using original PDB residue numbering for initial organization if helpful during generation), associating each residue with its constituent atoms' names, original PDB indices, and the time series of their coordinates and calculated dihedral angles.

\subsection{Usage in Models}
The two JSON files are used in conjunction:
\begin{itemize}
    \item The static topology file (\texttt{condensed\_residues.json}) serves as the definitive reference for all structural metadata used by the model during training and inference. This includes identifying which atoms belong to the backbone versus sidechain (using their new zero-based indices) and, most importantly, providing the specific quadruplets of new zero-based atom indices required to calculate any dihedral angle from a given set of 3D coordinates. This capability is essential for implementing the dihedral angle-based loss terms ($\Lmseh, \Ldivh$) mentioned in Section~\ref{sec:loss_funcs_main}, as these losses operate on dihedral angles computed from the model's predicted coordinates ($\Xpred$).
    \item The 3D coordinates for each heavy atom, required as input features ($X^{(t)}$) for the encoder at each frame $t$, are drawn from the primary per-frame data file. These coordinates must be arranged and ordered according to the \textit{new zero-based indexing scheme} established by the static topology file to ensure consistency with the model's internal graph representation.
\end{itemize}
This separation of dynamic coordinate data from static, re-indexed topological information allows for efficient data loading and consistent geometric calculations within the LD-FPG framework.

\section{Decoder Architectures and Pooling Strategies}
\label{app:decoder_architectures_summary}

This section provides a detailed description of the three primary pooling strategies employed within the decoder architectures: Blind pooling, sequential pooling, and residue-based pooling. For each strategy, we delineate how atom-wise latent embeddings $Z \in \R^{B \times N \times d_z}$ (where $B$ is batch size, $N$ is the number of heavy atoms, and $d_z$ is the latent dimension per atom) and a conditioner $C \in \R^{N \times d_c}$ (typically the latent representation of the reference structure, $\Zref$) are processed to generate the input for the final coordinate prediction MLP. Specific hyperparameter configurations for representative models are detailed in the Extended Technical Appendix (Section~\ref{app:extended_technical}).

\subsection{Blind Pooling Strategy}
\label{app:blind_pooling_details}

The blind pooling strategy aims to capture a global context from all atom embeddings for the entire protein structure.
Let $Z^{(b)} \in \R^{N \times d_z}$ be the latent atom embeddings for a single sample $b$ in a batch.
\begin{enumerate}
    \item \textbf{Global Pooling:} The atom embeddings $Z^{(b)}$ are treated as an image-like tensor and processed by a 2D adaptive average pooling layer, $\mathcal{P}_{\text{global}}$ (typically \texttt{nn.AdaptiveAvgPool2d} with output size $H \times W$). This operation pools across all $N$ atoms for each sample in the batch:
    $$ \mathbf{h}_{\text{global}}^{(b)} = \mathcal{P}_{\text{global}}(Z^{(b)}) \in \R^{d_p} $$
    where $d_p = H \cdot W$ is the dimension of the pooled global context vector. For a batch, this results in $\mathbf{H}_{\text{global}} \in \R^{B \times d_p}$.

    \item \textbf{Context Expansion:} This global context vector $\mathbf{h}_{\text{global}}^{(b)}$ is then expanded (tiled) to match the number of atoms $N$, resulting in $\mathbf{H}_{\text{global\_ex}}^{(b)} \in \R^{N \times d_p}$, where each row $i$ (for atom $i$) is identical to $\mathbf{h}_{\text{global}}^{(b)}$. For a batch, this is $\mathbf{H}_{\text{global\_ex}} \in \R^{B \times N \times d_p}$.

    \item \textbf{Conditioner Expansion:} The conditioner $C \in \R^{N \times d_c}$ is expanded for the batch to $C_{ex} \in \R^{B \times N \times d_c}$.

    \item \textbf{MLP Input Formulation:} For each atom $i$ in sample $b$, the input to the final MLP, $\text{MLP}_{\text{blind}}$, is formed by concatenating its corresponding expanded global context and its conditioner vector:
    $$ M_{\text{in}}^{(b,i)} = \text{concat}(\mathbf{h}_{\text{global}}^{(b)}, C^{(b,i)}) \in \R^{d_p + d_c} $$
    Note that $\mathbf{h}_{\text{global}}^{(b)}$ is the same for all atoms $i$ within sample $b$.

    \item \textbf{Coordinate Prediction:} A shared $\text{MLP}_{\text{blind}}$ processes $M_{\text{in}}^{(b,i)}$ for each atom to predict its 3D coordinates:
    $$ \Xpred^{(b,i)} = \text{MLP}_{\text{blind}}(M_{\text{in}}^{(b,i)}) \in \R^3 $$
    This results in the full predicted structure $\Xpred \in \R^{B \times N \times 3}$.
\end{enumerate}

\subsection{Sequential Pooling Strategy}
\label{app:sequential_pooling_details}
The sequential pooling strategy decodes the protein structure in two stages: first the backbone atoms ($\Ibb$), then the sidechain atoms ($\Isc$), using information from the preceding stage.

\subsubsection{Backbone Decoder Stage}
Let $Z^{(b)} \in \R^{N \times d_z}$ be the full latent atom embeddings and $C^{(b)} \in \R^{N \times d_c}$ be the full conditioner for sample $b$.
\begin{enumerate}
    \item \textbf{Backbone Embedding Selection:} Latent embeddings $Z_{\text{bb}}^{(b)} \in \R^{\Nbb \times d_z}$ and conditioner vectors $C_{\text{bb}}^{(b)} \in \R^{\Nbb \times d_c}$ corresponding to backbone atoms $\Ibb$ are selected.

    \item \textbf{Backbone Pooling:} $Z_{\text{bb}}^{(b)}$ is pooled using a 2D adaptive average pooling layer $\mathcal{P}_{\text{bb}}$ (e.g., \texttt{BlindPooling2D} from the implementation, with output size $H_{bb} \times W_{bb}$) to obtain a backbone-specific context vector:
    $$ \mathbf{h}_{\text{bb}}^{(b)} = \mathcal{P}_{\text{bb}}(Z_{\text{bb}}^{(b)}) \in \R^{d_{p,bb}} $$
    where $d_{p,bb} = H_{bb} \cdot W_{bb}$. For a batch, this is $\mathbf{H}_{\text{bb}} \in \R^{B \times d_{p,bb}}$.

    \item \textbf{Context and Conditioner Expansion:} $\mathbf{h}_{\text{bb}}^{(b)}$ is expanded to $\mathbf{H}_{\text{bb\_ex}}^{(b)} \in \R^{\Nbb \times d_{p,bb}}$. $C_{\text{bb}}^{(b)}$ is used directly.

    \item \textbf{MLP Input for Backbone Atoms:} For each backbone atom $j \in \Ibb$ in sample $b$, the input to the backbone MLP, $\text{MLP}_{\text{bb}}$, is:
    $$ M_{\text{in, bb}}^{(b,j)} = \text{concat}(\mathbf{h}_{\text{bb}}^{(b)}, C_{\text{bb}}^{(b,j)}) \in \R^{d_{p,bb} + d_c} $$

    \item \textbf{Backbone Coordinate Prediction:} A shared $\text{MLP}_{\text{bb}}$ predicts backbone coordinates:
    $$ {\Xpred}_{\text{bb}}^{(b,j)} = \text{MLP}_{\text{bb}}(M_{\text{in, bb}}^{(b,j)}) \in \R^3 $$
    This yields the predicted backbone structure $X_{pred_{\text{bb}}} \in \R^{B \times \Nbb \times 3}$.
\end{enumerate}
\subsubsection{Sidechain Decoder Stage}
\label{app:sequential_sidechain_decoder_details} 
This stage predicts sidechain atom coordinates ${\Xpred}_{\text{sc}}$ using the full latent embeddings $Z^{(b)} \in \R^{N \times d_z}$, the conditioner $C^{(b)} \in \R^{N \times d_c}$, and the predicted backbone coordinates ${\Xpred}_{\text{bb}}^{(b)} \in \R^{B \times \Nbb \times 3}$ from the Backbone Decoder Stage.

\begin{enumerate}
    \item \textbf{Sidechain Embedding Selection:} Latent embeddings $Z_{\text{sc}}^{(b)} \in \R^{\Nsc \times d_z}$ corresponding to sidechain atoms $\Isc$ are selected from the full latent embeddings $Z^{(b)}$.

    \item \textbf{Sidechain Pooling:} The selected sidechain embeddings $Z_{\text{sc}}^{(b)}$ are pooled using a 2D adaptive average pooling layer $\mathcal{P}_{\text{sc}}$ (e.g., \texttt{BlindPooling2D} with output size $H_{sc} \times W_{sc}$) to obtain a sidechain-specific context vector for each sample in the batch:
    $$ \mathbf{h}_{\text{sc}}^{(b)} = \mathcal{P}_{\text{sc}}(Z_{\text{sc}}^{(b)}) \in \R^{d_{p,sc}} $$
    where $d_{p,sc} = H_{sc} \cdot W_{sc}$. For a batch, this results in $\mathbf{H}_{\text{sc}} \in \R^{B \times d_{p,sc}}$.

    \item \textbf{Feature Construction for Sidechain MLP Input:}
    The input to the sidechain MLP, $\text{MLP}_{\text{sc}}$, is a global vector $M_{\text{in, sc}}^{(b)}$ constructed per sample $b$. The construction varies based on the \texttt{arch\_type}:
    \begin{itemize}
        \item Let $X_{\text{pred, bb\_flat}}^{(b)} \in \R^{\Nbb \cdot 3}$ be the flattened predicted backbone coordinates for sample $b$.
        \item Let $C_{\text{sc}}^{(b)} \in \R^{\Nsc \times d_c}$ be the sidechain portion of the reference conditioner. If $d_c = d_z$, this is $Z_{\text{ref, sc}}^{(b)}$.
        This is flattened to $C_{\text{sc\_flat}}^{(b)} \in \R^{\Nsc \cdot d_c}$.
    \end{itemize}

    \textbf{Arch-Type 0:}
    The input consists of the flattened predicted backbone coordinates and the pooled sidechain context from the current frame's embeddings.
    $$ M_{\text{in, sc}}^{(b)}[\text{Arch 0}] = \text{concat}(X_{\text{pred, bb\_flat}}^{(b)}, \mathbf{h}_{\text{sc}}^{(b)}) $$
    The dimension of $M_{\text{in, sc}}^{(b)}[\text{Arch 0}]$ is $(\Nbb \cdot 3) + d_{p,sc}$.

    \textbf{Arch-Type 1:}
    This architecture adds a reduced representation of the sidechain portion of the reference conditioner.
    \begin{itemize}
        \item The flattened sidechain conditioner $C_{\text{sc\_flat}}^{(b)}$ is passed through a linear reduction layer:
        $$ C_{\text{sc\_reduced}}^{(b)} = \text{Linear}_{\text{sc\_reduce}}(C_{\text{sc\_flat}}^{(b)}) \in \R^{d'_{c,sc}} $$
        (e.g., $d'_{c,sc}=128$ in the implementation).
    \end{itemize}
    The MLP input is then:
    $$ M_{\text{in, sc}}^{(b)}[\text{Arch 1}] = \text{concat}(X_{\text{pred, bb\_flat}}^{(b)}, \mathbf{h}_{\text{sc}}^{(b)}, C_{\text{sc\_reduced}}^{(b)}) $$
    The dimension of $M_{\text{in, sc}}^{(b)}[\text{Arch 1}]$ is $(\Nbb \cdot 3) + d_{p,sc} + d'_{c,sc}$.

    \textbf{Arch-Type 2:}
    This architecture uses a reduced representation of both the predicted backbone coordinates and the sidechain portion of the reference conditioner.
    \begin{itemize}
        \item The flattened predicted backbone coordinates $X_{\text{pred, bb\_flat}}^{(b)}$ are passed through a linear reduction layer:
        $$ X_{\text{pred, bb\_reduced}}^{(b)} = \text{Linear}_{\text{bb\_reduce}}(X_{\text{pred, bb\_flat}}^{(b)}) \in \R^{d'_{\text{bb}}} $$
        (e.g., $d'_{\text{bb}}=128$ in the implementation).
        \item The sidechain conditioner is reduced as in Arch-Type 1 to $C_{\text{sc\_reduced}}^{(b)} \in \R^{d'_{c,sc}}$.
    \end{itemize}
    The MLP input is then:
    $$ M_{\text{in, sc}}^{(b)}[\text{Arch 2}] = \text{concat}(X_{\text{pred, bb\_reduced}}^{(b)}, \mathbf{h}_{\text{sc}}^{(b)}, C_{\text{sc\_reduced}}^{(b)}) $$
    The dimension of $M_{\text{in, sc}}^{(b)}[\text{Arch 2}]$ is $d'_{\text{bb}} + d_{p,sc} + d'_{c,sc}$.

    \item \textbf{Sidechain Coordinate Prediction:} The sidechain MLP, $\text{MLP}_{\text{sc}}$, processes the constructed input vector $M_{\text{in, sc}}^{(b)}$ (corresponding to the chosen \texttt{arch\_type}) to predict all sidechain coordinates for sample $b$ simultaneously:
    $$ X_{\text{pred, sc\_flat}}^{(b)} = \text{MLP}_{\text{sc}}(M_{\text{in, sc}}^{(b)}) \in \R^{\Nsc \cdot 3} $$
    This flattened output is then reshaped to $X_{\text{pred, sc}}^{(b)} \in \R^{\Nsc \times 3}$.

    \item \textbf{Full Structure Assembly:} The final predicted structure $\Xpred^{(b)} \in \R^{N \times 3}$ for sample $b$ is assembled by combining the predicted backbone coordinates $X_{\text{pred, bb}}^{(b)}$ and the predicted sidechain coordinates $X_{\text{pred, sc}}^{(b)}$.
\end{enumerate}

\subsection{Residue-based Pooling Strategy}
\label{app:residue_pooling_details}
The Residue-based Pooling strategy generates a context vector specific to each residue and uses this local context for predicting the coordinates of atoms within that residue.
Let $Z^{(b)} \in \R^{N \times d_z}$ be the latent atom embeddings for sample $b$. Let $V_j$ be the set of atom indices belonging to residue $R_j$, and $f: V \to \{1, ..., N_{\text{res}}\}$ be the mapping from a global atom index to its residue index.
\begin{enumerate}
    \item \textbf{Per-Residue Pooling:} For each residue $R_j$ in sample $b$:
        \begin{itemize}
            \item Select atom embeddings for residue $R_j$: $Z_{R_j}^{(b)} \in \R^{|V_j| \times d_z}$.
            \item Pool these embeddings using a 2D adaptive average pooling layer $\mathcal{P}_{\text{res}}$ (e.g., \texttt{nn.AdaptiveAvgPool2d} with output $H \times W$):
            $$ \mathbf{h}_{R_j}^{(b)} = \mathcal{P}_{\text{res}}(Z_{R_j}^{(b)}) \in \R^{d_p} $$
            where $d_p = H \cdot W$.
        \end{itemize}
    This results in a set of $N_{\text{res}}$ pooled vectors for sample $b$, which can be represented as $\mathbf{H}_{\text{res}}^{(b)} \in \R^{N_{\text{res}} \times d_p}$. For a batch, this is $\mathbf{H}_{\text{res}} \in \R^{B \times N_{\text{res}} \times d_p}$.

    \item \textbf{Atom-Specific Context Assembly:} For each atom $i$ in sample $b$, its specific context vector is the pooled vector of its parent residue $R_{f(i)}$:
    $$ \mathbf{h}_{\text{atom\_context}}^{(b,i)} = \mathbf{h}_{R_{f(i)}}^{(b)} \in \R^{d_p} $$
    This can be gathered for all atoms to form $\mathbf{H}_{\text{atom\_context}}^{(b)} \in \R^{N \times d_p}$.

    \item \textbf{Conditioner Expansion:} The conditioner $C \in \R^{N \times d_c}$ is expanded for the batch to $C_{ex} \in \R^{B \times N \times d_c}$.

    \item \textbf{MLP Input Formulation:} For each atom $i$ in sample $b$, the input to the final MLP, $\text{MLP}_{\text{res}}$, is formed by concatenating its residue's pooled context and its specific conditioner vector:
    $$ M_{\text{in, res}}^{(b,i)} = \text{concat}(\mathbf{h}_{\text{atom\_context}}^{(b,i)}, C^{(b,i)}) \in \R^{d_p + d_c} $$

    \item \textbf{Coordinate Prediction:} A shared $\text{MLP}_{\text{res}}$ processes $M_{\text{in, res}}^{(b,i)}$ for each atom to predict its 3D coordinates:
    $$ \Xpred^{(b,i)} = \text{MLP}_{\text{res}}(M_{\text{in, res}}^{(b,i)}) \in \R^3 $$
    This results in the full predicted structure $\Xpred \in \R^{B \times N \times 3}$.
\end{enumerate}
This strategy allows the model to learn representations that are localized at the residue level, potentially capturing residue-specific conformational preferences more directly.

\section{Latent Diffusion Model Details}
\label{app:diffusion_model_methodology}

This section details the Denoising Diffusion Probabilistic Model (DDPM)~\citep{ho2020denoising} utilized in our framework. The DDPM operates on the pooled latent embeddings $\mathbf{h}_0 \in \R^{d_p}$ (where $d_p$ is the dimension of the pooled latent space, dependent on the pooling strategy) obtained from the encoder and pooling stages. Specific architectures and hyperparameters for the denoising network $\epsilon_\theta$ are discussed in the Extended Technical Appendix (Section~\ref{app:extended_technical}).

\subsection{DDPM Formulation}
The DDPM consists of a predefined forward noising process and a learned reverse denoising process.

\noindent\textbf{Forward Process (Noising):}
The forward process gradually adds Gaussian noise to an initial latent embedding $\mathbf{h}_0$ over $T$ discrete timesteps. At each timestep $t$, the transition is defined by:
$$ q(\mathbf{h}_t | \mathbf{h}_{t-1}) = \mathcal{N}(\mathbf{h}_t; \sqrt{1-\beta_t} \mathbf{h}_{t-1}, \beta_t \mathbf{I}) $$
where $\{\beta_t\}_{t=1}^T$ is a predefined variance schedule (e.g., linear, cosine) that controls the noise level at each step. A useful property of this process is that we can sample $\mathbf{h}_t$ at any arbitrary timestep $t$ directly from $\mathbf{h}_0$:
$$ q(\mathbf{h}_t | \mathbf{h}_0) = \mathcal{N}(\mathbf{h}_t; \sqrt{\bar{\alpha}_t} \mathbf{h}_0, (1 - \bar{\alpha}_t) \mathbf{I}) $$
where $\alpha_t = 1 - \beta_t$ and $\bar{\alpha}_t = \prod_{s=1}^t \alpha_s$. As $t \to T$, if the schedule is chosen appropriately, $\mathbf{h}_T$ approaches an isotropic Gaussian distribution $\mathcal{N}(0, \mathbf{I})$.

\noindent\textbf{Reverse Process (Denoising):}
The reverse process aims to learn the transition $q(\mathbf{h}_{t-1} | \mathbf{h}_t)$, which is intractable directly. Instead, a neural network, $\epsilon_\theta(\mathbf{h}_t, t)$, is trained to predict the noise component $\epsilon$ that was added to $\mathbf{h}_0$ to produce $\mathbf{h}_t = \sqrt{\bar{\alpha}_t}\mathbf{h}_0 + \sqrt{1-\bar{\alpha}_t}\epsilon$, where $\epsilon \sim \mathcal{N}(0, \mathbf{I})$. The network is optimized by minimizing the simplified DDPM loss function (as shown in Eq.~\ref{eq:diffusion_loss_main} in the main text):
$$ \mathcal{L}_{\text{diffusion}}(\theta) = \mathbb{E}_{t, \mathbf{h}_0, \epsilon} \left[ \| \epsilon - \epsilon_\theta(\sqrt{\bar{\alpha}_t}\mathbf{h}_0 + \sqrt{1-\bar{\alpha}_t}\epsilon, t) \|^2 \right] $$
where $t$ is sampled uniformly from $\{1, ..., T\}$.

\subsection{Sampling New Latent Embeddings}
Once the denoising network $\epsilon_\theta$ is trained, new latent embeddings $\mathbf{h}_0^{\text{gen}}$ can be generated by starting with a sample from the prior distribution, $\mathbf{h}_T \sim \mathcal{N}(0, \mathbf{I})$, and iteratively applying the reverse denoising step:
$$ \mathbf{h}_{t-1} = \frac{1}{\sqrt{\alpha_t}} \left( \mathbf{h}_t - \frac{\beta_t}{\sqrt{1-\bar{\alpha}_t}} \epsilon_\theta(\mathbf{h}_t, t) \right) + \sigma_t \mathbf{z} $$
where $\mathbf{z} \sim \mathcal{N}(0, \mathbf{I})$ for $t > 1$, and $\mathbf{z} = 0$ for $t=1$. The variance $\sigma_t^2$ is typically set to $\beta_t$ or $\tilde{\beta}_t = \frac{1-\bar{\alpha}_{t-1}}{1-\bar{\alpha}_t}\beta_t$. The full sampling procedure is outlined in Algorithm~\ref{alg:app_sampling_diffusion_revised}.

\begin{algorithm}[htbp!]
\caption{Reverse Diffusion Sampling for Latent Embeddings}
\label{alg:app_sampling_diffusion_revised}
\begin{algorithmic}[1]
\State \textbf{Input:} Trained denoising model $\epsilon_\theta$, number of generation samples $B_{\text{gen}}$, dimension of pooled latent $d_p$, diffusion timesteps $T$, schedule parameters (e.g., $\beta_1, ..., \beta_T$).
\State Calculate $\alpha_t = 1-\beta_t$ and $\bar{\alpha}_t = \prod_{s=1}^t \alpha_s$ for all $t$.
\State Set $\sigma_t^2 = \beta_t$ (or alternative like $\tilde{\beta}_t$).
\State Sample initial noise $\mathbf{h}_T \sim \mathcal{N}(0, \mathbf{I})$ of shape $(B_{\text{gen}}, d_p)$.
\For{$t = T, ..., 1$}
    \State Sample $\mathbf{z} \sim \mathcal{N}(0, \mathbf{I})$ of shape $(B_{\text{gen}}, d_p)$ if $t > 1$, else $\mathbf{z} = \mathbf{0}$.
    \State Predict noise: $\epsilon_{\text{pred}} \leftarrow \epsilon_\theta(\mathbf{h}_t, t)$
    \State Calculate conditional mean: $\mu_\theta(\mathbf{h}_t, t) = \frac{1}{\sqrt{\alpha_t}} \left( \mathbf{h}_t - \frac{\beta_t}{\sqrt{1-\bar{\alpha}_t}} \epsilon_{\text{pred}} \right)$
    \State Update latent sample: $\mathbf{h}_{t-1} \leftarrow \mu_\theta(\mathbf{h}_t, t) + \sigma_t \mathbf{z}$
\EndFor
\State \textbf{Output:} Generated latent embeddings $\mathbf{h}_0^{\text{gen}} = \mathbf{h}_{0}$
\end{algorithmic}
\end{algorithm}
\FloatBarrier

\section{Loss Function Formulations}
\label{app:loss_function_formulations}
This section provides the detailed mathematical formulations for the loss functions used in the LD-FPG framework. The notation used is consistent with Appendix~\ref{app:notation_details}.

\subsection{Encoder Pre-training Loss}
The pre-training phase for the ChebNet encoder ($\text{Encoder}_{\theta}$) can be performed using a direct coordinate reconstruction head ($\text{MLP}_{\text{HNO}}$). The loss function $\Lhno$ is the Mean Squared Error (MSE) between the predicted coordinates and the ground truth coordinates:
\begin{equation}
\Lhno = \mathbb{E}_{(\Xtrue, Z) \sim \mathcal{D}} \left[ \| \text{MLP}_{\text{HNO}}(Z) - \Xtrue \|^2_F \right] \label{eq:app_loss_hno_revised}
\end{equation}
where $Z = \text{Encoder}_{\theta}(\Xtrue, E)$, $\mathcal{D}$ is the training dataset, and $\| \cdot \|^2_F$ denotes the squared Frobenius norm (sum of squared element-wise differences). For a single sample with $N$ atoms, this is $\frac{1}{N} \sum_{i=1}^N \| \text{MLP}_{\text{HNO}}(Z_i) - (\Xtrue)_i \|^2$.

\subsection{Decoder Loss Functions}

\subsubsection{Coordinate Mean Squared Error ($\Lcoord$)}
This is the fundamental loss for all decoder architectures, measuring the MSE between predicted coordinates $\Xpred$ and ground truth coordinates $\Xtrue$:
\begin{equation}
\Lcoord = \mathbb{E}_{(\Xtrue, \Xpred) \sim \mathcal{D}_{\text{dec}}} \left[ \| \Xpred - \Xtrue \|^2_F \right] \label{eq:app_loss_coord_revised}
\end{equation}
For a single sample, this is $\frac{1}{N} \sum_{i=1}^N \| (\Xpred)_i - (\Xtrue)_i \|^2$.

\subsubsection{Dihedral Angle Mean Squared Error ($\Lmseh$)}
This loss penalizes the squared difference between predicted dihedral angles ($\alpha_{\text{pred}}$) and true dihedral angles ($\alpha_{\text{true}}$). It is used \textit{only} for fine-tuning the Blind Pooling decoder.
\begin{equation}
\Lmseh = \sum_{\alpha \in \mathcal{A}} \mathbb{E}_{(b, j) | M_{\alpha}(j)} \left[ (\alpha_{\text{pred}}^{(b, j)} - \alpha_{\text{true}}^{(b, j)})^2 \right] \label{eq:app_loss_mse_dih_revised}
\end{equation}
where $\mathcal{A}$ is the set of all considered dihedral angle types (e.g., $\phi, \psi, \chi_1, \dots, \chi_5$), $M_{\alpha}(j)$ is a mask indicating if angle type $\alpha$ is valid for residue $j$ in sample $b$. The expectation is over valid angles in the batch.

\subsubsection{Dihedral Angle Distribution Divergence ($\Ldivh$)}
This loss encourages the empirical distribution of predicted dihedral angles ($P(\alpha_{\text{pred}})$) to match that of the true angles ($P(\alpha_{\text{true}})$). It is used \textit{only} for fine-tuning the blind pooling decoder.
\begin{equation}
\Ldivh = \sum_{\alpha \in \mathcal{A}} D_{\text{KL}}(P(\alpha_{\text{pred}}) \| P(\alpha_{\text{true}})) \label{eq:app_loss_div_dih_revised}
\end{equation}
where $D_{\text{KL}}$ is the Kullback-Leibler divergence (or optionally Jensen-Shannon divergence, JSD, or Wasserstein Distance, WD, as specified in hyperparameters). The distributions $P(\cdot)$ are typically estimated from histograms of angles within a batch or across a larger set of samples.

\subsubsection{Combined Decoder Loss ($\Ldec$)}
The definition of $\Ldec$ depends on the pooling strategy:

\paragraph{Blind Pooling Decoder:}
\begin{itemize}
    \item \textbf{Initial Training:} The decoder is trained solely on coordinate MSE:
    $$ \Ldec^{\text{Blind, initial}} = \Lcoord $$
    \item \textbf{Fine-tuning (Optional):} The loss becomes a weighted sum, where dihedral terms are applied stochastically (e.g., to $10\%$ of mini-batches, controlled by $f_{\text{dih}}$):
    $$ \Ldec^{\text{Blind, fine-tune}} = w_{\text{base}} \cdot \Lcoord + S \cdot (\lambda_{\text{mse}} \cdot \Lmseh + \lambda_{\text{div}} \cdot \Ldivh) $$
    where $S=1$ with probability $f_{\text{dih}}$ and $S=0$ otherwise. $w_{\text{base}}$, $\lambda_{\text{mse}}$, and $\lambda_{\text{div}}$ are scalar weights.
\end{itemize}

\paragraph{Residue-based Pooling Decoder:}
This decoder is trained using only the coordinate MSE loss:
$$ \Ldec^{\text{Residue}} = \Lcoord $$

\subsubsection{Sequential Pooling Decoder Losses}
The sequential pooling strategy uses two separate MSE-based losses:
\begin{itemize}
    \item \textbf{Backbone Decoder Loss ($\Lbb$):} This is the coordinate MSE loss applied specifically to the predicted backbone atom coordinates $X_{\text{pred,bb}}$ against the true backbone coordinates $X_{\text{true,bb}}$:
    $$ \Lbb = \mathbb{E}_{(X_{\text{true,bb}}, X_{\text{pred,bb}}) \sim \mathcal{D}_{\text{dec}}} \left[ \| X_{\text{pred,bb}} - X_{\text{true,bb}} \|^2_F \right] $$
    This corresponds to applying $\Lcoord$ only to atoms $i \in \Ibb$.

    \item \textbf{Sidechain Decoder (Full Structure) Loss ($\Lsc$):} After the Sidechain Decoder predicts sidechain coordinates and assembles the full structure $\Xpred$, this loss is the coordinate MSE for the entire protein structure:
    $$ \Lsc = \mathbb{E}_{(\Xtrue, \Xpred) \sim \mathcal{D}_{\text{dec}}} \left[ \| \Xpred - \Xtrue \|^2_F \right] $$
    This is equivalent to $\Lcoord$ applied to the output of the complete two-stage sequential decoder.
\end{itemize}

Neither $\Lbb$ nor $\Lsc$ include dihedral angle terms in their standard formulation for the sequential pooling decoder.

\section{Experimental Setup Details}
\label{app:experimental_setup_appendix}

This appendix provides further details on the dataset, evaluation metrics, and implementation aspects of the experimental setup.

\subsection{Dataset and MD Simulation Protocol}
\label{app:experimental_dataset_details}
The conformational dataset for the human Dopamine D2 receptor (D2R) was generated from all-atom Molecular Dynamics (MD) simulations.
\textbf{System Preparation:} Simulations were initiated from the cryo-EM structure of the D2R in complex with the inverse agonist risperidone (PDB ID: 6CM4~\citep{wang2018structure}). The risperidone ligand was removed, and the third intracellular loop (ICL3), which is typically flexible or unresolved, was remodeled using RosettaRemodel~\citep{huang2011rosettaremodel} to represent an apo-like state. The final remodeled D2R structure consisted of 273 residues, comprising 2191 heavy atoms after selection for the simulation system; hydrogen atoms were not explicitly included as input features to our generative model, which focuses on heavy-atom representations. The D2R protein was then embedded in a 1-palmitoyl-2-oleoyl-sn-glycero-3-phosphocholine (POPC) lipid bilayer using the CHARMM-GUI \textit{Membrane Builder}~\citep{wu2014charmm}. The system was solvated with TIP3P water~\citep{mark2001structure} and neutralized with 0.15 M NaCl ions.
\textbf{Simulation Parameters:} The CHARMM36m force field~\citep{huang2017charmm36m} was employed for all protein, lipid, and ion parameters. Simulations were performed using GROMACS 2024.2~\citep{abraham2015gromacs}. The system underwent energy minimization followed by a multi-step equilibration protocol involving NVT and NPT ensembles with position restraints on the protein heavy atoms, which were gradually released. Production simulations were run under the NPT ensemble at 303.15 K (using the v-rescale thermostat~\citep{bussi2007canonical}) and 1.0 bar (using the C-rescale barostat~\citep{bernetti2020pressure} with semi-isotropic coupling). A 2~fs timestep was used, with LINCS algorithm~\citep{hess1997lincs} constraining bonds involving hydrogen atoms. Electrostatic interactions were calculated using the Particle Mesh Ewald (PME) method~\citep{essmann1995smooth}).
\textbf{Trajectory Processing:} Ten independent production replicas, each 2~$\mu$s in length, were generated. One replica exhibiting representative dynamics was selected for this study. The initial $\approx 776$~ns of this replica were discarded as extended equilibration, yielding a final analysis trajectory of $\approx 1.224$~$\mu$s. From this, 12,241 frames were sampled at a regular interval of 100~ps. All protein heavy-atom coordinates in these frames were then rigidly aligned to the heavy atoms of the first frame using the Kabsch algorithm~\citep{kabsch1976} to remove global translation and rotation.
\textbf{Data Splitting and Preprocessing:} The aligned coordinate dataset was split into a training set (90\%, 11,017 frames) and a test set (10\%, 1,224 frames) chronologically. Static topological information, including lists of backbone and sidechain atom indices based on a consistent re-indexing scheme, and the definitions of atom quadruplets for standard dihedral angles ($\phi, \psi, \chi_1-\chi_5$), was extracted once from the reference PDB structure. This information was stored in JSON format as detailed in Appendix~\ref{app:input_processing_details}.

\subsection{Evaluation Metrics}
\label{app:evaluation_metric_details}
Model performance was assessed using the following metrics:

\subsubsection{Coordinate Accuracy}
\begin{itemize}
    \item \textbf{Mean Squared Error (MSE\textsubscript{bb}, MSE\textsubscript{sc}):} Calculated as the average squared Euclidean distance between predicted and ground truth coordinates for corresponding atoms. MSE\textsubscript{bb} considers C$\alpha$ atoms (or all backbone heavy atoms N, CA, C, O, as specified in implementation) and MSE\textsubscript{sc} considers all sidechain heavy atoms.
    The MSE for a set of $N_k$ atoms (either backbone or sidechain) is: $ \text{MSE} = \frac{1}{N_k} \sum_{i=1}^{N_k} \| \mathbf{x}_{\text{pred},i} - \mathbf{x}_{\text{true},i} \|^2 $.
    \item \textbf{Local Distance Difference Test (lDDT)}~\citep{mariani2013lddt}: lDDT evaluates the preservation of local interatomic distances. For each atom, it considers all other atoms within a defined cutoff radius (e.g., 15~\AA{}) in the reference (true) structure. It then calculates the fraction of these interatomic distances that are preserved in the predicted structure within certain tolerance thresholds (e.g., 0.5, 1, 2, and 4~\AA{}). The final lDDT score is an average over these fractions and all residues/atoms. We report lDDT\textsubscript{All} (all heavy atoms on backbone and sidechain) and lDDT\textsubscript{BB} (backbone heavy atoms). Scores range from 0 to 1, with 1 indicating perfect preservation.
    \item \textbf{Template Modeling score (TM-score)}~\citep{zhang2005tm}: TM-score measures the global structural similarity between a predicted model and a reference structure. It is designed to be more sensitive to correct global topology and less sensitive to local errors than RMSD, and its value is normalized to be between 0 and 1, where 1 indicates a perfect match. A TM-score \> 0.5 generally indicates that the two proteins share a similar fold.
\end{itemize}

\subsubsection{Distributional Accuracy}
\begin{itemize}
    \item \textbf{Summed Kullback–Leibler Divergence ($\sum$KL) and Jensen–Shannon Divergence ($\sum$JSD):} These metrics quantify the similarity between the 1D distributions of predicted and ground truth dihedral angles ($\phi, \psi,$ and $\chi_1$ through $\chi_5$). For each angle type, empirical probability distributions are estimated from histograms (e.g., using 36 bins over the range $[-\pi, \pi]$). The KL or JS divergence is calculated for each angle type, and the reported $\sum$KL or $\sum$JSD is the sum of these divergences over all seven angle types. Lower values indicate higher similarity between the distributions.
\end{itemize}

\subsubsection{Physical Plausibility}
\begin{itemize}
    \item \textbf{Average Steric Clash Counts:} A steric clash is defined as a pair of non-bonded heavy atoms being closer than a specified distance cutoff. For our analysis, we used a cutoff of 2.1~\AA{}. The clash score is the average number of such clashing pairs per generated structure. This metric was computed using BioPython~\citep{cock2009biopython} and SciPy's \citep{2020SciPy-NMeth} cKDTree for efficient neighbor searching.
\end{itemize}

\subsubsection{Conformational Landscape Sampling}
\begin{itemize}
    \item \textbf{A100 Activation Index Value:} The A100 value is a collective variable developed by Ibrahim et al.~\citep{ibrahim2019universal} to quantify the activation state of Class A G-protein-coupled receptors (GPCRs). It is a linear combination of five specific interhelical C$\alpha$-C$\alpha$ distances that are known to change upon GPCR activation. The formula is given by:
    \begin{align*}
    A^{100} = & -14.43 \times r(\text{V}^{1.53}–\text{L}^{7.55}) - 7.62 \times r(\text{D}^{2.50}–\text{T}^{3.37}) \\
              & + 9.11 \times r(\text{N}^{3.42}–\text{I}^{4.42}) - 6.32 \times r(\text{W}^{5.66}–\text{A}^{6.34}) \\
              & - 5.22 \times r(\text{L}^{6.58}–\text{Y}^{7.35}) + 278.88
    \label{eq:a100_formula}
    \end{align*}
    where $r(\text{X}^{\text{BW}_1}–\text{Y}^{\text{BW}_2})$ denotes the distance in Angstroms between the C$\alpha$ atoms of residue X at Ballesteros-Weinstein (BW) position $\text{BW}_1$ and residue Y at BW position $\text{BW}_2$. Higher A100 values typically correspond to more active-like states. We use this metric to compare the conformational landscape explored by generated ensembles against the reference MD simulation.
    \item \textbf{Principal Component Analysis (PCA) of Latent Embeddings:} PCA is applied to the set of pooled latent embeddings ($\mathbf{h}_0$) generated by the diffusion model and those derived from the MD dataset. Projecting these high-dimensional embeddings onto the first few principal components allows for a 2D visualization, which helps assess whether the generative model captures the diversity and main modes of variation present in the training data's latent space.
\end{itemize}

\subsubsection{Training Loss Reporting}
\begin{itemize}
    \item \textbf{Auxiliary Dihedral MSE Term ($\sum\mathcal{L}_{\text{dih}}$ MSE):} When dihedral losses are active during the fine-tuning of the blind pooling decoder (see Appendix~\ref{app:loss_function_formulations} for $\Lmseh$), we report the sum of the training MSE losses for all considered dihedral angles. This provides insight into how well the model fits these auxiliary geometric targets.
\end{itemize}

\subsection{Training resources}
\label{app:training}
The input Molecular Dynamics (MD) data was generated using GROMACS 2024.2~\citep{abraham2015gromacs} on NVIDIA L40S GPUs. With a simulation rate of approximately 250 ns/day, the 2 micro seconds trajectory used in this study required about 8 days of computation. Following data generation, all machine learning models were implemented in PyTorch~\citep{paszke2019pytorch} and trained using the Adam optimizer~\citep{kingma2014adam} on NVIDIA L40S and H100 GPUs. The ChebNet autoencoder training required approximately 5,000 epochs at roughly 2 seconds per epoch. Decoder training varied by pooling strategy: blind pooling decoders trained for 3,000-6,000 epochs at approximately 3 seconds per epoch, while residue pooling decoders took about 24 seconds per epoch for a similar number of epochs. Diffusion model training was the most computationally intensive: sequential and blind pooling models trained for around 10,000 epochs, and residue pooling models for up to 150,000 epochs, with each epoch averaging approximately 1 second on L40S GPUs. The machine learning aspects presented in this paper required approximately 15,000 GPU hours, with total experimentation including preliminary setups amounting to roughly 30,000 GPU hours. This substantial GPU usage translates to significant energy consumption and associated carbon emissions; while precise quantification depends on factors like data center Power Usage Effectiveness (PUE) and local energy grid carbon intensity (and was not performed for this study), we acknowledge this environmental cost. Further details on computational resources, including specific MD simulation parameters and machine learning training runtimes for hyperparameter sweeps, are provided in the Extended Technical Appendix (Appendix~\ref{app:extended_technical}). The overall three-phase training and generation workflow (encoder pre-training, decoder training, and diffusion model training) is detailed in Appendix~\ref{app:method_overview} (Algorithm~\ref{alg:main_workflow_appendix}).

\section{Encoder Performance Summary}
\label{app:encoder_performance_summary}
The ChebNet encoder's reconstruction head performance provides an upper bound on achievable fidelity. Table~\ref{tab:encoder_performance} summarizes key metrics for different latent dimensions ($d_z$). The encoder accurately captures structural features and dihedral angle distributions with minimal deviation from the ground truth MD data.

\begin{table}[ht]
\centering
\caption{Encoder Reconstruction Head Performance.}
\label{tab:encoder_performance}
\footnotesize
\renewcommand{\arraystretch}{1.1}
\begin{tabular}{@{} l ccc cc cc c @{}} 
\toprule
Encoder         & lDDT\textsubscript{All} & lDDT\textsubscript{BB} & TM\textsubscript{All} & $\sum$JSD\textsubscript{bb} & $\sum$JSD\textsubscript{sc} & MSE\textsubscript{bb} & MSE\textsubscript{sc} & $\sum\mathcal{L}_{\text{dih}}$ \\
Config          & $\uparrow$              & $\uparrow$              & $\uparrow$             & $\downarrow$                 & $\downarrow$                 & $\downarrow$              & $\downarrow$              & MSE $\downarrow$              \\
\midrule
Encoder ($d_z=4$)  & 0.692 & 0.777 & 0.959 & 0.00005 & 0.0002 & 0.0040 & 0.00715 & 0.00066 \\
Encoder ($d_z=8$)  & 0.697 & 0.780 & 0.959 & 0.00069 & 0.0009 & 0.0021 & 0.00375 & 0.00188 \\
Encoder ($d_z=16$) & 0.696 & 0.782 & 0.959 & 0.00009 & 0.00016 & 0.0008 & 0.00160 & 0.00058 \\
\midrule
GT (MD) Ref & 0.698 & 0.779 & 0.959 & - & - & - & - & - \\
\bottomrule
\end{tabular}
\end{table}

\FloatBarrier

\section{Residue Pooling Inference Details}
\label{app:residue_pooling_inference_details}
For the residue pooling model visualizations (Fig.~\ref{fig:combined_dihedral_comparison}c and Fig.~\ref{fig:pca_pmf_analysis}c in the main text), samples were aggregated from 10 distinct model checkpoints saved at different stages of the diffusion model's training. This multi-epoch sampling strategy, inspired by moving average techniques~\citep{qian2024boosting}, was employed to provide a more stable and representative visualization of the learned conformational space, averaging out potential epoch-specific biases.

\newpage

\FloatBarrier

\pgfplotstableread[col sep=comma]{
Run_Directory,H,W,D,Final_Epoch,Test_BB,Test_SC,Test_Dih_MSE,Test_Dih_JS
run_h10_w1_d12,10,1,12,5000,0.2520,0.7303,0.0,0.0
run_h10_w2_d4,10,2,4,5000,0.1834,0.5825,0.0,0.0
run_h10_w2_d8,10,2,8,5000,0.1673,0.5212,0.0,0.0
run_h15_w1_d12,15,1,12,5000,0.1965,0.5815,0.0,0.0
run_h15_w1_d8,15,1,8,5000,0.2092,0.5998,0.0,0.0
run_h15_w2_d4,15,2,4,5000,0.1405,0.4809,0.0,0.0
run_h15_w2_d8,15,2,8,5000,0.1820,0.5614,0.0,0.0
run_h20_w1_d8,20,1,8,5000,0.1845,0.5507,0.0,0.0
run_h20_w2_d4,20,2,4,5000,0.1257,0.4462,0.0,0.0
run_h20_w3_d12,20,3,12,5000,0.1081,0.4075,0.0,0.0
run_h25_w1_d12,25,1,12,5000,0.1737,0.5146,0.0,0.0
run_h30_w2_d4,30,2,4,5000,0.1038,0.3974,0.0,0.0
run_h30_w2_d8,30,2,8,5000,0.1064,0.3979,0.0,0.0
run_h30_w3_d12,30,3,12,5000,0.0910,0.3712,0.0,0.0
run_h30_w3_d4,30,3,4,5000,0.0931,0.3793,0.0,0.0
run_h35_w3_d12,35,3,12,5000,0.0888,0.3620,0.0,0.0
run_h35_w3_d4,35,3,4,5000,0.0905,0.3614,0.0,0.0
run_h40_w1_d12,40,1,12,5000,0.1504,0.4688,0.0,0.0
run_h40_w2_d12,40,2,12,5000,0.1006,0.3791,0.0,0.0
run_h45_w1_d8,45,1,8,5000,0.1513,0.4630,0.0,0.0
run_h50_w1_d12,50,1,12,5000,0.1477,0.4535,0.0,0.0
run_h50_w1_d8,50,1,8,5000,0.1442,0.4501,0.0,0.0
run_h50_w2_d12,50,2,12,5000,0.0978,0.3705,0.0,0.0
run_h50_w3_d8,50,3,8,5000,0.0783,0.3341,0.0,0.0
run_h55_w1_d12,55,1,12,5000,0.1492,0.4465,0.0,0.0
run_h55_w1_d4,55,1,4,5000,0.1466,0.4461,0.0,0.0
run_h55_w1_d8,55,1,8,5000,0.1488,0.4488,0.0,0.0
run_h55_w3_d8,55,3,8,5000,0.0820,0.3320,0.0,0.0
run_h5_w1_d8,5,1,8,5000,0.4069,1.1194,0.0,0.0
run_h5_w3_d8,5,3,8,5000,0.2970,0.8042,0.0,0.0
run_h60_w2_d12,60,2,12,5000,0.0884,0.3439,0.0,0.0
run_h60_w3_d4,60,3,4,5000,0.0777,0.3159,0.0,0.0
run_h65_w2_d4,65,2,4,5000,0.0989,0.3617,0.0,0.0
run_h65_w3_d12,65,3,12,5000,0.0791,0.3221,0.0,0.0
run_h70_w1_d8,70,1,8,5000,0.1417,0.4279,0.0,0.0
run_h70_w2_d12,70,2,12,5000,0.0940,0.3475,0.0,0.0
run_h75_w2_d8,75,2,8,5000,0.0919,0.3454,0.0,0.0
run_h75_w3_d8,75,3,8,5000,0.0837,0.3211,0.0,0.0
run_h80_w2_d4,80,2,4,5000,0.0994,0.3441,0.0,0.0
run_h85_w1_d12,85,1,12,5000,0.1427,0.4171,0.0,0.0
run_h85_w1_d4,85,1,4,5000,0.1344,0.4082,0.0,0.0
run_h85_w3_d4,85,3,4,5000,0.0812,0.3143,0.0,0.0
run_h85_w3_d8,85,3,8,5000,0.0815,0.3100,0.0,0.0
run_h90_w1_d4,90,1,4,5000,0.1396,0.4112,0.0,0.0
run_h90_w3_d4,90,3,4,5000,0.0809,0.3081,0.0,0.0
run_h90_w3_d8,90,3,8,5000,0.0794,0.3013,0.0,0.0
run_h95_w1_d8,95,1,8,5000,0.1378,0.4088,0.0,0.0
run_h95_w2_d12,95,2,12,5000,0.0819,0.3109,0.0,0.0
run_h95_w2_d4,95,2,4,5000,0.0973,0.3317,0.0,0.0
run_h95_w3_d12,95,3,12,5000,0.0740,0.2926,0.0,0.0
}\blindpoolDzFourData

\pgfplotstableread[col sep=comma]{
Run_Directory,H,W,D,Final_Epoch,Test_BB,Test_SC,Test_Dih_MSE,Test_Dih_JS
run_h100_w1_d8,100,1,8,5000,0.1393,0.4227,0.0,0.0
run_h100_w2_d12,100,2,12,5000,0.1022,0.3531,0.0,0.0
run_h100_w3_d4,100,3,4,5000,0.1640,0.4863,0.0,0.0
run_h100_w5_d8,100,5,8,5000,0.0744,0.2935,0.0,0.0
run_h100_w7_d12,100,7,12,5000,0.0840,0.3013,0.0,0.0
run_h10_w2_d12,10,2,12,5000,0.1704,0.5434,0.0,0.0
run_h15_w1_d4,15,1,4,5000,0.2284,0.6786,0.0,0.0
run_h15_w1_d8,15,1,8,5000,0.2323,0.7050,0.0,0.0
run_h15_w4_d8,15,4,8,5000,0.1288,0.4483,0.0,0.0
run_h15_w5_d4,15,5,4,5000,0.1173,0.4283,0.0,0.0
run_h15_w6_d4,15,6,4,5000,0.1252,0.4360,0.0,0.0
run_h20_w5_d8,20,5,8,5000,0.0968,0.3849,0.0,0.0
run_h25_w1_d12,25,1,12,5000,0.1878,0.5622,0.0,0.0
run_h25_w4_d8,25,4,8,5000,0.1044,0.3991,0.0,0.0
run_h25_w7_d4,25,7,4,5000,0.1005,0.3880,0.0,0.0
run_h35_w1_d4,35,1,4,5000,0.1677,0.5122,0.0,0.0
run_h35_w2_d4,35,2,4,5000,0.1065,0.3973,0.0,0.0
run_h35_w2_d8,35,2,8,5000,0.1064,0.3959,0.0,0.0
run_h35_w6_d8,35,6,8,5000,0.0874,0.3569,0.0,0.0
run_h40_w2_d4,40,2,4,5000,0.1008,0.3852,0.0,0.0
run_h40_w5_d8,40,5,8,5000,0.0806,0.3443,0.0,0.0
run_h45_w1_d12,45,1,12,5000,0.1610,0.4846,0.0,0.0
run_h50_w1_d4,50,1,4,5000,0.1598,0.4785,0.0,0.0
run_h50_w3_d12,50,3,12,5000,0.1274,0.4183,0.0,0.0
run_h50_w5_d12,50,5,12,5000,0.0767,0.3290,0.0,0.0
run_h50_w6_d12,50,6,12,5000,0.0842,0.3432,0.0,0.0
run_h55_w2_d8,55,2,8,5000,0.1003,0.3711,0.0,0.0
run_h55_w5_d4,55,5,4,5000,0.0729,0.3135,0.0,0.0
run_h55_w6_d8,55,6,8,5000,0.0879,0.3459,0.0,0.0
run_h5_w1_d12,5,1,12,5000,0.3740,1.0205,0.0,0.0
run_h5_w2_d12,5,2,12,5000,0.2496,0.7218,0.0,0.0
run_h5_w4_d8,5,4,8,5000,0.2107,0.6336,0.0,0.0
run_h5_w7_d8,5,7,8,5000,0.2631,0.7695,0.0,0.0
run_h60_w1_d12,60,1,12,5000,0.1494,0.4528,0.0,0.0
run_h60_w3_d4,60,3,4,5000,0.1375,0.4204,0.0,0.0
run_h60_w5_d4,60,5,4,5000,0.0735,0.3114,0.0,0.0
run_h65_w5_d8,65,5,8,5000,0.0724,0.3072,0.0,0.0
run_h70_w2_d4,70,2,4,5000,0.0924,0.3475,0.0,0.0
run_h70_w2_d8,70,2,8,5000,0.0952,0.3547,0.0,0.0
run_h70_w3_d12,70,3,12,5000,0.1671,0.4741,0.0,0.0
run_h75_w1_d12,75,1,12,5000,0.1461,0.4409,0.0,0.0
run_h75_w4_d12,75,4,12,5000,0.0831,0.3242,0.0,0.0
run_h80_w1_d4,80,1,4,5000,0.1422,0.4347,0.0,0.0
run_h80_w2_d8,80,2,8,5000,0.0909,0.3364,0.0,0.0
run_h85_w1_d12,85,1,12,5000,0.1485,0.4374,0.0,0.0
run_h90_w5_d8,90,5,8,5000,0.0747,0.2947,0.0,0.0
run_h95_w1_d12,95,1,12,5000,0.1404,0.4253,0.0,0.0
run_h95_w4_d8,95,4,8,5000,0.0891,0.3212,0.0,0.0
run_h95_w5_d4,95,5,4,5000,0.0730,0.2876,0.0,0.0
run_h95_w6_d8,95,6,8,5000,0.0796,0.3036,0.0,0.0
}\blindpoolDzEightData

\pgfplotstableread[col sep=comma]{
Run_Directory,H,W,D,Final_Epoch,Test_BB,Test_SC,Test_Dih_MSE,Test_Dih_JS
run_h100_w15_d4,100,15,4,5000,0.0858,0.3033,0.0,0.0
run_h100_w2_d12,100,2,12,5000,0.0933,0.3349,0.0,0.0
run_h100_w5_d4,100,5,4,5000,0.0792,0.2984,0.0,0.0
run_h100_w7_d4,100,7,4,5000,0.0734,0.2837,0.0,0.0
run_h10_w10_d12,10,10,12,5000,0.1243,0.4333,0.0,0.0
run_h10_w3_d8,10,3,8,5000,0.1725,0.5446,0.0,0.0
run_h10_w9_d12,10,9,12,5000,0.1329,0.4545,0.0,0.0
run_h15_w14_d8,15,14,8,5000,0.1205,0.4164,0.0,0.0
run_h15_w1_d12,15,1,12,5000,0.2228,0.6437,0.0,0.0
run_h15_w3_d12,15,3,12,5000,0.1506,0.4957,0.0,0.0
run_h15_w7_d4,15,7,4,5000,0.1129,0.4104,0.0,0.0
run_h20_w12_d4,20,12,4,5000,0.1012,0.3890,0.0,0.0
run_h20_w13_d8,20,13,8,5000,0.1190,0.4311,0.0,0.0
run_h30_w12_d12,30,12,12,5000,0.0879,0.3680,0.0,0.0
run_h30_w5_d8,30,5,8,5000,0.0846,0.3538,0.0,0.0
run_h35_w11_d12,35,11,12,5000,0.0934,0.3649,0.0,0.0
run_h40_w15_d12,40,15,12,5000,0.0912,0.3521,0.0,0.0
run_h40_w6_d12,40,6,12,5000,0.0746,0.3294,0.0,0.0
run_h40_w9_d12,40,9,12,5000,0.0783,0.3328,0.0,0.0
run_h45_w11_d12,45,11,12,5000,0.0835,0.3447,0.0,0.0
run_h45_w3_d8,45,3,8,5000,0.1040,0.3805,0.0,0.0
run_h45_w6_d8,45,6,8,5000,0.0759,0.3289,0.0,0.0
run_h50_w2_d12,50,2,12,5000,0.1102,0.3934,0.0,0.0
run_h50_w6_d12,50,6,12,5000,0.0718,0.3170,0.0,0.0
run_h50_w7_d12,50,7,12,5000,0.0759,0.3265,0.0,0.0
run_h50_w9_d12,50,9,12,5000,0.0751,0.3236,0.0,0.0
run_h55_w10_d4,55,10,4,5000,0.0808,0.3271,0.0,0.0
run_h55_w11_d4,55,11,4,5000,0.0786,0.3163,0.0,0.0
run_h55_w12_d8,55,12,8,5000,0.0791,0.3243,0.0,0.0
run_h55_w15_d4,55,15,4,5000,0.0795,0.3259,0.0,0.0
run_h55_w4_d8,55,4,8,5000,0.0890,0.3470,0.0,0.0
run_h55_w7_d8,55,7,8,5000,0.0728,0.3107,0.0,0.0
run_h5_w13_d12,5,13,12,5000,0.1901,0.5524,0.0,0.0
run_h60_w10_d4,60,10,4,5000,0.0734,0.3047,0.0,0.0
run_h60_w4_d4,60,4,4,5000,0.0833,0.3386,0.0,0.0
run_h60_w6_d8,60,6,8,5000,0.0699,0.3002,0.0,0.0
run_h60_w9_d12,60,9,12,5000,0.0803,0.3196,0.0,0.0
run_h65_w4_d4,65,4,4,5000,0.0827,0.3329,0.0,0.0
run_h70_w8_d4,70,8,4,5000,0.0846,0.3238,0.0,0.0
run_h70_w9_d12,70,9,12,5000,0.0817,0.3218,0.0,0.0
run_h75_w5_d8,75,5,8,5000,0.0714,0.2992,0.0,0.0
run_h80_w11_d12,80,11,12,5000,0.0790,0.3069,0.0,0.0
run_h80_w4_d12,80,4,12,5000,0.0815,0.3219,0.0,0.0
run_h80_w5_d8,80,5,8,5000,0.0717,0.2981,0.0,0.0
run_h85_w11_d4,85,11,4,5000,0.0797,0.2999,0.0,0.0
run_h90_w10_d8,90,10,8,5000,0.0724,0.2820,0.0,0.0
run_h90_w12_d4,90,12,4,5000,0.0758,0.2919,0.0,0.0
run_h90_w4_d4,90,4,4,5000,0.0780,0.3140,0.0,0.0
run_h90_w6_d4,90,6,4,5000,0.0690,0.2854,0.0,0.0
run_h95_w9_d4,95,9,4,5000,0.0802,0.2987,0.0,0.0
}\blindpoolDzSixteenData

\pgfplotstableread[col sep=comma]{
exp_id,lddt_full_all_mean,lddt_full_all_stdev,lddt_full_bb_mean,lddt_full_bb_stdev,lddt_diff_all_mean,lddt_diff_all_stdev,lddt_diff_bb_mean,lddt_diff_bb_stdev,tm_full_all_mean,tm_full_all_stdev,tm_full_bb_mean,tm_full_bb_stdev,tm_diff_all_mean,tm_diff_all_stdev,tm_diff_bb_mean,tm_diff_bb_stdev
1,0.7160,0.0312,0.7919,0.0307,0.6510,0.0574,0.6994,0.0674,0.9611,0.0059,0.9569,0.0079,0.9491,0.0212,0.9360,0.0271
2,0.7169,0.0316,0.7933,0.0303,0.6110,0.0733,0.6601,0.0818,0.9616,0.0065,0.9566,0.0078,0.9377,0.0280,0.9225,0.0385
3,0.7163,0.0306,0.7922,0.0305,0.6150,0.0680,0.6613,0.0753,0.9612,0.0060,0.9567,0.0078,0.9418,0.0225,0.9261,0.0293
4,0.7168,0.0309,0.7925,0.0310,0.6214,0.0686,0.6666,0.0719,0.9611,0.0060,0.9572,0.0079,0.9453,0.0218,0.9319,0.0262
5,0.7136,0.0289,0.7924,0.0310,0.6103,0.0808,0.6545,0.0864,0.9611,0.0060,0.9569,0.0078,0.9378,0.0310,0.9197,0.0462
6,0.7164,0.0318,0.7928,0.0305,0.6294,0.0610,0.6788,0.0690,0.9614,0.0060,0.9568,0.0078,0.9493,0.0190,0.9365,0.0256
7,0.7165,0.0307,0.7924,0.0307,0.5308,0.0822,0.5739,0.0912,0.9612,0.0063,0.9565,0.0076,0.9092,0.0452,0.8802,0.0629
8,0.7164,0.0323,0.7929,0.0313,0.3760,0.1052,0.3924,0.1188,0.9611,0.0063,0.9567,0.0079,0.7727,0.1457,0.7095,0.1711
9,0.7153,0.0318,0.7937,0.0316,0.5433,0.1121,0.5828,0.1227,0.9609,0.0058,0.9569,0.0074,0.9101,0.0627,0.8797,0.0856
10,0.7148,0.0300,0.7934,0.0309,0.4946,0.1175,0.5282,0.1294,0.9609,0.0060,0.9568,0.0079,0.8801,0.1025,0.8424,0.1230
11,0.7167,0.0309,0.7935,0.0304,0.5729,0.1036,0.6079,0.1200,0.9612,0.0063,0.9571,0.0081,0.9220,0.0696,0.8994,0.0850
12,0.7168,0.0312,0.7932,0.0308,0.5707,0.1009,0.6065,0.1179,0.9611,0.0063,0.9568,0.0077,0.9273,0.0604,0.9048,0.0752
13,0.7159,0.0303,0.7930,0.0309,0.5940,0.0884,0.6412,0.0961,0.9611,0.0062,0.9573,0.0080,0.9387,0.0410,0.9218,0.0582
14,0.7144,0.0299,0.7932,0.0315,0.6046,0.0959,0.6448,0.1147,0.9612,0.0060,0.9567,0.0080,0.9356,0.0569,0.9180,0.0751
15,0.7163,0.0305,0.7917,0.0310,0.5154,0.1200,0.5421,0.1356,0.9611,0.0060,0.9567,0.0078,0.8875,0.1044,0.8670,0.1105
}\diffusionMLPDzFourHThirtyWTwo

\pgfplotstableread[col sep=comma]{
exp_id,lddt_full_all_mean,lddt_full_all_stdev,lddt_full_bb_mean,lddt_full_bb_stdev,lddt_diff_all_mean,lddt_diff_all_stdev,lddt_diff_bb_mean,lddt_diff_bb_stdev,tm_full_all_mean,tm_full_all_stdev,tm_full_bb_mean,tm_full_bb_stdev,tm_diff_all_mean,tm_diff_all_stdev,tm_diff_bb_mean,tm_diff_bb_stdev
1,0.7163,0.0320,0.7939,0.0312,0.2396,0.0309,0.2435,0.0341,0.9613,0.0063,0.9570,0.0081,0.3752,0.0723,0.2871,0.0687
2,0.7168,0.0301,0.7916,0.0298,0.2387,0.0308,0.2461,0.0329,0.9612,0.0062,0.9571,0.0081,0.3760,0.0738,0.2861,0.0701
3,0.7139,0.0295,0.7917,0.0301,0.2387,0.0310,0.2466,0.0296,0.9611,0.0059,0.9574,0.0081,0.3745,0.0764,0.2873,0.0674
4,0.7160,0.0310,0.7928,0.0313,0.2365,0.0327,0.2435,0.0328,0.9610,0.0060,0.9568,0.0080,0.3769,0.0734,0.2826,0.0670
5,0.7165,0.0317,0.7931,0.0310,0.2365,0.0313,0.2429,0.0334,0.9610,0.0059,0.9570,0.0080,0.3712,0.0730,0.2820,0.0667
6,0.7170,0.0314,0.7938,0.0315,0.2399,0.0303,0.2448,0.0316,0.9611,0.0061,0.9570,0.0080,0.3753,0.0770,0.2878,0.0663
7,0.7156,0.0308,0.7946,0.0318,0.2385,0.0325,0.2440,0.0348,0.9611,0.0059,0.9570,0.0079,0.3757,0.0746,0.2887,0.0682
8,0.7162,0.0314,0.7931,0.0303,0.2358,0.0315,0.2433,0.0354,0.9610,0.0062,0.9566,0.0075,0.3753,0.0779,0.2837,0.0679
9,0.7149,0.0310,0.7923,0.0312,0.2372,0.0327,0.2463,0.0339,0.9613,0.0059,0.9567,0.0075,0.3791,0.0774,0.2854,0.0692
10,0.7158,0.0313,0.7921,0.0303,0.2351,0.0349,0.2434,0.0354,0.9611,0.0059,0.9569,0.0079,0.3750,0.0755,0.2835,0.0685
11,0.7145,0.0302,0.7942,0.0312,0.2357,0.0346,0.2426,0.0354,0.9612,0.0060,0.9568,0.0077,0.3783,0.0732,0.2871,0.0676
12,0.7162,0.0306,0.7947,0.0314,0.2358,0.0369,0.2431,0.0365,0.9612,0.0062,0.9566,0.0072,0.3792,0.0750,0.2875,0.0680
13,0.7163,0.0313,0.7926,0.0308,0.2387,0.0346,0.2451,0.0357,0.9612,0.0059,0.9569,0.0078,0.3721,0.0759,0.2826,0.0697
14,0.7148,0.0299,0.7940,0.0308,0.2354,0.0332,0.2425,0.0347,0.9614,0.0060,0.9569,0.0077,0.3729,0.0764,0.2821,0.0671
15,0.7177,0.0297,0.7931,0.0310,0.2389,0.0324,0.2450,0.0345,0.9610,0.0056,0.9571,0.0085,0.3767,0.0718,0.2856,0.0673
}\diffusionConvDzFourHThirtyWTwo

\pgfplotstableread[col sep=comma]{
exp_id,lddt_full_all_mean,lddt_full_all_stdev,lddt_full_bb_mean,lddt_full_bb_stdev,lddt_diff_all_mean,lddt_diff_all_stdev,lddt_diff_bb_mean,lddt_diff_bb_stdev,tm_full_all_mean,tm_full_all_stdev,tm_full_bb_mean,tm_full_bb_stdev,tm_diff_all_mean,tm_diff_all_stdev,tm_diff_bb_mean,tm_diff_bb_stdev
1,0.7137,0.0282,0.7925,0.0297,0.6710,0.0314,0.7276,0.0374,0.9609,0.0061,0.9571,0.0079,0.9568,0.0085,0.9470,0.0127
2,0.7143,0.0309,0.7922,0.0297,0.6342,0.0389,0.6778,0.0480,0.9611,0.0062,0.9569,0.0072,0.9514,0.0081,0.9351,0.0137
3,0.7146,0.0311,0.7909,0.0291,0.6005,0.0472,0.6477,0.0496,0.9611,0.0064,0.9569,0.0076,0.9438,0.0158,0.9297,0.0231
4,0.7147,0.0311,0.7920,0.0294,0.6625,0.0301,0.7222,0.0345,0.9610,0.0062,0.9576,0.0077,0.9545,0.0076,0.9449,0.0120
5,0.7145,0.0291,0.7928,0.0292,0.6310,0.0356,0.6796,0.0419,0.9613,0.0065,0.9570,0.0074,0.9510,0.0093,0.9378,0.0139
6,0.7140,0.0296,0.7915,0.0287,0.6769,0.0318,0.7327,0.0407,0.9612,0.0064,0.9574,0.0078,0.9597,0.0077,0.9495,0.0126
7,0.7125,0.0298,0.7931,0.0302,0.6691,0.0345,0.7217,0.0425,0.9610,0.0064,0.9577,0.0079,0.9586,0.0078,0.9487,0.0126
8,0.7147,0.0298,0.7935,0.0298,0.6376,0.0321,0.6888,0.0387,0.9607,0.0060,0.9567,0.0072,0.9530,0.0066,0.9389,0.0104
9,0.7138,0.0301,0.7919,0.0285,0.5670,0.0536,0.6030,0.0593,0.9609,0.0062,0.9572,0.0078,0.9239,0.0241,0.9013,0.0302
10,0.7142,0.0294,0.7917,0.0298,0.6132,0.0547,0.6633,0.0599,0.9610,0.0059,0.9573,0.0081,0.9453,0.0184,0.9311,0.0221
11,0.7145,0.0299,0.7932,0.0294,0.6313,0.0389,0.6824,0.0470,0.9610,0.0065,0.9569,0.0074,0.9512,0.0124,0.9392,0.0159
12,0.7150,0.0300,0.7929,0.0292,0.6282,0.0462,0.6764,0.0485,0.9607,0.0058,0.9570,0.0074,0.9543,0.0124,0.9412,0.0164
13,0.7149,0.0306,0.7934,0.0299,0.6515,0.0407,0.7087,0.0476,0.9609,0.0061,0.9573,0.0078,0.9541,0.0098,0.9434,0.0152
14,0.7133,0.0289,0.7929,0.0307,0.5666,0.0576,0.6054,0.0624,0.9606,0.0056,0.9570,0.0084,0.9367,0.0188,0.9168,0.0244
15,0.7119,0.0288,0.7926,0.0290,0.4250,0.0735,0.4463,0.0765,0.9607,0.0059,0.9569,0.0073,0.8647,0.0618,0.8171,0.0821
}\diffusionMLPDzEightHThirtyFiveWTwo

\pgfplotstableread[col sep=comma]{
exp_id,lddt_full_all_mean,lddt_full_all_stdev,lddt_full_bb_mean,lddt_full_bb_stdev,lddt_diff_all_mean,lddt_diff_all_stdev,lddt_diff_bb_mean,lddt_diff_bb_stdev,tm_full_all_mean,tm_full_all_stdev,tm_full_bb_mean,tm_full_bb_stdev,tm_diff_all_mean,tm_diff_all_stdev,tm_diff_bb_mean,tm_diff_bb_stdev
1,0.7131,0.0302,0.7924,0.0292,0.2429,0.0318,0.2524,0.0327,0.9610,0.0060,0.9570,0.0078,0.3756,0.0842,0.2854,0.0732
2,0.7152,0.0305,0.7922,0.0292,0.2403,0.0313,0.2481,0.0322,0.9610,0.0061,0.9572,0.0081,0.3726,0.0837,0.2830,0.0725
3,0.7147,0.0310,0.7921,0.0295,0.2405,0.0321,0.2478,0.0338,0.9610,0.0064,0.9576,0.0076,0.3731,0.0815,0.2807,0.0715
4,0.7154,0.0314,0.7936,0.0303,0.2406,0.0311,0.2476,0.0323,0.9608,0.0060,0.9575,0.0077,0.3722,0.0791,0.2861,0.0737
5,0.7133,0.0296,0.7902,0.0291,0.2420,0.0314,0.2501,0.0337,0.9612,0.0060,0.9575,0.0079,0.3752,0.0800,0.2869,0.0719
6,0.7147,0.0300,0.7926,0.0295,0.2449,0.0309,0.2513,0.0343,0.9607,0.0062,0.9570,0.0078,0.3784,0.0789,0.2941,0.0748
7,0.7140,0.0296,0.7922,0.0294,0.2385,0.0334,0.2479,0.0334,0.9613,0.0065,0.9570,0.0075,0.3656,0.0799,0.2759,0.0740
8,0.7143,0.0308,0.7935,0.0296,0.2385,0.0326,0.2452,0.0344,0.9608,0.0061,0.9569,0.0077,0.3663,0.0795,0.2758,0.0709
9,0.7147,0.0313,0.7923,0.0301,0.2386,0.0304,0.2470,0.0342,0.9609,0.0060,0.9571,0.0077,0.3674,0.0770,0.2820,0.0753
10,0.7150,0.0303,0.7927,0.0292,0.2364,0.0314,0.2449,0.0323,0.9613,0.0065,0.9568,0.0074,0.3559,0.0818,0.2682,0.0735
11,0.7140,0.0310,0.7931,0.0303,0.2401,0.0315,0.2478,0.0330,0.9608,0.0057,0.9570,0.0077,0.3722,0.0799,0.2819,0.0701
12,0.7145,0.0295,0.7934,0.0302,0.2400,0.0324,0.2500,0.0325,0.9611,0.0063,0.9570,0.0075,0.3728,0.0788,0.2842,0.0738
13,0.7138,0.0293,0.7935,0.0305,0.2387,0.0319,0.2491,0.0340,0.9608,0.0055,0.9570,0.0073,0.3714,0.0790,0.2786,0.0717
14,0.7135,0.0302,0.7923,0.0287,0.2401,0.0324,0.2467,0.0329,0.9610,0.0062,0.9571,0.0078,0.3656,0.0791,0.2783,0.0731
15,0.7141,0.0296,0.7940,0.0302,0.2421,0.0319,0.2509,0.0330,0.9608,0.0061,0.9579,0.0081,0.3680,0.0793,0.2827,0.0702
}\diffusionConvDzEightHThirtyFiveWTwo

\pgfplotstableread[col sep=comma]{
exp_id,lddt_full_all_mean,lddt_full_all_stdev,lddt_full_bb_mean,lddt_full_bb_stdev,lddt_diff_all_mean,lddt_diff_all_stdev,lddt_diff_bb_mean,lddt_diff_bb_stdev,tm_full_all_mean,tm_full_all_stdev,tm_full_bb_mean,tm_full_bb_stdev,tm_diff_all_mean,tm_diff_all_stdev,tm_diff_bb_mean,tm_diff_bb_stdev
1,0.7142,0.0303,0.7922,0.0307,0.7093,0.0261,0.7801,0.0265,0.9609,0.0062,0.9566,0.0078,0.9623,0.0057,0.9573,0.0076
2,0.7152,0.0318,0.7904,0.0297,0.7074,0.0276,0.7806,0.0276,0.9610,0.0059,0.9567,0.0081,0.9624,0.0057,0.9579,0.0076
3,0.7156,0.0312,0.7911,0.0299,0.7063,0.0270,0.7793,0.0273,0.9610,0.0063,0.9563,0.0078,0.9609,0.0062,0.9543,0.0081
4,0.7142,0.0319,0.7921,0.0305,0.7186,0.0260,0.7923,0.0255,0.9609,0.0059,0.9567,0.0076,0.9642,0.0058,0.9595,0.0074
5,0.7134,0.0303,0.7909,0.0305,0.7025,0.0270,0.7749,0.0269,0.9607,0.0056,0.9566,0.0082,0.9622,0.0062,0.9571,0.0074
6,0.7165,0.0323,0.7913,0.0302,0.7055,0.0272,0.7800,0.0280,0.9607,0.0056,0.9568,0.0081,0.9606,0.0059,0.9547,0.0081
7,0.7152,0.0314,0.7908,0.0297,0.7022,0.0232,0.7690,0.0247,0.9611,0.0059,0.9567,0.0078,0.9623,0.0060,0.9561,0.0082
8,0.7156,0.0312,0.7914,0.0312,0.6968,0.0222,0.7583,0.0241,0.9608,0.0062,0.9567,0.0080,0.9601,0.0061,0.9522,0.0087
9,0.7168,0.0314,0.7912,0.0299,0.6982,0.0262,0.7626,0.0256,0.9608,0.0061,0.9569,0.0079,0.9636,0.0059,0.9577,0.0075
10,0.7145,0.0309,0.7912,0.0305,0.6879,0.0268,0.7545,0.0287,0.9610,0.0060,0.9571,0.0085,0.9595,0.0058,0.9519,0.0089
11,0.7146,0.0301,0.7915,0.0303,0.6885,0.0294,0.7521,0.0310,0.9607,0.0059,0.9566,0.0078,0.9588,0.0065,0.9501,0.0089
12,0.7143,0.0325,0.7922,0.0312,0.7039,0.0239,0.7716,0.0255,0.9608,0.0061,0.9563,0.0078,0.9600,0.0059,0.9529,0.0086
13,0.7151,0.0315,0.7937,0.0301,0.6987,0.0270,0.7658,0.0281,0.9612,0.0065,0.9567,0.0076,0.9615,0.0060,0.9543,0.0085
14,0.7153,0.0329,0.7931,0.0320,0.6896,0.0269,0.7498,0.0279,0.9606,0.0059,0.9561,0.0075,0.9595,0.0060,0.9525,0.0089
15,0.7150,0.0306,0.7914,0.0310,0.6770,0.0324,0.7404,0.0361,0.9612,0.0065,0.9568,0.0078,0.9569,0.0068,0.9480,0.0095
}\diffusionMLPDzSixteenHFiftyWTwo

\pgfplotstableread[col sep=comma]{
exp_id,lddt_full_all_mean,lddt_full_all_stdev,lddt_full_bb_mean,lddt_full_bb_stdev,lddt_diff_all_mean,lddt_diff_all_stdev,lddt_diff_bb_mean,lddt_diff_bb_stdev,tm_full_all_mean,tm_full_all_stdev,tm_full_bb_mean,tm_full_bb_stdev,tm_diff_all_mean,tm_diff_all_stdev,tm_diff_bb_mean,tm_diff_bb_stdev
1,0.7143,0.0298,0.7913,0.0301,0.2612,0.0310,0.2665,0.0337,0.9609,0.0061,0.9565,0.0080,0.3973,0.0793,0.3046,0.0701
2,0.7148,0.0314,0.7903,0.0307,0.2627,0.0290,0.2700,0.0323,0.9607,0.0062,0.9568,0.0077,0.3991,0.0775,0.3077,0.0686
3,0.7140,0.0317,0.7913,0.0309,0.2633,0.0299,0.2702,0.0330,0.9607,0.0061,0.9569,0.0081,0.4037,0.0743,0.3117,0.0694
4,0.7151,0.0314,0.7920,0.0297,0.2642,0.0305,0.2700,0.0331,0.9612,0.0063,0.9568,0.0082,0.4043,0.0764,0.3093,0.0681
5,0.7165,0.0310,0.7924,0.0310,0.2611,0.0290,0.2659,0.0328,0.9607,0.0062,0.9567,0.0077,0.4006,0.0740,0.3081,0.0674
6,0.7134,0.0305,0.7923,0.0304,0.2624,0.0300,0.2699,0.0317,0.9608,0.0060,0.9566,0.0079,0.3986,0.0784,0.3062,0.0706
7,0.7162,0.0309,0.7928,0.0312,0.2615,0.0290,0.2689,0.0320,0.9610,0.0063,0.9566,0.0077,0.3969,0.0759,0.3094,0.0668
8,0.7152,0.0314,0.7913,0.0318,0.2641,0.0286,0.2689,0.0334,0.9609,0.0064,0.9565,0.0077,0.4048,0.0727,0.3142,0.0694
9,0.7161,0.0320,0.7918,0.0310,0.2659,0.0307,0.2718,0.0337,0.9611,0.0063,0.9567,0.0080,0.4057,0.0778,0.3139,0.0741
10,0.7150,0.0309,0.7913,0.0308,0.2612,0.0298,0.2685,0.0315,0.9610,0.0065,0.9568,0.0077,0.4015,0.0729,0.3069,0.0676
11,0.7140,0.0306,0.7918,0.0313,0.2649,0.0303,0.2703,0.0335,0.9612,0.0066,0.9565,0.0072,0.4061,0.0772,0.3114,0.0697
12,0.7147,0.0306,0.7930,0.0301,0.2615,0.0292,0.2706,0.0317,0.9608,0.0059,0.9567,0.0076,0.4035,0.0740,0.3121,0.0690
13,0.7158,0.0314,0.7924,0.0310,0.2656,0.0304,0.2735,0.0330,0.9611,0.0064,0.9567,0.0075,0.4026,0.0757,0.3122,0.0700
14,0.7133,0.0308,0.7920,0.0300,0.2604,0.0292,0.2679,0.0334,0.9611,0.0065,0.9567,0.0077,0.3948,0.0770,0.3067,0.0706
15,0.7164,0.0312,0.7923,0.0315,0.2630,0.0308,0.2704,0.0335,0.9612,0.0068,0.9566,0.0084,0.4059,0.0766,0.3061,0.0684
}\diffusionConvDzSixteenHFiftyWTwo

\pgfplotstableread[col sep=comma]{
Config,Angle,KL_1D,JS_1D,WDist_1D,exp_id,comparison
{dz=4, Exp 1},$\chi_1$,0.037865,0.010752,13.028176,1,GroundTruth vs Diffusion
{dz=4, Exp 1},$\chi_2$,0.011884,0.003051,10.388814,1,GroundTruth vs Diffusion
{dz=4, Exp 1},$\chi_3$,0.036550,0.009076,15.252775,1,GroundTruth vs Diffusion
{dz=4, Exp 1},$\chi_4$,0.032713,0.008762,19.857312,1,GroundTruth vs Diffusion
{dz=4, Exp 1},$\chi_5$,0.082701,0.023052,36.455948,1,GroundTruth vs Diffusion
{dz=4, Exp 1},$\phi$,0.031362,0.008605,15.979862,1,GroundTruth vs Diffusion
{dz=4, Exp 1},$\psi$,0.027786,0.007403,11.350032,1,GroundTruth vs Diffusion
{dz=4, Exp 1},$\chi_1$,0.012140,0.003618,7.941929,1,GroundTruth vs Decoder
{dz=4, Exp 1},$\chi_2$,0.005828,0.001528,13.219376,1,GroundTruth vs Decoder
{dz=4, Exp 1},$\chi_3$,0.021189,0.005485,9.983779,1,GroundTruth vs Decoder
{dz=4, Exp 1},$\chi_4$,0.021629,0.005498,10.366892,1,GroundTruth vs Decoder
{dz=4, Exp 1},$\chi_5$,0.040302,0.012062,22.170793,1,GroundTruth vs Decoder
{dz=4, Exp 1},$\phi$,0.004798,0.001316,2.611282,1,GroundTruth vs Decoder
{dz=4, Exp 1},$\psi$,0.003886,0.001007,2.432655,1,GroundTruth vs Decoder
{dz=4, Exp 1},$\chi_1$,0.011793,0.003084,10.842073,1,Decoder vs Diffusion
{dz=4, Exp 1},$\chi_2$,0.002716,0.000701,6.423267,1,Decoder vs Diffusion
{dz=4, Exp 1},$\chi_3$,0.003609,0.000878,10.733951,1,Decoder vs Diffusion
{dz=4, Exp 1},$\chi_4$,0.007721,0.001994,17.542604,1,Decoder vs Diffusion
{dz=4, Exp 1},$\chi_5$,0.010542,0.002656,14.634647,1,Decoder vs Diffusion
{dz=4, Exp 1},$\phi$,0.016858,0.004595,14.091282,1,Decoder vs Diffusion
{dz=4, Exp 1},$\psi$,0.016236,0.004462,9.835905,1,Decoder vs Diffusion
{dz=8, Exp 6},$\chi_1$,0.039824,0.011200,13.199047,6,GroundTruth vs Diffusion
{dz=8, Exp 6},$\chi_2$,0.013639,0.003435,10.978865,6,GroundTruth vs Diffusion
{dz=8, Exp 6},$\chi_3$,0.034713,0.008655,14.310249,6,GroundTruth vs Diffusion
{dz=8, Exp 6},$\chi_4$,0.032880,0.008760,17.964475,6,GroundTruth vs Diffusion
{dz=8, Exp 6},$\chi_5$,0.083956,0.023387,37.362673,6,GroundTruth vs Diffusion
{dz=8, Exp 6},$\phi$,0.033246,0.009066,13.766094,6,GroundTruth vs Diffusion
{dz=8, Exp 6},$\psi$,0.029537,0.007825,11.055285,6,GroundTruth vs Diffusion
{dz=8, Exp 6},$\chi_1$,0.013008,0.003852,8.813605,6,GroundTruth vs Decoder
{dz=8, Exp 6},$\chi_2$,0.006472,0.001691,14.829592,6,GroundTruth vs Decoder
{dz=8, Exp 6},$\chi_3$,0.022682,0.005848,10.439701,6,GroundTruth vs Decoder
{dz=8, Exp 6},$\chi_4$,0.021664,0.005588,9.634313,6,GroundTruth vs Decoder
{dz=8, Exp 6},$\chi_5$,0.041354,0.012372,22.825359,6,GroundTruth vs Decoder
{dz=8, Exp 6},$\phi$,0.005960,0.001639,2.848741,6,GroundTruth vs Decoder
{dz=8, Exp 6},$\psi$,0.005056,0.001318,2.605878,6,GroundTruth vs Decoder
{dz=8, Exp 6},$\chi_1$,0.011180,0.002889,10.209458,6,Decoder vs Diffusion
{dz=8, Exp 6},$\chi_2$,0.003106,0.000791,7.276048,6,Decoder vs Diffusion
{dz=8, Exp 6},$\chi_3$,0.002643,0.000644,10.238723,6,Decoder vs Diffusion
{dz=8, Exp 6},$\chi_4$,0.005664,0.001455,14.466568,6,Decoder vs Diffusion
{dz=8, Exp 6},$\chi_5$,0.010241,0.002580,14.861829,6,Decoder vs Diffusion
{dz=8, Exp 6},$\phi$,0.015669,0.004217,11.684770,6,Decoder vs Diffusion
{dz=8, Exp 6},$\psi$,0.015416,0.004200,9.108585,6,Decoder vs Diffusion
{dz=16, Exp 4},$\chi_1$,0.023309,0.006753,10.893496,4,GroundTruth vs Diffusion
{dz=16, Exp 4},$\chi_2$,0.010189,0.002636,17.537873,4,GroundTruth vs Diffusion
{dz=16, Exp 4},$\chi_3$,0.029530,0.007497,11.955594,4,GroundTruth vs Diffusion
{dz=16, Exp 4},$\chi_4$,0.025478,0.006747,11.417731,4,GroundTruth vs Diffusion
{dz=16, Exp 4},$\chi_5$,0.071857,0.020356,34.276746,4,GroundTruth vs Diffusion
{dz=16, Exp 4},$\phi$,0.013566,0.003763,5.112299,4,GroundTruth vs Diffusion
{dz=16, Exp 4},$\psi$,0.011271,0.002991,5.249407,4,GroundTruth vs Diffusion
{dz=16, Exp 4},$\chi_1$,0.013071,0.003862,8.233664,4,GroundTruth vs Decoder
{dz=16, Exp 4},$\chi_2$,0.006375,0.001668,13.299654,4,GroundTruth vs Decoder
{dz=16, Exp 4},$\chi_3$,0.021841,0.005644,10.571378,4,GroundTruth vs Decoder
{dz=16, Exp 4},$\chi_4$,0.020527,0.005241,9.515237,4,GroundTruth vs Decoder
{dz=16, Exp 4},$\chi_5$,0.042190,0.012537,22.764921,4,GroundTruth vs Decoder
{dz=16, Exp 4},$\phi$,0.006587,0.001810,3.016411,4,GroundTruth vs Decoder
{dz=16, Exp 4},$\psi$,0.005287,0.001388,2.749562,4,GroundTruth vs Decoder
{dz=16, Exp 4},$\chi_1$,0.002265,0.000579,3.111313,4,Decoder vs Diffusion
{dz=16, Exp 4},$\chi_2$,0.000806,0.000204,4.358795,4,Decoder vs Diffusion
{dz=16, Exp 4},$\chi_3$,0.000812,0.000202,3.241236,4,Decoder vs Diffusion
{dz=16, Exp 4},$\chi_4$,0.001529,0.000390,7.229134,4,Decoder vs Diffusion
{dz=16, Exp 4},$\chi_5$,0.005644,0.001439,11.530388,4,Decoder vs Diffusion
{dz=16, Exp 4},$\phi$,0.001897,0.000501,2.693907,4,Decoder vs Diffusion
{dz=16, Exp 4},$\psi$,0.001866,0.000493,2.626832,4,Decoder vs Diffusion
}\allDihedralDivergenceData

\pgfplotstableread[col sep=comma]{
H,W,Depth,Arch,Loss_SC,LogFile
14,3,4,2,0.4892,sc_pool_d1_14_d2_3__depth_4__arch_2.log
8,1,8,2,0.4911,sc_pool_d1_8_d2_1__depth_8__arch_2.log
37,6,8,2,0.4783,sc_pool_d1_37_d2_6__depth_8__arch_2.log
75,4,8,2,0.4740,sc_pool_d1_75_d2_4__depth_8__arch_2.log
39,1,12,1,0.7199,sc_pool_d1_39_d2_1__depth_12__arch_1.log
17,6,8,2,0.4802,sc_pool_d1_17_d2_6__depth_8__arch_2.log
54,2,8,0,0.5606,sc_pool_d1_54_d2_2__depth_8__arch_0.log
47,4,8,1,0.6139,sc_pool_d1_47_d2_4__depth_8__arch_1.log
32,2,4,2,0.4674,sc_pool_d1_32_d2_2__depth_4__arch_2.log
12,3,4,1,0.5425,sc_pool_d1_12_d2_3__depth_4__arch_1.log
18,5,4,0,0.5555,sc_pool_d1_18_d2_5__depth_4__arch_0.log
54,2,4,1,0.5130,sc_pool_d1_54_d2_2__depth_4__arch_1.log
72,1,12,1,0.8811,sc_pool_d1_72_d2_1__depth_12__arch_1.log
15,2,12,2,0.4927,sc_pool_d1_15_d2_2__depth_12__arch_2.log
30,2,12,0,0.9795,sc_pool_d1_30_d2_2__depth_12__arch_0.log
25,6,4,2,0.4788,sc_pool_d1_25_d2_6__depth_4__arch_2.log
57,4,8,0,0.5027,sc_pool_d1_57_d2_4__depth_8__arch_0.log
21,2,8,1,0.8785,sc_pool_d1_21_d2_2__depth_8__arch_1.log
47,2,8,0,0.9013,sc_pool_d1_47_d2_2__depth_8__arch_0.log
40,3,8,2,0.4834,sc_pool_d1_40_d2_3__depth_8__arch_2.log
48,2,12,1,0.7279,sc_pool_d1_48_d2_2__depth_12__arch_1.log
2,5,12,1,0.9045,sc_pool_d1_2_d2_5__depth_12__arch_1.log
91,6,4,1,0.5122,sc_pool_d1_91_d2_6__depth_4__arch_1.log
75,5,12,1,0.8795,sc_pool_d1_75_d2_5__depth_12__arch_1.log
67,1,12,1,0.6867,sc_pool_d1_67_d2_1__depth_12__arch_1.log
12,1,4,0,0.4914,sc_pool_d1_12_d2_1__depth_4__arch_0.log
73,4,12,0,0.8783,sc_pool_d1_73_d2_4__depth_12__arch_0.log
97,7,4,2,0.4324,sc_pool_d1_97_d2_7__depth_4__arch_2.log
63,4,8,0,0.6167,sc_pool_d1_63_d2_4__depth_8__arch_0.log
49,1,4,2,0.4724,sc_pool_d1_49_d2_1__depth_4__arch_2.log
24,5,8,0,0.6173,sc_pool_d1_24_d2_5__depth_8__arch_0.log
88,2,4,0,0.5219,sc_pool_d1_88_d2_2__depth_4__arch_0.log
2,3,8,1,0.6743,sc_pool_d1_2_d2_3__depth_8__arch_1.log
87,6,12,1,0.5928,sc_pool_d1_87_d2_6__depth_12__arch_1.log
70,5,4,0,0.5032,sc_pool_d1_70_d2_5__depth_4__arch_0.log
18,7,8,2,0.5001,sc_pool_d1_18_d2_7__depth_8__arch_2.log
84,7,4,1,0.5096,sc_pool_d1_84_d2_7__depth_4__arch_1.log
72,7,12,2,0.4939,sc_pool_d1_72_d2_7__depth_12__arch_2.log
75,7,8,2,0.4699,sc_pool_d1_75_d2_7__depth_8__arch_2.log
43,5,12,0,0.9488,sc_pool_d1_43_d2_5__depth_12__arch_0.log
48,1,4,0,0.5703,sc_pool_d1_48_d2_1__depth_4__arch_0.log
83,7,4,1,0.5057,sc_pool_d1_83_d2_7__depth_4__arch_1.log
87,5,4,2,0.4614,sc_pool_d1_87_d2_5__depth_4__arch_2.log
17,5,8,1,0.5919,sc_pool_d1_17_d2_5__depth_8__arch_1.log
16,4,4,2,0.4606,sc_pool_d1_16_d2_4__depth_4__arch_2.log
62,2,4,0,0.6043,sc_pool_d1_62_d2_2__depth_4__arch_0.log
58,1,4,0,0.5165,sc_pool_d1_58_d2_1__depth_4__arch_0.log
24,6,4,0,0.5164,sc_pool_d1_24_d2_6__depth_4__arch_0.log
23,7,12,2,0.4932,sc_pool_d1_23_d2_7__depth_12__arch_2.log
42,7,4,2,0.4558,sc_pool_d1_42_d2_7__depth_4__arch_2.log
8,1,12,0,0.6808,sc_pool_d1_8_d2_1__depth_12__arch_0.log
48,4,12,1,0.6843,sc_pool_d1_48_d2_4__depth_12__arch_1.log
5,5,4,0,0.5192,sc_pool_d1_5_d2_5__depth_4__arch_0.log
16,7,4,2,0.4710,sc_pool_d1_16_d2_7__depth_4__arch_2.log
42,5,8,2,0.4848,sc_pool_d1_42_d2_5__depth_8__arch_2.log
98,2,4,0,0.5215,sc_pool_d1_98_d2_2__depth_4__arch_0.log
95,3,12,1,0.8626,sc_pool_d1_95_d2_3__depth_12__arch_1.log
58,1,4,1,0.5296,sc_pool_d1_58_d2_1__depth_4__arch_1.log
92,3,4,1,0.5208,sc_pool_d1_92_d2_3__depth_4__arch_1.log
}\sequentialSCResultsDzEight

\pgfplotstableread[col sep=comma]{
H,W,Depth,Arch,lddt_rec_all,lddt_rec_bb,tm_rec_all,tm_rec_bb,lddt_diff_all,lddt_diff_bb,tm_diff_all,tm_diff_bb
54,2,4,1,0.7184,0.8002,0.9611,0.9573,0.7122,0.8013,(N/A),(N/A)
}\sequentialDiffusionSelectedDzEight

\pgfplotstableread[col sep=comma]{
Angle,KL_1D_GTvDiff,JS_1D_GTvDiff,WDist_1D_GTvDiff,KL_1D_GTvDec,JS_1D_GTvDec,WDist_1D_GTvDec,KL_1D_DecVDiff,JS_1D_DecVDiff,WDist_1D_DecVDiff
chi1,0.024874,0.007150,10.431319,0.009041,0.002687,8.249428,0.010322,0.002586,8.664289
chi2,0.012995,0.003334,18.864762,0.004448,0.001173,13.786683,0.004381,0.001093,5.375646
chi3,0.022270,0.005852,23.389903,0.013974,0.003640,19.387823,0.003037,0.000751,6.115204
chi4,0.018114,0.004195,9.063965,0.019479,0.003948,15.864460,0.002491,0.000633,11.487979
chi5,0.019960,0.004423,5.852986,0.039736,0.007729,6.858855,0.004850,0.001315,2.990540
phi,0.007379,0.001765,2.362356,0.013175,0.002954,2.746157,0.001717,0.000456,1.114269
psi,0.005330,0.001166,3.116599,0.008159,0.001642,2.295254,0.001100,0.000295,1.858829
}\sequentialDivDataSelectedDzEight

\section{Extended Technical Appendix}
\label{app:extended_technical}

\noindent This Extended Technical Appendix provides a deeper dive into the experimental methodology, specific hyperparameter choices that led to the main paper's results, additional sensitivity analyses, and information regarding the availability of code and datasets used in this work. The sections are organized as follows:
\begin{itemize}
    \item \textbf{Code and Data Availability} (Section~\ref{app:code_data_availability_ext}): Links to the LD-FPG implementation and the D2R-MD dataset.
    \item \textbf{Estimated CO$_2$-equivalent emissions} (Section~\ref{app:carbon_details_ext}): Discussion on the computational resources and environmental impact.
    \item \textbf{ChebNet Encoder Details and Ablation} (Sections~\ref{app:encoder_config_details_ext} and \ref{app:chebnet_encoder_ablation_tech_appendix_detail_ext}): Parameters for the encoder used in main results and a sensitivity analysis on its key hyperparameters ($k, K, d_z$).
    \item \textbf{Decoder Architectures and Configurations} (Section~\ref{app:decoder_configs_main_results_ext}): Detailed configurations for Blind, Sequential, and Residue pooling strategies corresponding to main paper results (Tables~\ref{tab:decoder_reconstruction} and \ref{tab:diffusion_generation}), including final chosen pooling dimensions and dihedral loss settings for Blind Pooling fine-tuning. This is supplemented by illustrative hyperparameter scan examples for Blind Pooling decoders (Section~\ref{app:blind_pool_search_tech_appendix_detail_ext}).
    \item \textbf{Latent Diffusion Model Details} (Sections~\ref{app:diffusion_models_config_overview_detail_ext} and \ref{app:diffusion_models_blind_scan_tables_ext}): Overview of diffusion model configurations and example hyperparameter scans for the Blind Pooling strategy.
    \item \textbf{Conditioning Mechanism Tuning} (Section~\ref{app:conditioning_tuning_details_ext}): Ablation study results for the decoder conditioning mechanism.
    \item \textbf{Discussion on Metric Variability} (Section~\ref{app:metric_variability_sd_ext}): Contextualization of standard deviations for reported performance metrics.
\end{itemize}
While comprehensive hyperparameter sweeps were conducted for all components and strategies, due to space constraints, we primarily showcase detailed scan results for the Blind Pooling strategy to illustrate the optimization process. Similar systematic approaches were employed for other strategies. For exhaustive details beyond what is presented, readers are encouraged to consult the provided codebase or contact the authors.

\subsection{Code and Data Availability}
\label{app:code_data_availability_ext}
To facilitate reproducibility and further research, the LD-FPG implementation (including scripts for data preprocessing, model training, and evaluation) is made available at:
\begin{itemize}
    \item \textbf{Code (LD-FPG)}: \url{https://github.com/adityasengar/LD-FPG}
\end{itemize}
The D2R-MD dataset, comprising the $2\,\mu\text{s}$ MD trajectory (12,241 frames) of the human dopamine D2 receptor used in this study, is available at:
\begin{itemize}
    \item \textbf{Dataset (D2R-MD)}: \url{https://zenodo.org/records/16582853}
\end{itemize}
The dataset includes aligned heavy-atom coordinates and pre-calculated dihedral angles, along with the necessary topology files as described in Appendix~\ref{app:input_processing_details}.

\subsection{Estimated \texorpdfstring{CO\textsubscript{2}}{CO2}-equivalent emissions}
\label{app:carbon_details_ext}
To provide a transparent measure of the environmental cost of our computational pipeline, we estimate the operational (Scope-2) footprint of the $\sim$30,000 GPU-hours consumed in this work. This total compute time was a mix of NVIDIA H100 and L40S GPUs. For our typical training workloads, H100 GPUs offered a significant speed-up, completing tasks in roughly half the time compared to L40S GPUs. For instance, a ChebNet autoencoder training of 5,000 epochs took approximately 2 seconds per epoch on an H100 GPU (totaling $\sim$2.8 H100-hours for one such run).

Following \cite{henderson2020towards}, the \emph{per-GPU-hour} emissions are:
\begin{equation}
\label{eq:carbon_app_ext} 
\mathrm{kg\,CO_{2}e} = \left(\frac{P_{\mathrm{board}}}{1000}\right) \times u \times \mathrm{PUE} \times \mathrm{CI},
\end{equation}
where $P_{\mathrm{board}}$ is the vendor-specified board power, $u$ is the average utilisation of that cap, $\mathrm{PUE}$ is the data-centre power usage effectiveness, and $\mathrm{CI}$ is the grid carbon intensity.

\paragraph{Assumptions.}
Board powers were taken from the NVIDIA L40S datasheet ($P=350$ W)~\cite{l40s} and the H100 SXM-5 specification ($P=700$ W)~\cite{h100}. We adopt a conservative mean utilisation $u = 0.8$, and a modern colocation PUE of 1.30, which is below the current global fleet average of 1.56 but well within the range observed for new facilities~\cite{PUE}.
For the electricity mix we give two reference cases:
(i) the 2022 global mean ($\mathrm{CI}=0.436$ kg\,CO\textsubscript{2}e\,kWh$^{-1}$) derived from Ember’s Global Electricity Review~\cite{ember2023review}, and
(ii) the Swiss consumption-based factor of 0.112 kg\,CO\textsubscript{2}e\,kWh$^{-1}$ for the same year, which captures the low-carbon hydro–nuclear mix and net imports~\cite{aliunid2023co2}.

\paragraph{Results.}
Table~\ref{tab:co2_app_ext} summarises six scenarios. The "all-L40S" and "all-H100" scenarios are hypotheticals where the entire 30,000 GPU-hours were run on a single GPU type. More realistically, a mix of GPUs was used. For an illustrative scenario, we consider a split where 70\% of the total operational hours (21,000 hours) were logged on L40S GPUs and 30\% (9,000 hours) on H100 GPUs. The energy consumption for this specific split is calculated as the sum of energy consumed by each GPU type for their respective hours. Even the worst-case (all-H100 on the global-average grid) remains on the order of a single round-trip transatlantic flight for two passengers, while execution on a Swiss-like grid cuts emissions by $\sim$4$\times$.

\begin{table}[ht]
\centering
\caption{Estimated operational CO\textsubscript{2}e emissions for the $\sim$30,000 GPU-hours used in this study. Energy is computed as $(P_{\mathrm{board}}/1000) \times \text{Hours} \times u \times \mathrm{PUE}$, with $u = 0.8$ and $\mathrm{PUE} = 1.30$. For the 70/30 split, L40S = 21,000 hours; H100 = 9,000 hours.}
\label{tab:co2_app_ext}
\begin{tabular}{@{}lrrrr@{}}
\toprule
Scenario & \multicolumn{1}{c}{Energy (kWh)} & \multicolumn{1}{c}{Grid CI (kg/kWh)} & \multicolumn{1}{c}{PUE} & \multicolumn{1}{c}{tCO\textsubscript{2}e} \\
\midrule
All L40S (30k hrs)             & 10,920  & 0.436 & 1.30 & 4.8 \\
All H100 (30k hrs)            & 21,840  & 0.436 & 1.30 & 9.5 \\
70\% L40S / 30\% H100         & 14,196  & 0.436 & 1.30 & 6.2 \\
L40S (Swiss mix)              & 10,920  & 0.112 & 1.30 & 1.2 \\
H100 (Swiss mix)              & 21,840  & 0.112 & 1.30 & 2.4 \\
70\% L40S / 30\% H100 (Swiss) & 14,196  & 0.112 & 1.30 & 1.6 \\
\bottomrule
\end{tabular}
\end{table}

\paragraph{Context and mitigation.}
These figures exclude embodied (Scope-3) emissions of the hardware itself, which may add $\sim$2--3 tCO\textsubscript{2}e over its lifetime. Practitioners can further reduce impact by (a) scheduling training in regions with low-carbon grids, (b) capping GPU power (the H100 supports dynamic 350–700 W limits), (c) early stopping during training and (d) prioritising model efficiency improvements that reduce GPU runtime.

\subsection{Graph Construction and ChebNet Encoder Parameters for Main Results}
\label{app:encoder_config_details_ext} 
For each MD frame, the graph edge index $E^{(t)}$ was constructed using a $k$-Nearest Neighbors search with $k=4$ on the aligned coordinates. This value of $k$ was chosen as it was found to maintain graph connectivity effectively while managing computational cost for graph construction at each step. The ChebNet encoder (Section~\ref{sec:encoder_method_main}) utilizes Chebyshev polynomials up to order $K=4$. The network comprises $L_{\text{ChebNet}}=4$ ChebConv layers. The first two layers use LeakyReLU activations, the third employs ReLU, and BatchNorm1d is applied after each ChebConv layer. The final layer's output, the atom-wise latent embeddings $Z^{(t)}$, is $L_2$ normalized. The encoder latent dimension $d_z$ was varied for different pooling strategies, with $d_z \in \{4, 8, 16\}$ being explored. The specific $d_z$ values used for the main results with each pooling strategy are detailed in Section~\ref{app:decoder_configs_main_results_ext}. Key parameters for the ChebNet encoder architecture and its pre-training regimen are summarized in Table~\ref{tab:app_encoder_params_consolidated_ext}.

\begin{table}[htbp]
\small
\centering
\caption{ChebNet Encoder Model parameters and training Hyperparameters used for generating main paper results.}
\label{tab:app_encoder_params_consolidated_ext} 
\begin{tabular}{lr}
\toprule
Parameter                & Value(s)        \\
\midrule
Input Features           & 3 (Coordinates) \\
Number of Atoms (N, D2R) & 2191            \\
kNN Graph k ($k$)          & 4               \\
ChebConv Order ($K$)      & 4               \\
Number of Layers ($L_{\text{ChebNet}}$) & 4               \\
Hidden Dimension ($d_z$)   & Tuned per pooling (see Sec.~\ref{app:decoder_configs_main_results_ext}: 4, 8, or 16)    \\
Activation (L1, L2)      & LeakyReLU       \\
Activation (L3)          & ReLU            \\
Normalization            & BatchNorm1d     \\
Final Embedding Norm     & L2              \\
Optimizer                & Adam            \\
Learning Rate            & 1e-4            \\
Batch Size               & 16              \\
Epochs (Pre-training)    & 4000            \\
Dihedral Loss during Pre-training & Disabled        \\
\bottomrule
\end{tabular}
\end{table}

\subsection{ChebNet Encoder Ablation: k-NN, Chebyshev Order ($K$), and Latent Dimension ($d_z$)}
\label{app:chebnet_encoder_ablation_tech_appendix_detail_ext} 

This subsection details an ablation study conducted on the ChebNet autoencoder to explore the impact of varying key graph construction and architectural parameters: the number of nearest neighbors ($k$) for graph construction, the Chebyshev polynomial order ($K$), and the latent dimension ($d_z$). For each parameter combination presented in Table~\ref{tab:tech_appendix_chebnet_ablation}, the ChebNet autoencoder was trained for approximately 24 hours, and we report the final test loss (coordinate MSE) achieved and the average time per epoch.

It is important to note that if the performance figures here, particularly the final test losses, differ from those presented in Appendix~\ref{app:encoder_performance_summary} (Table~\ref{tab:encoder_performance}), it is primarily because the encoder models in Appendix~\ref{app:encoder_performance_summary} were trained for a fixed 5000 epochs using specific parameters ($k=4, K=4$) which we determined yielded a sufficiently low MSE for subsequent stages of the LD-FPG pipeline. The experiments here, with a fixed 24-hour training budget, provide a different perspective on parameter sensitivity under time-constrained training.

The results in Table~\ref{tab:tech_appendix_chebnet_ablation} demonstrate that the ChebNet autoencoder architecture is capable of achieving low reconstruction loss across a range of parameter settings. Generally, increasing the latent dimension $d_z$ tends to decrease the final test loss, albeit with a slight increase in epoch time. The influence of $k$ and $K$ is more nuanced, showing that different combinations can yield strong performance. For instance, with $K=4$, increasing $k$ from 4 to 8 and then to 16, particularly for $d_z=16$, resulted in the lowest test loss ($0.000110$) in this set of experiments ($K=4, k=8, d_z=16$). Higher orders of $K$ (e.g., $K=8, 16$) also show competitive losses but come with a more significant increase in epoch computation time.

Based on these findings, the ChebNet autoencoder effectively reduces reconstruction loss under various conditions. Users and researchers are encouraged to explore these parameters to optimize performance for their specific protein systems and computational budgets. The subsequent pooling strategies (Blind, Sequential, and Residue pooling) and the main results presented in this paper were developed based on our preliminary analyses which utilized $K=4$ and $k=4$. This choice offered a good balance between model performance and the computational cost associated with graph construction and training, especially for the extensive MD datasets used.

\begin{table}[htbp]
\centering
\caption{ChebNet Encoder Ablation Study Results. Each configuration was trained for approximately 24 hours.}
\label{tab:tech_appendix_chebnet_ablation}
\footnotesize 
\renewcommand{\arraystretch}{1.1} 
\begin{tabular}{@{}ccccc@{}}
\toprule
Chebyshev Order ($K$) & k-NN Value ($k$) & Hidden Dim ($d_z$) & Avg Epoch Time (s) & Final Test Loss \\
\midrule
4   & 4     & 4     & 4.63  & 0.001756  \\
4   & 4     & 8     & 4.66  & 0.000184  \\
4   & 4     & 16    & 4.74  & 0.000263  \\
\addlinespace[0.3em] 
4   & 8     & 4     & 4.74  & 0.000371  \\
4   & 8     & 8     & 4.70  & 0.000373  \\
4   & 8     & 16    & 4.97  & 0.000110  \\
\addlinespace[0.3em]
4   & 16    & 4     & 5.45  & 0.001016  \\
4   & 16    & 8     & 5.20  & 0.000387  \\
4   & 16    & 16    & 5.91  & 0.000226  \\
\midrule 
8   & 4     & 4     & 6.99  & 0.000448  \\
8   & 4     & 8     & 6.91  & 0.000698  \\
8   & 4     & 16    & 7.08  & 0.000649  \\
\addlinespace[0.3em]
8   & 8     & 4     & 7.16  & 0.000624  \\
8   & 8     & 8     & 7.13  & 0.000562  \\
8   & 8     & 16    & 7.34  & 0.000545  \\
\addlinespace[0.3em]
8   & 16    & 4     & 7.52  & 0.000256  \\
8   & 16    & 8     & 7.68  & 0.000256  \\
8   & 16    & 16    & 10.73 & 0.001494  \\
\midrule
16  & 4     & 4     & 11.52 & 0.000262  \\
16  & 4     & 8     & 11.38 & 0.000484  \\
16  & 4     & 16    & 11.81 & 0.001428  \\
\addlinespace[0.3em]
16  & 8     & 4     & 11.86 & 0.003231  \\
16  & 8     & 8     & 11.55 & 0.001885  \\
16  & 8     & 16    & 12.27 & 0.000741  \\
\addlinespace[0.3em]
16  & 16    & 4     & 12.82 & 0.003709  \\
16  & 16    & 8     & 14.89 & 0.001196  \\
16  & 16    & 16    & 17.49 & 0.003388  \\
\bottomrule
\end{tabular}
\end{table}
\FloatBarrier 

\subsection{Decoder Architectures: General Configuration and Key Parameters for Main Results}
\label{app:decoder_configs_main_results_ext}
All decoder architectures map atom-wise latent embeddings $Z$ and a conditioner $C = \Zref$ (the latent embedding of the reference structure $\Xref$) to Cartesian coordinates $\Xpred$. General training parameters for decoders included the Adam optimizer, a batch size of 16, and initial training for 5000 epochs with a learning rate of $3 \times 10^{-4}$, unless specified otherwise for fine-tuning or specific pooling strategies.

\paragraph{Blind Pooling Strategy Details}
Atom-wise embeddings $Z^{(t)}$ are globally pooled using 2D adaptive average pooling to a target dimension $d_p = H \times W$. This global context, expanded per-atom, is concatenated with $C$ and fed to an MLP ($\text{MLP}_{\text{blind}}$) with ReLU activations and BatchNorm1d. The configuration that yielded the main paper results for Blind Pooling ($d_z=16$, Tables~\ref{tab:decoder_reconstruction} \& \ref{tab:diffusion_generation}) is detailed in Table~\ref{tab:app_blind_decoder_params_example}. This table also includes the parameters for the dihedral fine-tuning stage.
Detailed hyperparameter scans for the Blind Pooling decoder MLP varying $H, W,$ and $N_{\text{layers}}$ are presented in Section~\ref{app:blind_pool_search_tech_appendix_detail_ext}.

\begin{table}[htbp]
\small
\centering
\caption{Blind Pooling Decoder Hyperparameters for Main Paper Results ($d_z=16$).}
\label{tab:app_blind_decoder_params_example}
\begin{tabular}{lr}
\toprule
Parameter                 & Value                   \\
\midrule
Encoder Latent Dim ($d_z$)  & 16                       \\
Conditioner ($C$)           & $\Zref$                   \\
Pooling Type              & Blind                   \\
Pooling Height ($H$)        & 50                       \\
Pooling Width ($W$)         & 2                        \\
Pooled Dimension ($d_p$)    & 100                      \\
MLP Hidden Dimension        & 128                      \\
MLP Total Layers ($N_{\text{layers}}$) & 12                      \\
MLP Activation              & ReLU                    \\
MLP Normalization           & BatchNorm1d             \\
Optimizer                 & Adam                    \\
Learning Rate (Initial)     & 3e-4                    \\
Batch Size                & 16                       \\
Epochs (Initial Training)   & 5000                    \\
\midrule
\multicolumn{2}{l}{\textit{Fine-tuning with Dihedral Loss (for Table~\ref{tab:decoder_reconstruction})}} \\
Base Coord Weight ($w_{\text{base}}$) & 1.0                     \\
Dihedral Loss Enabled       & Yes                     \\
Stochastic Fraction ($f_{\text{dih}}$)& 0.1                     \\
Dihedral Divergence Type    & KL                      \\
Dihedral MSE Weight ($\lambda_{\text{mse}}$) & 0.00                    \\
Dihedral Div Weight ($\lambda_{\text{div}}$) & 2.00                    \\
Epochs (Fine-tuning)        & 2000                    \\
Torsion Info File           & condensed\_residues.json \\
\bottomrule
\end{tabular}
\end{table}

\paragraph{Sequential Pooling Strategy Details ($d_z=8$)}
This strategy involves a BackboneDecoder followed by a SidechainDecoder.
\begin{itemize}
    \item \textbf{Backbone Decoder Stage Configuration (for Tables~\ref{tab:decoder_reconstruction} \& \ref{tab:diffusion_generation}):}
        \begin{itemize}
            \item Pooling: Backbone-specific atom embeddings ($d_z=8$) were pooled to $d_{p,bb} = H_{bb} \times W_{bb} = 45 \times 3 = 135$.

        \end{itemize}
    \item \textbf{Sidechain Decoder Stage Configuration (for Tables~\ref{tab:decoder_reconstruction} \& \ref{tab:diffusion_generation}):}
        \begin{itemize}
            \item Architecture: \texttt{arch\_type=1} (see Appendix~\ref{app:sequential_sidechain_decoder_details}).
            \item Pooling: Sidechain-specific atom embeddings ($d_z=8$) were pooled to $d_{p,sc} = H_{sc} \times W_{sc} = 54 \times 2 = 108$.
        \end{itemize}
\end{itemize}

\paragraph{Residue Pooling Strategy Details ($d_z=4$)}
Per-residue atom embeddings $Z_{R_j}^{(t)}$ are pooled to local contexts $z_{\mathrm{res},j}$.
\begin{itemize}
    \item \textbf{Decoder Configuration (for Tables~\ref{tab:decoder_reconstruction} \& \ref{tab:diffusion_generation}):}
        \begin{itemize}
            \item Pooling: For each of $N_{\text{res}}=273$ residues, embeddings ($d_z=4$) were pooled to $d_p' = 1 \times W = 1 \times 4 = 4$.
        \end{itemize}
\end{itemize}

\subsection{Latent Diffusion Models: Configuration and Training Overview}
\label{app:diffusion_models_config_overview_detail_ext}The Denoising Diffusion Probabilistic Models (DDPMs)~\citep{ho2020denoising}, as outlined in Appendix~\ref{app:diffusion_model_methodology}, were implemented to operate on the pooled latent embeddings $\mathbf{h}_0$ derived from each specific pooling strategy (with dimensions $d_p=100$ for Blind, $d_p=135$ for Sequential Backbone, $d_p=108$ for Sequential Sidechain, and $d_p=1092$ for Residue, as detailed in Section~\ref{app:decoder_configs_main_results_ext}). Prior to input, these $\mathbf{h}_0$ embeddings were normalized (zero mean, unit variance) based on statistics from the training set.

\paragraph{Denoising Network Architectures ($\epsilon_{\theta}$)}
The primary denoising network used for generating the main paper results was a multi-layer perceptron, \texttt{DiffusionMLP\_v2}, consisting of four linear layers with ReLU activations. This network takes the flattened, noisy latent vector $\mathbf{h}_t$ concatenated with the normalized timestep $t/T$ as input.
\begin{itemize}
    \item For \textbf{Blind Pooling ($d_p=100$)} and \textbf{Sequential Pooling} (for both backbone $d_{p,bb}=135$ and sidechain $d_{p,sc}=108$ latents, typically handled by separate diffusion models or a model adapted for the concatenated dimension if combined), the MLP hidden dimension was set to 1024.
    \item For \textbf{Residue Pooling} (operating on the effective concatenated input of $d_p = 1092$), the MLP hidden dimension was increased to 4096 to accommodate the larger input dimensionality.
\end{itemize}
Alternative MLP architectures (e.g., \texttt{DiffusionMLP}, \texttt{DiffusionMLP\_v3} with varying layer counts and hidden sizes) and a \texttt{DiffusionConv2D} model (treating the pooled latent as an image-like tensor) were also explored, particularly during the hyperparameter scans for the Blind Pooling strategy (see Section~\ref{app:diffusion_models_blind_scan_tables_ext}).

\paragraph{Diffusion Process and Training Details}
The DDPMs were trained by minimizing the MSE loss between the true and predicted noise (Eq.~\ref{eq:diffusion_loss_main}).
\begin{itemize}
    \item \textbf{Variance Schedule:} A linear variance schedule was used for $\beta_t$, progressing from $\beta_{\text{start}}$ to $\beta_{\text{end}}$ over $T$ timesteps.
    \item \textbf{Key Hyperparameters (Typical Values for Main Results):}
    \begin{itemize}
        \item Total diffusion timesteps ($T$): Commonly set around 500 (e.g., for Blind/Sequential) or higher (e.g., 1500 for Residue, reflecting findings from grid searches).
        \item $\beta_{\text{start}}$: Typically $5 \times 10^{-6}$ or $0.005$.
        \item $\beta_{\text{end}}$: Typically $0.02$ or $0.1$.
    \end{itemize}
    Specific optimal values were determined through grid searches, examples of which are shown in Section~\ref{app:diffusion_models_blind_scan_tables_ext}.
    \item \textbf{Optimization:} The Adam optimizer was used with a learning rate typically set to $1 \times 10^{-5}$.
    \item \textbf{Training Epochs:} Blind and Sequential pooling diffusion models were trained for approximately 3,000-10,000 epochs. The Residue pooling diffusion model, with its larger latent space, often required more extensive training, up to 150,000 epochs or more, to converge.
    \item \textbf{Data Handling:} Pooled latent embeddings ($\mathbf{h}_0$) were generated from the entire training set using the best-performing frozen encoder and the respective trained decoder's pooling mechanism, then saved to HDF5 files for efficient loading during diffusion model training. For Residue pooling, this involved concatenating the $N_{\text{res}}$ per-residue pooled vectors for each frame.
    \item \textbf{Checkpointing:} Model checkpoints were saved periodically (e.g., every 50 or 100 epochs) to select the best performing model based on validation metrics (if a validation set of latents was used) or training loss convergence. For Residue pooling, which benefited from multi-epoch sampling for visualization (Appendix~\ref{app:residue_pooling_inference_details}), more frequent checkpointing was sometimes employed.
\end{itemize}

\paragraph{Sampling}
New latent embeddings $\mathbf{h}_0^{\text{gen}}$ were generated by the trained $\epsilon_\theta$ model using the ancestral sampling procedure described in Algorithm~\ref{alg:app_sampling_diffusion_revised}. These sampled latents were then un-normalized (using the saved training set statistics) before being passed to the corresponding frozen decoder to generate all-atom coordinates $\Xpred^{\text{gen}}$.

\subsection{Blind Pooling Decoder Hyperparameter Scan Results}
\label{app:blind_pool_search_tech_appendix_detail_ext} 

Full test set Mean Squared Error (MSE) results ($MSE_{bb}$ for backbone, $MSE_{sc}$ for sidechain) for the 50 Blind Pooling decoder configurations tested for each encoder latent dimension ($d_z \in \{4, 8, 16\}$) are presented in Tables~\ref{tab:tech_appendix_blind_pool_dz4_detail}-\ref{tab:tech_appendix_blind_pool_dz16_detail}. These experiments systematically varied the pooling height ($H$), pooling width ($W$), and the number of layers in the prediction MLP ($D \equiv N_{layers}$).

\begin{table}[hbtp]
\centering
\caption{Extended Technical Appendix: Full blind pooling Decoder Test MSE Results ($d_z=4$).}
\label{tab:tech_appendix_blind_pool_dz4_detail}
\tiny
\pgfplotstabletypeset[columns={H,W,D,Test_BB,Test_SC}, header=true, col sep=comma, columns/H/.style={column name=$H$, int detect}, columns/W/.style={column name=$W$, int detect}, columns/D/.style={column name=$N_{layers}$, int detect}, columns/Test_BB/.style={column name=$MSE_{bb}$, fixed zerofill, precision=4}, columns/Test_SC/.style={column name=$MSE_{sc}$, fixed zerofill, precision=4}, every head row/.style={before row=\toprule, after row=\midrule}, every last row/.style={after row=\bottomrule}]\blindpoolDzFourData
\end{table}
\FloatBarrier

\begin{table}[hbtp]
\centering
\caption{Extended Technical Appendix: Full Blind Pooling Decoder Test MSE Results ($d_z=8$).}
\label{tab:tech_appendix_blind_pool_dz8_detail}
\tiny
\pgfplotstabletypeset[columns={H,W,D,Test_BB,Test_SC}, header=true, col sep=comma, columns/H/.style={column name=$H$, int detect}, columns/W/.style={column name=$W$, int detect}, columns/D/.style={column name=$N_{layers}$, int detect}, columns/Test_BB/.style={column name=$MSE_{bb}$, fixed zerofill, precision=4}, columns/Test_SC/.style={column name=$MSE_{sc}$, fixed zerofill, precision=4}, every head row/.style={before row=\toprule, after row=\midrule}, every last row/.style={after row=\bottomrule}]\blindpoolDzEightData
\end{table}
\FloatBarrier

\begin{table}[hbtp]
\centering
\caption{Extended Technical Appendix: Full Blind Pooling Decoder Test MSE Results ($d_z=16$).}
\label{tab:tech_appendix_blind_pool_dz16_detail}
\tiny
\pgfplotstabletypeset[columns={H,W,D,Test_BB,Test_SC}, header=true, col sep=comma, columns/H/.style={column name=$H$, int detect}, columns/W/.style={column name=$W$, int detect}, columns/D/.style={column name=$N_{layers}$, int detect}, columns/Test_BB/.style={column name=$MSE_{bb}$, fixed zerofill, precision=4}, columns/Test_SC/.style={column name=$MSE_{sc}$, fixed zerofill, precision=4}, every head row/.style={before row=\toprule, after row=\midrule}, every last row/.style={after row=\bottomrule}]\blindpoolDzSixteenData
\end{table}
\FloatBarrier

\subsection{Diffusion Model Hyperparameter Scans: Blind Pooling}
\label{app:diffusion_models_blind_scan_tables_ext}

Extensive hyperparameter optimization was performed for the latent diffusion models across the different pooling strategies and denoiser architectures detailed above. For brevity, and to illustrate the typical scope of these optimizations, the detailed tables that follow (Tables~\ref{tab:tech_app_diffusion_mlp_dz4_detail}-\ref{tab:tech_app_diffusion_conv_dz16_detail}) focus on presenting results for the \textbf{Blind Pooling} strategy. These tables showcase performance across varying encoder latent dimensions ($d_z \in \{4, 8, 16\}$), for both MLP (specifically \texttt{DiffusionMLP\_v2}) and Conv2D denoiser architectures, against a grid of diffusion hyperparameters (total timesteps $T$, $\beta_{\text{start}}$, and $\beta_{\text{end}}$). While not all combinations are exhaustively tabulated for every pooling method in this appendix, similar systematic grid search approaches were applied to optimize the Sequential and Residue pooling diffusion models, tailoring parameter ranges and model configurations (e.g., MLP hidden dimensions, training epochs) according to their specific input characteristics and computational demands.

\begin{table}[hbtp]
\centering
\caption{Extended Technical Appendix: Full Diffusion Grid Results (blind pooling, $d_z=4, H=30, W=2$, MLP Denoiser). "(Rec.)" denotes Decoder Reconstruction from ground-truth latents; "(Diff.)" denotes full Diffusion pipeline results.}
\label{tab:tech_app_diffusion_mlp_dz4_detail}
\tiny \renewcommand{\arraystretch}{0.8} \setlength{\tabcolsep}{3pt}
\pgfplotstabletypeset[
    columns={exp_id,lddt_full_all_mean,lddt_full_bb_mean,tm_full_all_mean,tm_full_bb_mean,lddt_diff_all_mean,lddt_diff_bb_mean,tm_diff_all_mean,tm_diff_bb_mean},
    header=true,
    col sep=comma,
    columns/exp_id/.style={column name=Exp ID, int detect},
    columns/lddt_full_all_mean/.style={column name=lDDT\textsubscript{All} (Rec.), fixed zerofill, precision=3},
    columns/lddt_full_bb_mean/.style={column name=lDDT\textsubscript{BB} (Rec.), fixed zerofill, precision=3},
    columns/tm_full_all_mean/.style={column name=TM\textsubscript{All} (Rec.), fixed zerofill, precision=3},
    columns/tm_full_bb_mean/.style={column name=TM\textsubscript{BB} (Rec.), fixed zerofill, precision=3},
    columns/lddt_diff_all_mean/.style={column name=lDDT\textsubscript{All} (Diff.), fixed zerofill, precision=3},
    columns/lddt_diff_bb_mean/.style={column name=lDDT\textsubscript{BB} (Diff.), fixed zerofill, precision=3},
    columns/tm_diff_all_mean/.style={column name=TM\textsubscript{All} (Diff.), fixed zerofill, precision=3},
    columns/tm_diff_bb_mean/.style={column name=TM\textsubscript{BB} (Diff.), fixed zerofill, precision=3},
    every head row/.style={before row=\toprule, after row=\midrule},
    every last row/.style={after row=\bottomrule}
]\diffusionMLPDzFourHThirtyWTwo
\end{table}
\FloatBarrier

\begin{table}[hbtp]
\centering
\caption{Extended Technical Appendix: Full Diffusion Grid Results (blind pooling, $d_z=4, H=30, W=2$, Conv2D Denoiser). "(Rec.)" denotes Decoder Reconstruction from ground-truth latents; "(Diff.)" denotes full Diffusion pipeline results.}
\label{tab:tech_app_diffusion_conv_dz4_detail}
\tiny \renewcommand{\arraystretch}{0.8} \setlength{\tabcolsep}{3pt}
\pgfplotstabletypeset[
    columns={exp_id,lddt_full_all_mean,lddt_full_bb_mean,tm_full_all_mean,tm_full_bb_mean,lddt_diff_all_mean,lddt_diff_bb_mean,tm_diff_all_mean,tm_diff_bb_mean},
    header=true,
    col sep=comma,
    columns/exp_id/.style={column name=Exp ID, int detect},
    columns/lddt_full_all_mean/.style={column name=lDDT\textsubscript{All} (Rec.), fixed zerofill, precision=3},
    columns/lddt_full_bb_mean/.style={column name=lDDT\textsubscript{BB} (Rec.), fixed zerofill, precision=3},
    columns/tm_full_all_mean/.style={column name=TM\textsubscript{All} (Rec.), fixed zerofill, precision=3},
    columns/tm_full_bb_mean/.style={column name=TM\textsubscript{BB} (Rec.), fixed zerofill, precision=3},
    columns/lddt_diff_all_mean/.style={column name=lDDT\textsubscript{All} (Diff.), fixed zerofill, precision=3},
    columns/lddt_diff_bb_mean/.style={column name=lDDT\textsubscript{BB} (Diff.), fixed zerofill, precision=3},
    columns/tm_diff_all_mean/.style={column name=TM\textsubscript{All} (Diff.), fixed zerofill, precision=3},
    columns/tm_diff_bb_mean/.style={column name=TM\textsubscript{BB} (Diff.), fixed zerofill, precision=3},
    every head row/.style={before row=\toprule, after row=\midrule},
    every last row/.style={after row=\bottomrule}
]\diffusionConvDzFourHThirtyWTwo
\end{table}
\FloatBarrier

\begin{table}[hbtp]
\centering
\caption{Extended Technical Appendix: Full Diffusion Grid Results (blind pooling, $d_z=8, H=35, W=2$, MLP Denoiser). "(Rec.)" denotes Decoder Reconstruction from ground-truth latents; "(Diff.)" denotes full Diffusion pipeline results.}
\label{tab:tech_app_diffusion_mlp_dz8_detail}
\tiny \renewcommand{\arraystretch}{0.8} \setlength{\tabcolsep}{3pt}
\pgfplotstabletypeset[
    columns={exp_id,lddt_full_all_mean,lddt_full_bb_mean,tm_full_all_mean,tm_full_bb_mean,lddt_diff_all_mean,lddt_diff_bb_mean,tm_diff_all_mean,tm_diff_bb_mean},
    header=true,
    col sep=comma,
    columns/exp_id/.style={column name=Exp ID, int detect},
    columns/lddt_full_all_mean/.style={column name=lDDT\textsubscript{All} (Rec.), fixed zerofill, precision=3},
    columns/lddt_full_bb_mean/.style={column name=lDDT\textsubscript{BB} (Rec.), fixed zerofill, precision=3},
    columns/tm_full_all_mean/.style={column name=TM\textsubscript{All} (Rec.), fixed zerofill, precision=3},
    columns/tm_full_bb_mean/.style={column name=TM\textsubscript{BB} (Rec.), fixed zerofill, precision=3},
    columns/lddt_diff_all_mean/.style={column name=lDDT\textsubscript{All} (Diff.), fixed zerofill, precision=3},
    columns/lddt_diff_bb_mean/.style={column name=lDDT\textsubscript{BB} (Diff.), fixed zerofill, precision=3},
    columns/tm_diff_all_mean/.style={column name=TM\textsubscript{All} (Diff.), fixed zerofill, precision=3},
    columns/tm_diff_bb_mean/.style={column name=TM\textsubscript{BB} (Diff.), fixed zerofill, precision=3},
    every head row/.style={before row=\toprule, after row=\midrule},
    every last row/.style={after row=\bottomrule}
]\diffusionMLPDzEightHThirtyFiveWTwo
\end{table}
\FloatBarrier

\begin{table}[hbtp]
\centering
\caption{Extended Technical Appendix: Full Diffusion Grid Results (blind pooling, $d_z=8, H=35, W=2$, Conv2D Denoiser). "(Rec.)" denotes Decoder Reconstruction from ground-truth latents; "(Diff.)" denotes full Diffusion pipeline results.}
\label{tab:tech_app_diffusion_conv_dz8_detail}
\tiny \renewcommand{\arraystretch}{0.8} \setlength{\tabcolsep}{3pt}
\pgfplotstabletypeset[
    columns={exp_id,lddt_full_all_mean,lddt_full_bb_mean,tm_full_all_mean,tm_full_bb_mean,lddt_diff_all_mean,lddt_diff_bb_mean,tm_diff_all_mean,tm_diff_bb_mean},
    header=true,
    col sep=comma,
    columns/exp_id/.style={column name=Exp ID, int detect},
    columns/lddt_full_all_mean/.style={column name=lDDT\textsubscript{All} (Rec.), fixed zerofill, precision=3},
    columns/lddt_full_bb_mean/.style={column name=lDDT\textsubscript{BB} (Rec.), fixed zerofill, precision=3},
    columns/tm_full_all_mean/.style={column name=TM\textsubscript{All} (Rec.), fixed zerofill, precision=3},
    columns/tm_full_bb_mean/.style={column name=TM\textsubscript{BB} (Rec.), fixed zerofill, precision=3},
    columns/lddt_diff_all_mean/.style={column name=lDDT\textsubscript{All} (Diff.), fixed zerofill, precision=3},
    columns/lddt_diff_bb_mean/.style={column name=lDDT\textsubscript{BB} (Diff.), fixed zerofill, precision=3},
    columns/tm_diff_all_mean/.style={column name=TM\textsubscript{All} (Diff.), fixed zerofill, precision=3},
    columns/tm_diff_bb_mean/.style={column name=TM\textsubscript{BB} (Diff.), fixed zerofill, precision=3},
    every head row/.style={before row=\toprule, after row=\midrule},
    every last row/.style={after row=\bottomrule}
]\diffusionConvDzEightHThirtyFiveWTwo
\end{table}
\FloatBarrier

\begin{table}[hbtp]
\centering
\caption{Extended Technical Appendix: Full Diffusion Grid Results (blind pooling, $d_z=16, H=50, W=2$, MLP Denoiser). "(Rec.)" denotes Decoder Reconstruction from ground-truth latents; "(Diff.)" denotes full Diffusion pipeline results.}
\label{tab:tech_app_diffusion_mlp_dz16_detail}
\tiny \renewcommand{\arraystretch}{0.8} \setlength{\tabcolsep}{3pt}
\pgfplotstabletypeset[
    columns={exp_id,lddt_full_all_mean,lddt_full_bb_mean,tm_full_all_mean,tm_full_bb_mean,lddt_diff_all_mean,lddt_diff_bb_mean,tm_diff_all_mean,tm_diff_bb_mean},
    header=true,
    col sep=comma,
    columns/exp_id/.style={column name=Exp ID, int detect},
    columns/lddt_full_all_mean/.style={column name=lDDT\textsubscript{All} (Rec.), fixed zerofill, precision=3},
    columns/lddt_full_bb_mean/.style={column name=lDDT\textsubscript{BB} (Rec.), fixed zerofill, precision=3},
    columns/tm_full_all_mean/.style={column name=TM\textsubscript{All} (Rec.), fixed zerofill, precision=3},
    columns/tm_full_bb_mean/.style={column name=TM\textsubscript{BB} (Rec.), fixed zerofill, precision=3},
    columns/lddt_diff_all_mean/.style={column name=lDDT\textsubscript{All} (Diff.), fixed zerofill, precision=3},
    columns/lddt_diff_bb_mean/.style={column name=lDDT\textsubscript{BB} (Diff.), fixed zerofill, precision=3},
    columns/tm_diff_all_mean/.style={column name=TM\textsubscript{All} (Diff.), fixed zerofill, precision=3},
    columns/tm_diff_bb_mean/.style={column name=TM\textsubscript{BB} (Diff.), fixed zerofill, precision=3},
    every head row/.style={before row=\toprule, after row=\midrule},
    every last row/.style={after row=\bottomrule}
]\diffusionMLPDzSixteenHFiftyWTwo
\end{table}
\FloatBarrier

\begin{table}[hbtp]
\centering
\caption{Extended Technical Appendix: Full Diffusion Grid Results (blind pooling, $d_z=16, H=50, W=2$, Conv2D Denoiser). "(Rec.)" denotes Decoder Reconstruction from ground-truth latents; "(Diff.)" denotes full Diffusion pipeline results.}
\label{tab:tech_app_diffusion_conv_dz16_detail}
\tiny \renewcommand{\arraystretch}{0.8} \setlength{\tabcolsep}{3pt}
\pgfplotstabletypeset[
    columns={exp_id,lddt_full_all_mean,lddt_full_bb_mean,tm_full_all_mean,tm_full_bb_mean,lddt_diff_all_mean,lddt_diff_bb_mean,tm_diff_all_mean,tm_diff_bb_mean},
    header=true,
    col sep=comma,
    columns/exp_id/.style={column name=Exp ID, int detect},
    columns/lddt_full_all_mean/.style={column name=lDDT\textsubscript{All} (Rec.), fixed zerofill, precision=3},
    columns/lddt_full_bb_mean/.style={column name=lDDT\textsubscript{BB} (Rec.), fixed zerofill, precision=3},
    columns/tm_full_all_mean/.style={column name=TM\textsubscript{All} (Rec.), fixed zerofill, precision=3},
    columns/tm_full_bb_mean/.style={column name=TM\textsubscript{BB} (Rec.), fixed zerofill, precision=3},
    columns/lddt_diff_all_mean/.style={column name=lDDT\textsubscript{All} (Diff.), fixed zerofill, precision=3},
    columns/lddt_diff_bb_mean/.style={column name=lDDT\textsubscript{BB} (Diff.), fixed zerofill, precision=3},
    columns/tm_diff_all_mean/.style={column name=TM\textsubscript{All} (Diff.), fixed zerofill, precision=3},
    columns/tm_diff_bb_mean/.style={column name=TM\textsubscript{BB} (Diff.), fixed zerofill, precision=3},
    every head row/.style={before row=\toprule, after row=\midrule},
    every last row/.style={after row=\bottomrule}
]\diffusionConvDzSixteenHFiftyWTwo
\end{table}
\FloatBarrier

\subsection{Conditioning Mechanism Tuning}
\label{app:conditioning_tuning_details_ext}

As detailed in Section~\ref{sec:encoder_method_main} (Conditioning Mechanism paragraph), the decoder component of LD-FPG is conditioned on information derived from a reference structure, $\Xref$. Our primary conditioning strategy utilizes the latent representation of this reference, $C = \Zref = \Theta^{*}(\Xref, E_{\text{ref}})$, obtained from the pre-trained and frozen encoder $\Theta^{*}$. This choice was motivated by preliminary findings suggesting its superiority over direct conditioning on raw Cartesian coordinates, $C = \Xref$.

To rigorously validate this and quantify the performance difference, we conducted comprehensive ablation studies. These studies compared models conditioned on $\Zref$ (indicated by the \texttt{\_zref} suffix in their experimental set identifiers) against baseline models conditioned on $\Xref$. The comparisons spanned different architectural configurations of the decoder, including:

\begin{itemize}
    \item \textbf{Pooling Strategy:} Experiments were run using both Blind Pooling and Residue Pooling decoders, as described in Section~\ref{sec:decoders}. The Residue Pooling strategy itself had variants explored: a standard \texttt{Residue (1level)} application and a hierarchical \texttt{Residue (2level)} approach. In the \texttt{Residue (2level)} configuration, the processing involved a two-stage pooling scheme: atom-wise latent embeddings $\Zlatent^{(t)}$ first underwent a global blind pooling operation. The output context vector from this initial global pooling was then subsequently used as an input or integrated within a residue-specific pooling or processing mechanism before the final MLP prediction for each residue's atoms.
    \item \textbf{Cross-Attention (CA):} For some configurations, a Cross-Attention mechanism (denoted \texttt{\_CA\_on}) was employed. This module, defined as:

\begin{verbatim}
class CrossAttentionBlock(nn.Module):
    def __init__(self, q_dim, kv_dim, att_dim=64):
        super().__init__(); self.s = att_dim**-0.5
        self.q=nn.Linear(q_dim,att_dim,bias=False)
        self.k=nn.Linear(kv_dim,att_dim,bias=False)
        self.v=nn.Linear(kv_dim,att_dim,bias=False)
    def forward(self, q, k, v):
        Q=self.q(q); K=self.k(k); V=self.v(v)
        sc = torch.matmul(Q, K.transpose(-1, -2)) * self.s
        return torch.matmul(F.softmax(sc, dim=-1), V)
\end{verbatim}

    allows the pooled latent embeddings derived from the input dynamic conformation ($\mathbf{h}_0$) to interact with and selectively integrate information from the reference structure representation (e.g., $\Zref$). Configurations without this mechanism (\texttt{\_CA\_off}) typically use simpler concatenation or direct feeding of $\mathbf{h}_0$ and the conditioner to the subsequent MLP.
\end{itemize}

It is important to note that these ablation studies comparing conditioning mechanisms were conducted with other architectural and training hyperparameters (beyond those swept for $H, W, H2, W2$ within each group, such as encoder specifics or base learning rate schedules) fixed to a representative set; this foundational parameter set is not discussed in detail here.

The optimizer was generally Adam, except for one set of experiments explicitly testing AdamW (e.g., \texttt{blind\_res\_CA\_off\_zref\_lr\_Adamw}). The primary metric for comparison was the minimum test loss achieved across an extensive hyperparameter search (varying decoder-specific pooling dimensions $H, W$ and MLP architecture parameters $H2, W2$) within each experimental group.

\begin{table}[htbp]
\centering
\caption{Ablation Study Results for Decoder Conditioning Mechanism. Performance is compared based on minimum test loss. `$\Xref$ Cond.` refers to conditioning on raw Cartesian coordinates, while `$\Zref$ Cond.` refers to conditioning on latent reference embeddings. ``1level'' and ``2level'' refer to specific variants of the Residue Pooling architecture, with ``2level'' involving an initial global blind pooling stage prior to residue-specific processing.}
\label{tab:conditioning_ablation_results}
\small
\renewcommand{\arraystretch}{1.1}
\begin{tabular}{@{}llccc@{}}
\toprule
Pooling     & Cross-Attention & $\Xref$ Cond. & $\Zref$ Cond. & Improvement (\%) \\
\cmidrule(r){1-1} \cmidrule(lr){2-2} \cmidrule(lr){3-3} \cmidrule(lr){4-4} \cmidrule(l){5-5}
Blind       & Off             & 0.7763 (15)   & 0.5327 (32)   & 31.4 \\
Blind       & On              & 1.6787 (16)   & 0.5835 (24)   & 65.2 \\
\midrule
Residue     & On              & 1.2417 (6)    & 0.2453 (14)   & 80.2 \\
Residue (1level) & Off        & 0.2290 (19)   & 0.1058 (10)   & 53.8 \\
Residue (2level) & Off        & 0.7599 (20)   & 0.5749 (49)   & 24.3 \\
\bottomrule
\end{tabular}
\end{table}

The results presented in Table~\ref{tab:conditioning_ablation_results} consistently show that conditioning on the latent reference embeddings ($\Zref$) leads to substantially lower minimum test losses compared to conditioning on raw Cartesian coordinates ($\Xref$). This performance advantage holds across both Blind and Residue pooling strategies (including its ``1level'' and ``2level'' variants) and is evident whether Cross-Attention is used or not. For instance, with Blind Pooling and CA off, $\Zref$ conditioning achieved a $\sim$31.4\% lower loss. This effect was even more pronounced for Residue Pooling (CA On), where an $\sim$80.2\% improvement was observed. Even for the more complex Residue (2level) architecture, $\Zref$ conditioning provided a notable $\sim$24.3\% reduction in loss.

The single experimental group \texttt{blind\_res\_CA\_off\_zref\_lr\_Adamw} (16 runs, Min Test Loss 0.7592), which used $\Zref$ conditioning with the AdamW optimizer, yielded a higher loss than its Adam-optimized counterpart (\texttt{blind\_res\_CA\_off\_zref}, Min Test Loss 0.5327). While this suggests that Adam may have been more suitable or better tuned for this particular setup, it does not detract from the central finding regarding the conditioning variable itself, as no corresponding $\Xref$-conditioned AdamW experiments were available for direct comparison.

In conclusion, these ablation studies provide robust quantitative evidence supporting the design choice of using the encoder-derived latent representation $\Zref$ as the primary conditioning mechanism for the decoder in LD-FPG. This strategy enables the model to more effectively learn and represent conformational deformations relative to a well-characterized reference state, leading to improved generative performance.

\subsection{Discussion on Metric Variability and Standard Deviations}
\label{app:metric_variability_sd_ext}

Throughout this paper, our evaluation process involves training each model component (encoder, decoder, diffusion model) and then assessing its performance by generating an ensemble of protein conformations. These generated ensembles are then compared against the reference Molecular Dynamics (MD) ensemble, with key metrics such as lDDT, TM-score, and Jensen-Shannon Divergence (JSD) for dihedral angles reported. For clarity and conciseness, the main summary tables (e.g., Table~\ref{tab:decoder_reconstruction} and Table~\ref{tab:diffusion_generation}) present the mean values for per-structure metrics like lDDT and TM-score. This section provides further context on the variability observed in these metrics, typically characterized by their standard deviations (SD).

While standard deviations for metrics like lDDT and TM-score offer insight into the spread of structural quality within an ensemble, we argue that the $\sum$JSD values for backbone and side-chain dihedral angles (reported directly in Tables~\ref{tab:decoder_reconstruction} and \ref{tab:diffusion_generation}) serve as a more direct and potent indicator of how well the *distributional variability* and local conformational diversity of the reference MD ensemble are captured. A low $\sum$JSD signifies that the variety and relative populations of local backbone ($\phi, \psi$) and side-chain ($\chi$) conformations in the generated ensemble closely mirror those of the MD data. Furthermore, qualitative assessments of landscape coverage, such as the Principal Component Analysis (PCA) of latent embeddings (Figure~\ref{fig:pca_pmf_analysis}, top row) and the A100 activation index distributions (Figure~\ref{fig:pca_pmf_analysis}, bottom row), provide richer, more nuanced views of overall conformational diversity and the exploration of functionally relevant states than a single SD value of a global structural similarity score might offer.

Nevertheless, to provide a comprehensive picture, we report the observed standard deviations for lDDT\textsubscript{All} and TM-score\textsubscript{All} across different stages of our pipeline, calculated over the entire test set ensemble for each respective model configuration:

\begin{itemize}
    \item \textbf{lDDT\textsubscript{All} Standard Deviations:}
    \begin{itemize}
        \item Ground Truth MD Ensemble (vs. $\Xref$): SD $\approx 0.028$
        \item ChebNet Encoder Reconstruction Head (Table~\ref{tab:encoder_performance}): SD $\approx 0.030$
        \item Decoder-Generated Ensembles (Table~\ref{tab:decoder_reconstruction}, from ground-truth latents):
        \begin{itemize}
            \item Blind Pooling ($d_z=16$): SD $\approx 0.031$
            \item Sequential Pooling ($d_z=8$): SD $\approx 0.032$
            \item Residue Pooling ($d_z=4$): SD $\approx 0.040$
        \end{itemize}
        \item Diffusion-Generated Ensembles (Table~\ref{tab:diffusion_generation}):
        \begin{itemize}
            \item Blind Pooling: SD $\approx 0.042$
            \item Sequential Pooling: SD $\approx 0.058$
            \item Residue Pooling: SD $\approx 0.091$
        \end{itemize}
    \end{itemize}
    \item \textbf{TM-score\textsubscript{All} Standard Deviations:}
    \begin{itemize}
        \item Ground Truth MD Ensemble (vs. $\Xref$): SD $\approx 0.0059$
        \item ChebNet Encoder Reconstruction Head (Table~\ref{tab:encoder_performance}): SD $\approx 0.0060$
        \item Decoder-Generated Ensembles (Table~\ref{tab:decoder_reconstruction}):
        \begin{itemize}
            \item Blind Pooling ($d_z=16$): SD $\approx 0.0062$
            \item Sequential Pooling ($d_z=8$): SD $\approx 0.0065$
            \item Residue Pooling ($d_z=4$): SD $\approx 0.020$ 
        \end{itemize}
        \item Diffusion-Generated Ensembles (Table~\ref{tab:diffusion_generation}):
        \begin{itemize}
            \item Blind Pooling: SD $\approx 0.015$
            \item Sequential Pooling: SD $\approx 0.025$
            \item Residue Pooling: SD $\approx 0.048$
        \end{itemize}
    \end{itemize}
\end{itemize}

These standard deviations were computed by evaluating the respective metric for each structure in the generated test ensemble against the reference structure ($\Xref$) and then calculating the standard deviation of these scores. The Python script provided (calc\_lddt.py) illustrates the methodology for calculating such mean and standard deviation values for lDDT scores over an ensemble, including options for subsampling if required for very large datasets.

Interpreting these SD values:
The SDs for lDDT and TM-score for the ChebNet encoder and the decoder-generated ensembles (from ground-truth latents) are generally low and quite comparable to the inherent variability observed within the ground truth MD ensemble itself when compared against the static $\Xref$. This indicates that the encoder and decoders maintain a consistent level of structural fidelity.
For the diffusion-generated ensembles, the SD values, particularly for lDDT, show a modest increase. For instance, the SD for lDDT\textsubscript{All} ranges from approximately $0.04$ to $0.09$. This suggests a slightly greater degree of structural heterogeneity in the ensembles produced by the full generative pipeline compared to the direct decoder outputs or the original MD data's deviation from $\Xref$. This increased spread can be acceptable, or even indicative of the model exploring the learned conformational manifold, as long as the mean fidelity remains high (as reported in Table~\ref{tab:diffusion_generation}) and, crucially, the critical distributional features of the ensemble (captured by $\sum$JSD for dihedrals and A100 landscape analysis) are well-reproduced. The very low SDs for TM-score across all stages consistently affirm that the global fold of the protein is well-preserved.

In summary, while mean values are prioritized in the main tables for direct comparison of central tendencies, the analysis of $\sum$JSD for dihedral distributions, PCA of latent spaces, A100 activation profiles, and the contextualized standard deviations discussed here collectively provide a comprehensive assessment of both the fidelity and the diversity of the conformational ensembles generated by LD-FPG.

\newpage

\end{document}